\documentclass[traditabstract]{aa}
\usepackage{float}
\usepackage{graphicx}
\usepackage{txfonts}
\usepackage{placeins}
\newcommand{\Ha} {H$\alpha$}  
\newcommand{\Hb} {H$\beta$}  
\newcommand{\oiii} {[O~III]}

\newcommand{\kms}{$\rm{\,km \,s}^{-1}$}

\begin{document}

   \title{The MURALES survey. IX.}
   \subtitle{MUSE observations and results for 3C radio sources at 0.3 $<$ z $<$ 0.82.}

  \author {Barbara Balmaverde \inst{1},  Alessandro Capetti\inst{1}, Alessio Guglielminotti Canun\inst{2,3}, Francesco Massaro\inst{3,1}, Alessandro Marconi\inst{4,5}}
\institute{
 INAF - Osservatorio
  Astrofisico di Torino, Via Osservatorio 20, I-10025 Pino Torinese,
  Italy 
  \and
  Instituto de Astronomía y Ciencias Planetarias, Universidad de Atacama, Copayapu 485, Copiapó, Chile
   \and
    Dipartimento di Fisica, Universit\`a degli Studi di
  Torino, Via Pietro Giuria 1, 10125 (Torino), Italy
  \and
    Dipartimento di Fisica e Astronomia, Universit\`a di Firenze, Via G. Sansone 1, 50019, Sesto F.no (Firenze), Italy
    \and
    INAF-Osservatorio Astrofisico di Arcetri, Largo E. Fermi 5, 50125, Firenze, Italy
}

  \abstract{This is the ninth paper in the MUse RAdio Loud Emission lines
  Snapshot (MURALES) project series, presenting the results of observations of 3C radio
  sources obtained with the VLT/MUSE integral field spectrograph. Here, we investigate the morphology and kinematics of the extended emission line regions (EELRs) of 26 radio galaxies at intermediate redshift $0.3 < z < 0.82$ and their relation with the radio ejecta. EELRs are observed in all but 4 of the 26 sources extending as far as 160 kpc from the host galaxy. They show different morphologies and kinematics usually related with the radio properties.
  The five sources associated with the smaller radio sources  show highly disturbed gas kinematics. Several EELRs have a filamentary morphology, often forming bubbles inflated by the radio lobes. The more extended EELRs usually have elongated morphologies and show well-ordered rotation.
We then confirm the general results obtained from the study of the source at lower redshift: radio galaxies are surrounded by super-disks, that is, structures of ionized gas on scales of many tens of kiloparsec, 
whose properties depend on a combination of the evolutionary state of the radio source and its  orientation with respect to the super-disks.}


 \titlerunning{Ionization properties of the emission line regions in 3C radio sources at 0.3 $< z <$ 0.82.} \authorrunning{Balmaverde et al.}
\maketitle

\section{Introduction}

Extended emission line regions (EELRs) are  spatially extended  reservoirs of warm  ($\sim$10$^{4}$ K)  ionized gas  associated with  AGNs, extending from the classical narrow-line region to scales of several tens of kiloparsec, and in some systems, beyond the host galaxy. They are commonly characterized by strong high-ionization lines such as the \oiii$\lambda5007$ line, and their morphology ranges from relatively ordered bi-conical structures to clumpy, filamentary, and irregular distributions associated with tidal features. Surveys have identified AGN-ionized clouds at projected distances >10 kpc, with emission detected out to $\sim$30–40 kpc in individual systems (e.g., \citealt{stockton02,keel12,keel17}). Their excitation is most often attributed to photoionization by the hard AGN radiation field, although shocks driven by AGN winds or radio jets, as well as star formation and tidal interactions, can contribute. EELRs therefore provide a unique means of tracing the geometry of escaping ionizing radiation and the distribution of gas on circumgalactic scales and offer complementary probes of the radiative and mechanical impact of AGN on their host galaxies.

 In the past years, we have been carrying out the MUse RAdio Loud Emission lines Snapshot (MURALES) survey of observations of 3C radio sources with the integral field spectrograph MUSE at the Very Large Telescope (VLT). The main aim of the survey is to investigate the AGN feedback effects 
by mapping the EELR around powerful radio galaxies.
We initially observed 37 3C radio  galaxies with $z<0.3$ and $\delta <20^\circ$ (\citealt{balmaverde19,balmaverde21}). These observations revealed warm ionized gas extended to several tens of kiloparsec in most objects, showing ordered rotational disks and complex or disturbed kinematics. 

A clear connection with the radio structures emerged.  The alignment between the radio jet and ionized gas emission  is rather weak, but the line emission can be explained as produced by an interaction between the radio lobes with a rotational gaseous disk. In \citealt{balmaverde22alias} (hereafter paper VI), we found that large-scale ionized gas is seen in all but two of the 26 FR~IIs. The observed emission line structures appear to be associated with gaseous super-disks that likely formed after a gas-rich merger.
Their ionized portion sometimes takes the form of elongated or filamentary structures, typically extending between 10 and 30 kpc, but reaching distances of $\sim$80 kpc.
In some cases, we observed the
formation of gas bubbles that were swept away by the expanding radio
lobe. When the jet escaped from the denser central region, it left behind a punctured emission line bubble that appears  with an arc-like
or S-shaped morphology. 
 The main conclusion is that the  different properties of the EELR can be explained with a combination of  the evolutionary state of the radio source and its  orientation with respect to the super-disks. 

We then extended MURALES to RGs of intermediate redshift (izRGs) observing 26 3C radio sources with $0.3 < z < 0.82$ (there are 29 such sources in the 3C with $\delta <20^\circ$, but the observations of 3 of them were not obtained due to scheduling constraints). The high-redshift threshold was selected so that the MUSE spectra included the [O~III]. From the point of view of their radio structure, all 26 izRGs have an FR~II radio morphology with a size exceeding $\sim$ 25 kpc, with the only exception of 3C~138, a compact steep-spectrum source with a triple structure extending for only $\sim$5 kpc \citep{akujor91}.

In \citet{capetti23} we explored the nuclear properties of these 26 izRGs with the main aim of deriving the optical spectroscopic classification based on diagnostic emission line ratios. 
At redshifts higher than $\sim$0.4, the \Ha\ and [N~II] emission lines are not covered by the MUSE spectra, and alternative diagnostic diagrams (with respect to the standard BPT diagrams, \citealt{baldwin81}) are needed to separate the different spectroscopic subclasses. \citeauthor{capetti23} derived a spectroscopic classification into star-forming high- and low-excitation galaxies (HEGs and LEGs) by only using the ratios of emission lines in the rest-frame UV and the blue portion of the spectra. This classification returned six LEGs and 20 HEGs, including in this class also the eight broad-line objects (BLOs) whose narrow emission line ratios are characteristic of the high-excitation gas of the HEG class. 

We here present and discuss the properties of the EELR in these 26 izRGs. 
The paper is organized as follows. In Sect. \ref{observation} we
present the observations and describe the data reduction. In Section \ref{results} we describe the  properties of the EELRs separating the sources based on their line morphology and kinematics.
In Sect. \ref{connection} we
explore the relation between the radio and line emission morphology and kinematics. The results are discussed in Sect. \ref{discussion}, where we also give our conclusions.
We adopted the following set of cosmological parameters: $H_{\rm
  o}=69.7$ \kms\ Mpc$^{-1}$ and $\Omega_m$=0.286
\citep{bennett14}. 

  \begin{table*}
    \caption{Main properties of the observed galaxies.}             
    \label{table:1}      
    \centering          
    \begin{tabular}{l c r | l r l | r | r r r}
        \hline\hline       
Name &\textit{z}  & Class & \multicolumn{3}{c}{Emission line} & Radio & \multicolumn{3}{c}{Position angle} \\
     &            &       &         & Size & Class  & Size (\arcsec) &  Radio & Line & Kin.\\ 
        \hline                    
3C~016   & 0.406  & LEG   & [O~II]  &    5\arcsec\ ( 27) & Rotator    &  75 & 30   &  --- & -51 \\  
3C~093   & 0.358  & BLO   & [O~III] &               ---  & Unresolved &  35 & 40   &  --- & --- \\
3C~099   & 0.426  & HEG   & [O~II]  &    5\arcsec\ ( 28) & Compact    &   5 & 45   &  30  & --- \\
3C~107   & 0.785  & LEG   & [O~II]  &    5\arcsec\ ( 38) & Bubble     &  20 & 15   &  30  & --- \\
3C~109   & 0.305  & BLO   & [O~III] &   17\arcsec\ (120) & Rotator    & 100 &-35   &  30  &  30 \\
3C~114   & 0.815  & HEG   & [O~III] &    3\arcsec\ ( 23) & Rotator    &  45 & 55   &  80  &  90 \\
3C~138   & 0.759  & BLO   & [O~III] &             ---    & Compact    &  -- & 80   &  --- & --- \\
3C~142.1 & 0.406  & HEG   & [O~II]  &    9\arcsec\ ( 64) & Bubble     & 110 &-50   & -70  & --- \\
3C~175   & 0.768  & BLO   & [O~II]  &    6\arcsec\ ( 45) & ---        &  10 & 55   &  10  & --- \\
3C~187   & 0.465  & HEG   & [O~III] &    3\arcsec\ ( 22) & Rotator    & 175 &-20   &  -5  &   5 \\
3C~207   & 0.684  & BLO   & [O~III] &    3\arcsec\ ( 23) & Rotator    &  12 &-80   &  --- &  -3 \\
3C~215   & 0.411  & BLO   & [O~III] &    7\arcsec\ ( 50) & Bubble     &  48 &-25   & -75  & --- \\
3C~225B  & 0.582  & LEG   & [O~II]  &    6\arcsec\ ( 45) & Comp./Rot. &   5 &-65   & -55  & -54 \\
3C~226   & 0.817  & HEG   & [O~II]  &    7\arcsec\ ( 55) & Rotator    &  30 &-35   &  25  &  -9 \\
3C~228   & 0.552  & HEG   & [O~II]  &    5\arcsec\ ( 35) & ---        &  45 & 10   &  35  & --- \\
3C~275   & 0.480  & HEG   & [O~III] &    2\arcsec\ ( 14) & Compact    &   5 & 50   &  45  & --- \\
3C~275.1 & 0.557  & BLO   & [O~III] &   22\arcsec\ (160) & Rotator    &  15 &-15   &  40  & -48 \\
3C~277.2 & 0.766  & HEG   & [O~III] &   17\arcsec\ (130) & Peculiar   &  55 & 60   &  75  & --- \\
3C~293.1 & 0.709  & LEG   & [O~II]  &    3\arcsec\ ( 23) & Rotator    &  55 & 45   &  --- &  72 \\
3C~306.1 & 0.441  & HEG   & [O~III] &    6\arcsec\ ( 45) & Rotator    &  95 &  0   &  40  &  34 \\
3C~327.1 & 0.462  & HEG   & [O~II]  &    4\arcsec\ ( 30) & ---        & 130 &-75   & -40  & --- \\
3C~334   & 0.555  & BLO   & [O~III] &   28\arcsec\ (200) & Bubble     &  15 &-50   & -35  & --- \\
3C~434   & 0.322  & LEG   & [O~II]  &        ---         & Unresolved &  50 & 80   &   0  & --- \\
3C~435   & 0.471  & LEG   & [O~II]  &   19\arcsec\ (110) & Bubble     &  45 & 55   &  45  & --- \\
3C~455   & 0.542  & HEG   & [O~II]  &    2\arcsec\ ( 12) & Compact    &   4 & 60   &  35  & --- \\
\hline                   
    \end{tabular}
    
    \medskip
    \small{Column description: (1) name, (2) redshift, (3) spectroscopic
      class, (4-6) emission line shown in the images, size of the EELR in arcsecond (and in kiloparsec), emission line class, (7) radio size, and (8 - 10) position angle of the radio, emission line, and
      kinematical major axis.}
\end{table*}

\section{Observations and data reduction}
\label{observation}

The 26 izRGs were observed between October 2020 and April 2021 with the MUSE spectrograph at the VLT as part of the program ID 0106.B-0564(A). In Table \ref{table:1} we report their redshifts, spectral classifications, and their main emission line properties. Additional information, such as radio and line luminosities, can be found in \citet{capetti23}. The observing times increased with redshift and ranged from 1274 to 2848 s (see \citet{capetti23} for more details on the observations). We used the ESO MUSE pipeline (version 1.6.2, \citealt{weilbacher20}) to obtain a fully reduced and calibrated data cube. The results obtained for 3C~313, a radio galaxy with a dual AGN, are presented elsewhere \citep{balmaverde26}. The MUSE field of view of 3C~435 includes a second 3C source (3C~435B; see \citealt{mccarthy89}) with a redshift higher than our distance threshold, z=0.865. We considered this source in our analysis.\footnote{We confirm the presence of strong Balmer absorption lines at zero redshift in its  spectrum. This result was discussed by \citet{mccarthy89} and is indicative of the superposition of a Galactic A star onto the host galaxy.}

The MUSE data have a spatial sampling of 0.2$\arcsec$ per spaxel. The spectral sampling is 1.25 $\AA$ per channel, while the instrumental resolving power varies from R $\approx$ 1770 at 4800 $\AA$ to R $\approx$ 3590 at 9300 $\AA$, corresponding to an instrumental FWHM of approximately 150-180 km s$^{-1}$.
 In each spaxel, we subtracted  the stellar continuum modeled with a linear function since the stellar absorption features were too faint to be fitted with a combination of stellar population models.
 We fitted the emission lines 
with mpfit
(Markwardt 2009) using a single-Gaussian component, except for the nuclear regions of the six BLOs, where we added a broad Balmer component.
 We forced all lines in the blue (namely
[Ne~V]$\lambda$3426, [O~II]$\lambda$$\lambda$3726, 3729, and [Ne~III]$\lambda$3870) and red 
portion (namely, \Hb\ and [O~III]$\lambda$$\lambda$4959,5007)  of the spectra to have the same velocity and velocity dispersion.
For all the emission lines in the spectrum, we produced flux, velocity, and velocity dispersion maps.
When present, we also fit the \Ha\ + [N~II]$\lambda6548,6584$ complex.

In the appendix we show the results obtained by fitting all spaxels in which a given line (shown in each panel title) has a signal-to-noise ratio higher than three. We derived the line surface brightness distribution, the gas velocity, and velocity dispersion. We also show the continuum image obtained in the line-free rest-frame region between 4500 and 4800 \AA  and superpose the radio contours (see Fig. \ref{fig1}). The registration between the MUSE and the radio images was performed using the location of the radio core when this was detected or by correcting the MUSE astrometry based on the location of sources visible in the MUSE images and in those of the
Panoramic Survey Telescope \& Rapid Response
System \citep{chambers16}. The radio images were obtained from NRAO VLA Archive Survey Images Page \footnote{available at https://www.vla.nrao.edu/astro/nvas/}. The main parameters of the images we used are reported in Table \ref{table:radio}.

\section{Results.}
\label{results}

\subsection{Gas nebulae and radio sources on galactic scales.}
\label{CSS}

The radio emission in most of the sources is very extended and reaches sizes of several hundred kiloparsec, for example,
$\sim 650$ kpc in 3C~187. However, in five objects (namely 3C~099, 3C~138, 3C~225B, 3C~275, and 3C~455), the radio source is smaller than $\sim 30$ kpc, similar to that of their host galaxy, and as small as $\sim 0\farcs7$ ($\sim 5$ kpc) in the case of 3C~138 \citep{akujor91}. 

These sources are also characterized by several common properties from the point of view of their gas properties: the EELR extends on a very similar scale to that of the radio emission, and they are very closely aligned to each other to within 20$^\circ$ (except for 3C~138, where the EELR is unresolved). The gas velocity dispersion is very high compared to the larger radio sources, with typical values of 400 \kms\, but it exceeds 1,000 \kms\ in 3C~138, the smaller radio source. Their gas velocity field is not ordered (with the possible exception of 3C~225B) and rather complex. These sources also show an increased velocity dispersion in the direction perpendicular to the radio jets.
A source with similar properties, namely 3C~456, was found in the $z<0.3$ sample presented in paper VI. 

\subsection{Bubbles and arcs of line emission.}
\label{bubbles}

\begin{figure*}
  \centering
\includegraphics[width=0.33\linewidth]{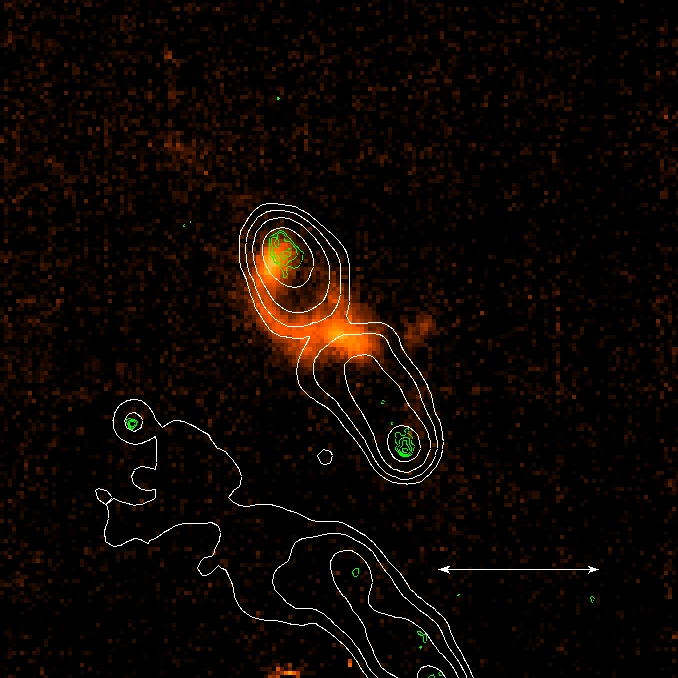}
\includegraphics[width=0.33\linewidth]{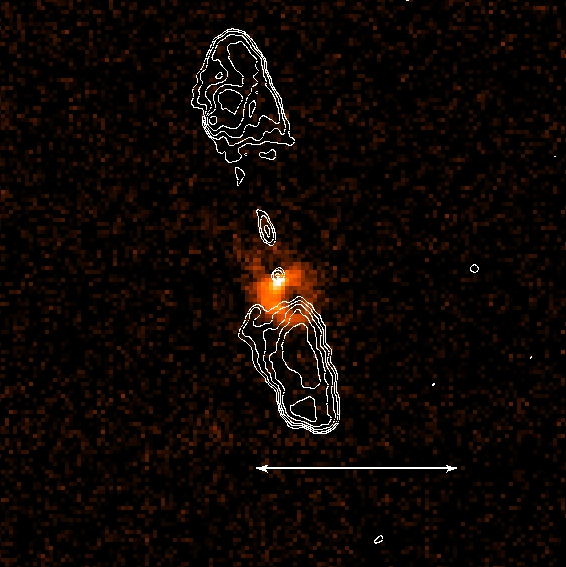}
\includegraphics[width=0.33\linewidth]{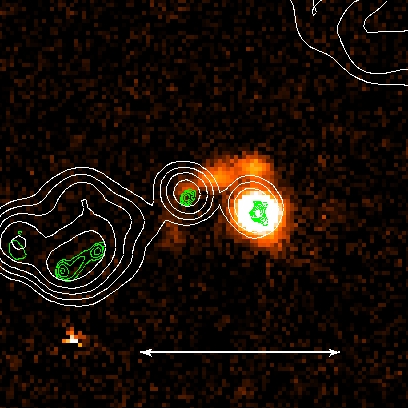}
\caption{Emission line images. We 
  superpose the radio isophotes for 3C~435A, 3C~107, and 3C~215. For 3C~215 and 3C435A we show radio contours at low (white) and high (green) resolution. The arrows in the three panels  are 10\arcsec\ long.}
    \label{fig2}
\end{figure*}

In Fig. \ref{fig2} (left panel) we show the [O~II] emission line image of 3C~435A and superpose the radio contours at two different spatial resolutions. The ionized gas forms a structure that almost completely surrounds the northern radio lobe. The extension of the gas structure is $\sim 7\arcsec$, corresponding to $\sim 40$ kpc. The emission line distribution is highly asymmetric: on the southern side, only a faint filament is detected. This asymmetry is reminiscent of what is observed in 3C~317 \citep{balmaverde18a}. The gas-projected velocity steadily increases away from the host galaxy, reaching $\sim 150$ \kms\ at its northern tip. 

We interpret this structure as a bubble of gas that is swept away by the expanding radio lobe, whose edges seen in projection form these high-brightness regions, similarly to what is observed in 3C~317 and 3C~277.3 \citep{capetti22}. The emission and absorption lines in the spectrum of the four close galaxies,  visible in the continuum image, confirm that they are companions at the same redshift of the RG. The high velocity dispersion $>$200 km $^{-1}$  disfavors a tidal-stream scenario, since the velocity dispersion measured in the gas streams of interacting galaxies
typically shows much lower values \citep{monreal10}.
Two further sources with similar properties, namely 3C~063 and 3C~196.1, were found in the $z<0.3$ sample (see paper VI and \citealt{jimenez22}).

A similar ionized-gas distribution is seen in another izRG, namely, 3C~107, where a bubble-like structure that is aligned with radio axis extends in the southern direction for $\sim 20$ kpc, with a fainter counterpart to the northern side (see Fig. \ref{fig2}, central panel). However, the radio structure is significantly larger ($\sim 60$ kpc) than the EELR in this source. 
In other sources, similar arc-like or S-shaped emission line structures are detected that are orientated along the radio emission, such as in 3C~142.1 and 3C~334. Similarly to 3C~107, the emission line region is smaller than the radio emission in these cases.  

A close connection between radio and an arc-like line emission is seen in 3C~215 (see Fig. \ref{fig2}, right panel). However, the line emission in this source is located along the boundary of the region in which the eastern radio jet bends sharply, which might be indicative of a strong interaction between the relativistic outflow and the external medium, similarly to what is observed in 3C~277.3.

\subsection{The peculiar case of 3C~277.2.}
\label{3c277}

\begin{figure}
  \centering
\includegraphics[width=1\linewidth,height=1\linewidth]
{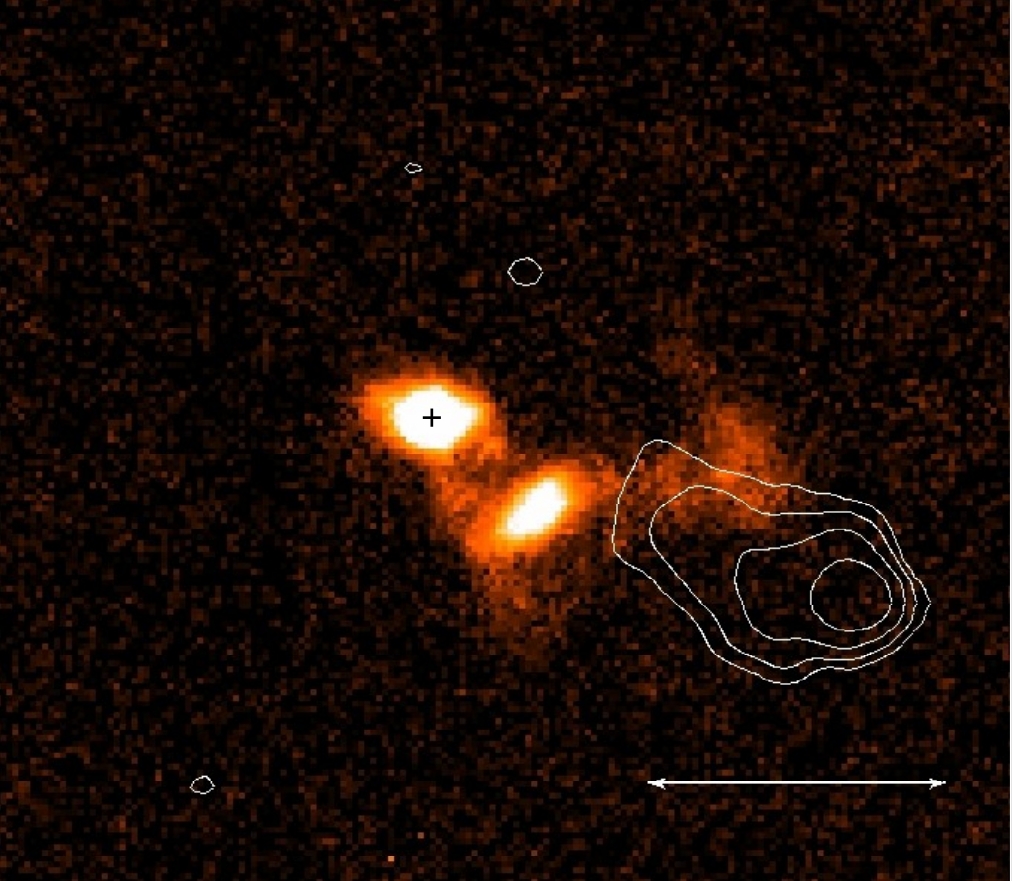}
\caption{Line image with the radio isophotes for 3C~277.2 superposed. The cross marks the location of the host. The arrow is 10\arcsec\ long.}
    \label{3c277fig}
\end{figure}

The source 3C~277.2 represents a unique example from the point of view of its line emission structure and gas velocity field. [O~III] line emission is detected out to $\sim 13\arcsec$ ($\sim$ 100 kpc) from the host galaxy. The gas distribution is highly asymmetrical and detected only in the southwestern region, showing a general arc-like but complex morphology (see Fig \ref{3c277fig}). The radio source is also strongly asymmetric, with the northeastern radio arm being more than twice as long as its opposite. This suggests that the SW lobe encounters regions of higher density that slow its expansion. However, the radio and line emission are not closely connected spatially.

The most notable property of this gas nebula is its velocity field. It shows a general redshift with respect to the host galaxy by more than 1,000 \kms. The velocity dispersion is also locally very high for such an off-nuclear regions, reaching values as high as 300-400 \kms. The origin of this structure is unclear. It might be gas swept by the radio ejecta, but it might also be a tidal stream of a gas remnant of the merger event. 

\subsection{Rotating disks and super-disks.}
\label{disks}

In several sources, large-scale structures of ionized gas is in apparently ordered rotation. 
This is seen in ten izRGs, namely 3C~016, 3C~109, 3C~114, 3C~187, 3C~207, 3C~225B, 3C~226, 3C~275.1, 3C~293.1, and
3C~306.1. Their sizes range from $\sim 16$ kpc in 3C~016 to $\sim 70$ kpc in 3C~275.1; five of them are larger than 30 kpc.
The full amplitude of the rotation curve is typically between 300 and 400 \kms. 
The most extreme example of ordered rotation is found in 3C~275.1, where the gas, despite its complex morphology, presents a very symmetric structure in velocity space: the velocity increases almost linearly out to the most distant regions located at $\sim$35 kpc, where the velocity reaches a full amplitude of $\sim 900$ \kms. The line width is very high in the central regions ($\sim 800$ \kms) because of a prominent high-velocity tail  associated with a nuclear outflow (Coloma Puga, in prep.), but it is generally consistent with the instrumental resolution of MUSE at radii larger than $\sim 10$ kpc. 

In paper VI we fit the 2D gas velocity field in the innermost regions with the {\sl kinemetry} software \citep{krajnovic06} to derive the gas kinematical axis. The distribution of the offsets between the radio and kinematical axis was found to be not significantly different from a uniform distribution, although
a notable paucity of low offset values ($<20^\circ$) is present, in particular considering the FR~II sources alone. 

We similarly derived the orientation of the kinematical major axis for the ten sources listed above (see Table \ref{table:1}). In general, the kinematical axis is well aligned with morphological axis, indicating that the disks rotate about their major axis. The distribution of offsets with respect to the radio axis is shown in Fig. \ref{offset2}. We confirm the two results obtained in paper VI, that is, that the offsets values cannot be distinguished from those extracted from a uniform distribution, and that the kinematical axis is aligned with the radio axis in only a few sources, likely because in the RGs in which the EELR is aligned with the radio axis, the gas kinematics is highly disturbed by the impact of the jets and does not show ordered rotation. 

It is interesting that in several sources, the kinematical axis changes significantly at different scales. For example, in 3C~114 it is at PA$\sim 48^\circ$ at a radius of $\sim$2\arcsec\,, and it increases to at PA$\sim 90^\circ$ at $\sim$ 4\arcsec, an indication of a significant warp.

\begin{figure}
  \centering
\includegraphics[width=1.\linewidth]{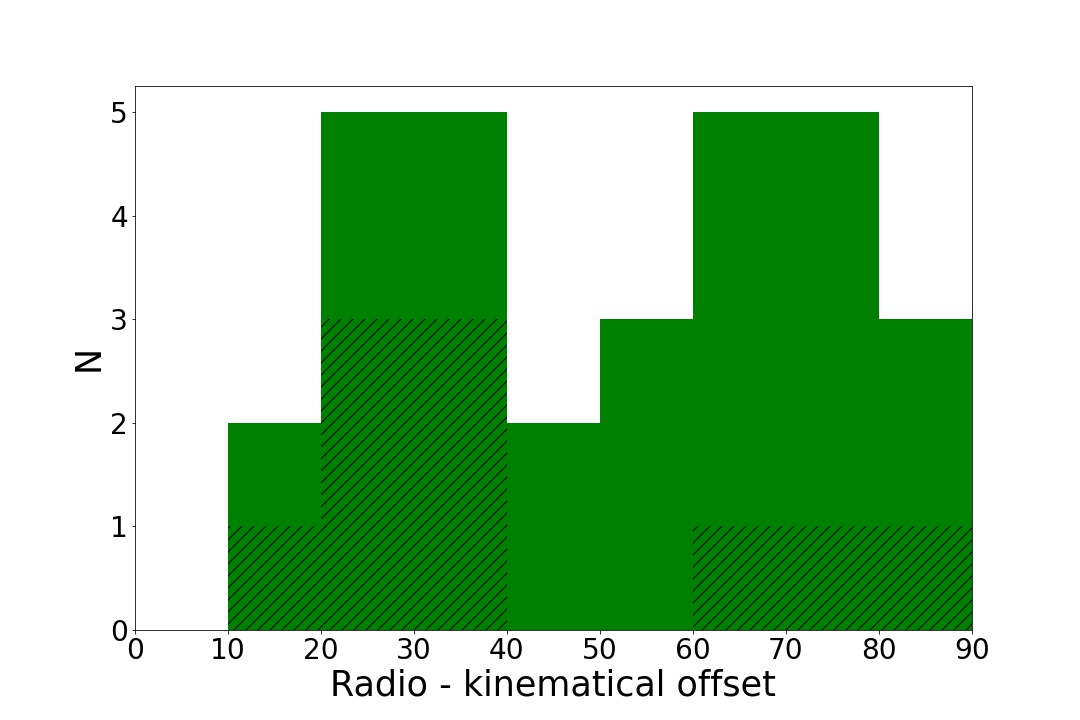}
\caption{Distribution of the offsets between the radio and kinematical major axis for the RGs with $0.3<z<0.82$ presented here (dashed histogram) combined with those with $z<0.3$ from paper VI.}
    \label{offset2}
\end{figure}

\section{The connection between line and radio emission.}
\label{connection}

In paper VI we discussed the relative orientation of the EELR with respect to the radio axis. We found that the distribution of the offsets is not significantly different from a uniform distribution, with just a marginal preference for large offsets between the radio and EELR axis, with 18 out of 29 sources showing an offset larger than 45$^\circ$.

For the izRGs the measurement of this offset is possible for 20 sources in which both line and radio are significantly extended (see Table \ref{table:1}). In these cases, an orientation measurement was performed on their farthest EELR regions and was obtained with a typical uncertainty of 5$^\circ$ (see paper VI for more details). Their distribution is shown in Fig. \ref{offset2}, where we also report the distribution of offsets for the sources with $z<0.3$, limited to the FR~IIs. With respect to those at low redshift, we find a larger fraction (12 out of 20 for $z>0.3$ against 6 out of 29 for $z<0.3$) of sources with an offset $<$ 30$^\circ$  degrees. This is due to the larger number of small sources (presented in Sect. \ref{CSS}) and gaseous bubbles (presented in Sect. \ref{bubbles}) that are all well aligned with the radio structure. 
The marginal preference for large offsets between radio and ionized gas seen at low redshift is no longer present, as there is an almost equal number of offsets that are larger and smaller than 45$^\circ$. 

We repeated this analysis be considering the EELR orientation closer to the AGN instead of the farthest EELR regions. This measurement is not possible for all sources because the emission line region in the center is often dominated by a nearly circular nuclear component. In seven of the sources for which we are able to obtain an EELR orientation at the large and small scale, the two measurements differ significantly (namely, 3C~142.1, 3C~175, 3C~225B, 3C~ 226, 3C~275, 3C~306, and 3C~334). We re-estimated the offset between the inner EELR and the radio. We found subtle differences in the offsets distribution with respect to that presented in Fig. \ref{offset2}, but the main results are unchanged.
We also searched for a connection between the size of the EELR and the offset, but found none. 

\begin{figure}
    \label{offset2}
  \centering
\includegraphics[width=1.\linewidth]{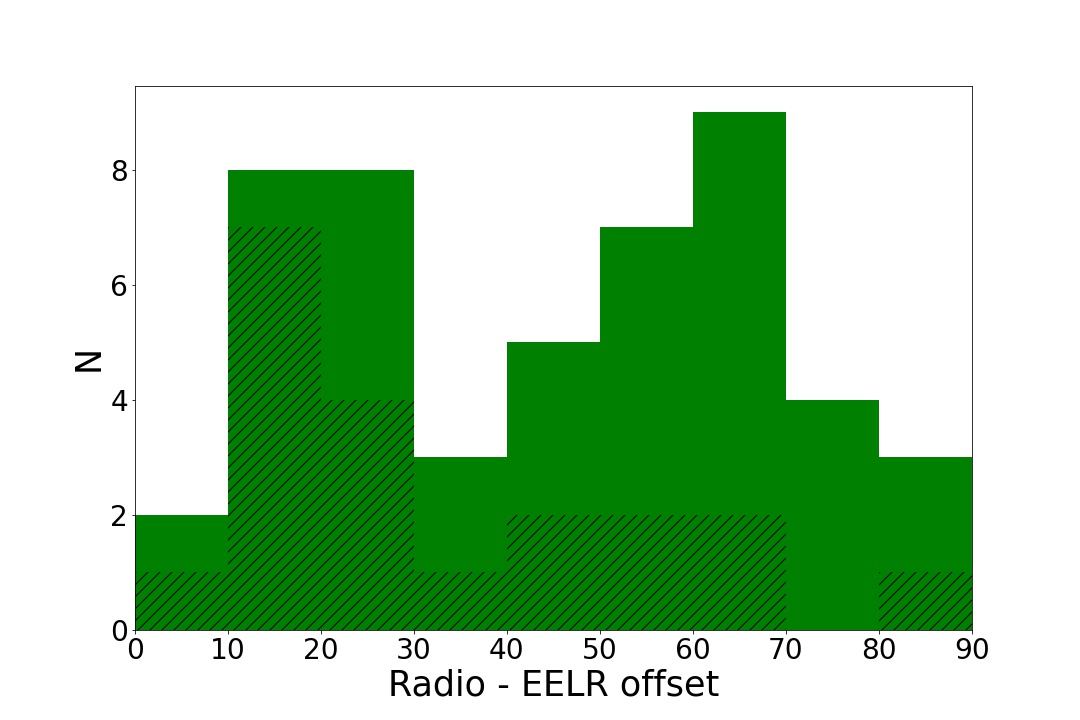}
\caption{Distribution of the offsets between the radio and EELR axis for the RGs with $0.3<z<0.82$ presented here (dashed histogram) combined with those with $z<0.3$ from paper VI.}
\end{figure}

\section{Discussion and conclusions.}
\label{discussion}

 In all but 4 of the 26 izRGs observed with MUSE, EELRs are observed to extend as far as 160 kpc from the host galaxy. In the sources in which the radio emission is confined within 30 kpc (namely, 3C~099, 3C~138, 3C~225B, 3C~275, and 3C~455), the EELR gas shows a complex velocity field and a large velocity dispersion, suggesting a strong jet-cloud interaction in the early growth phases of the radio source. 
These results agree with the early findings by \citet{gelderman94}, who reported that the lines profiles in compact radio sources 
are broader and complex: they suggested that this is the result of the jet interaction with the external medium. This conclusion was confirmed by the analysis of optical spectra of compact sources by \citet{kukreti24} and by the morphology of the line emission seen in high-resolution images \citep{axon00}. A clear signature of an expanding bubble produced by the interaction between the interstellar medium of
the galaxy and the radio jet has also been described 
 with unprecedented sensitivity and spatial resolution by JWST \citep{cresci23}.

We interpret these objects as the initial phases of the AGN activity, in which the radio source is confined within the galactic scale, where it expands into the dense interstellar medium of their hosts. 
The strong jet-cloud interaction in these sources also manifests itself as an increase in the velocity dispersion in the
direction perpendicular to the jets. This effect was also observed in nearby Seyfert galaxies with small-scale radio jets
and was interpreted as  due to the perturbation of the gas in the galactic disk through the jets  \citep{venturi21}.
An enhanced turbulent dispersion in the direction perpendicular to AGN jets was predicted by simulations performed by \citet{mukherjee18}, in which a relativistic jet interacted with a dense gaseous disk on a scale of a few kiloparsec.

In the extended sources, the EELR morphology is often connected with the radio properties. In several cases, the EELR takes the form of arcs or filaments, such as in 3C~435A, where these arcs surround the radio lobe. These structures are formed in the early phases of the growth of the radio source, the radio lobe is embedded in the denser gas regions, and its expansion leads to the formation of gas bubbles that are swept away by the expanding radio lobe. Although the asymmetric gas distribution and the presence of these companion galaxies might indicate an interaction or a merger scenario, the morphology and kinematics of the arc favor an interpretation in which the gas has been disturbed and swept up by the radio lobes. The absence of a corresponding southern gas bubble might result from projection effects and/or from an intrinsically asymmetric gas distribution. However, a contribution from tidal processes is disfavored, but cannot be excluded. 
We interpret this radio/optical association as a punctured gaseous bubble. The expansion of the radio source is caught at a later stage than in the sources described above, after it escaped the denser central regions and reached regions of lower gas density. 
The expansion of the radio jets then proceeds at a higher speed, leaving behind the gas bubble that formed in the initial phases. The lobe expansion speed in powerful radio galaxies was estimated with various methods (see, e.g., \citealt{Arshakian00,machalski09}) to be $\sim$0.05-0.1 of the speed of light. Among the four sources with a bubble like morphology, 3C~435A is associated with the smaller radio source, extending to a distance of $\sim 25$kpc, while the sizes of the other three lie between $\sim 60$ and $\sim$ 170 kpc. The timescale required for the radio lobes to expand to these distances can be estimated to be between 10$^6$-10$^7$ years.  We note that in other sources with similar arc-like or S-shaped emission line structures, the EELR is smaller than the radio emission. 

The following evolution of the gas bubble is governed by its internal motions, continuing in the expansion initially driven by the radio jets with measured values of hundreds of \kms.
The bubble expansion speed is lower by about two orders of magnitude than the expansion of the radio lobes, implying that it remains effectively unaltered for most of the lifetime of the radio source. Its size only grows by only a few kiloparsec over 10$^7$ years. This scenario accounts for the several arc-like gas structures seen in radio sources that are much more extended than their EELRs.

In other cases, the filaments appear to connect companion galaxies to the radio galaxies (see, e.g., 3C~16 or 3C~435A). These structures might be streams of  gas left over during the encounter process and ionized by the AGN. However, as already noted, the high-velocity dispersion measured in the gas streams disfavors a tidal-stream origin. 

Several EELRs instead show ordered rotation extending several tens of kiloparsec, forming structures that do not show a preferred axis (based on their morphological and kinematics) with respect to the radio axis. These results suggest an external origin of the gaseous super-disks. They might be formed, for example, by material acquired from a gas-rich merger. These large-scale disks can also be used to probe the gravitational potential to very large distances. 
Nonetheless, the innermost scale that we can probe is limited by the spatial resolution of the MUSE data that, for the izRGs, is several kiloparsec. We then explore the relative kinematical axis of the super-disks with respect to that of the jets and not of the gas kinematics in the immediate vicinity of the super massive black hole. The presence of disk warps is already seen in our data, but higher-resolution data, such as those that can be obtained by studying the gas properties with ALMA at the sub-arcsecond scale, are needed to further explore the connection between the gas and the radio jet orientation.   

Overall, the results we obtained for the izRGs confirm the findings of the study of the radio galaxies at lower redshift. The different properties of the EELRs can be explained with a combination of the evolutionary state of the radio source and its orientation with respect to the super-disks. A spectral aging and morphological analysis based on multi-frequency radio data would be very important to obtain independent information about the evolutionary stage of these sources and on their connection with the properties of their EELRs.

In forthcoming papers, we will evaluate the AGN feedback effects on the surrounding medium through outflows and ionizing radiation. In particular, we will explore the presence of AGN-driven outflows, following the analysis presented for the $z<0.3$ sources \citep{speranza21}. We will also address
the ionization properties of the EELRs and study the large-scale environment of radio galaxies in the search for companions, and we will testing the possibility that streams of ionized gas connect these galaxies to the central radio galaxies.

\section*{Data availability}
The reduced MUSE datacubes can be downloaded from the ESO archive (https://archive.eso.org/
eso/eso\_archive\_main.html).

\begin{acknowledgements}
We thank the anonymous referee for the useful comments that improved the paper. 
Based on observations made with ESO Telescopes at the La Silla Paranal Observatory under programme ID  0106.B-0564(A). The National Radio Astronomy
Observatory is a facility of the National Science Foundation operated under cooperative agreement by Associated Universities, Inc. The radio images were
retrieved from the NRAO VLA Archive Survey, (c) AUI/NRAO, available at http://archive.nrao.edu/nvas/.
BB acknowledges the prin INAF 2023 "The MURALES project: exploring AGN feedback in the most powerful radio loud active galactic nuclei".
Facilities: VLT:MUSE, VLA.
\end{acknowledgements}

\bibliographystyle{aa} 
\bibliography{my} 

\begin{appendix}
\onecolumn
\section{Radio, continuum, and line emission images of the izRGs.}
\FloatBarrier
\begin{figure*}

\includegraphics[width=0.21\linewidth]{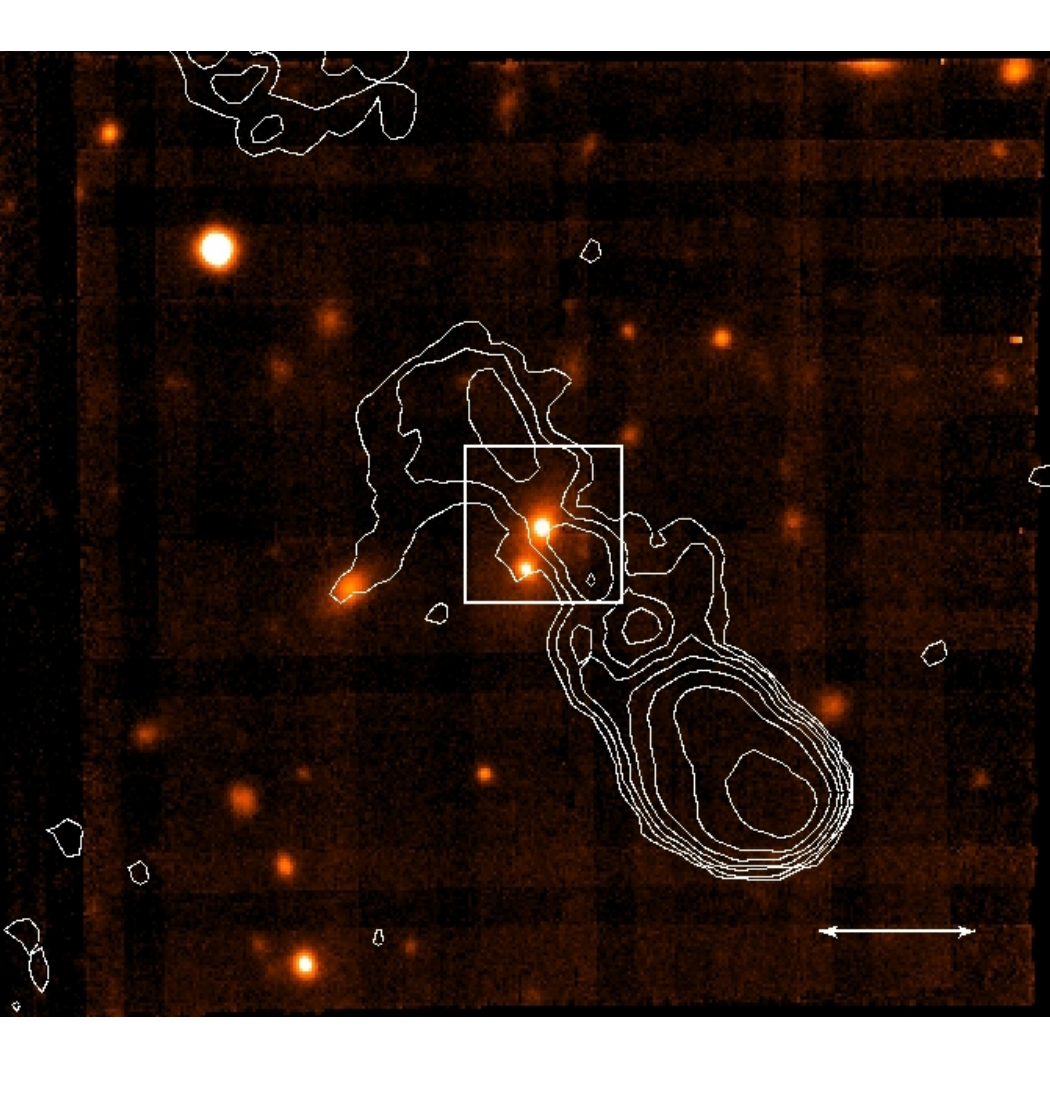}
\includegraphics[width=0.245\linewidth]{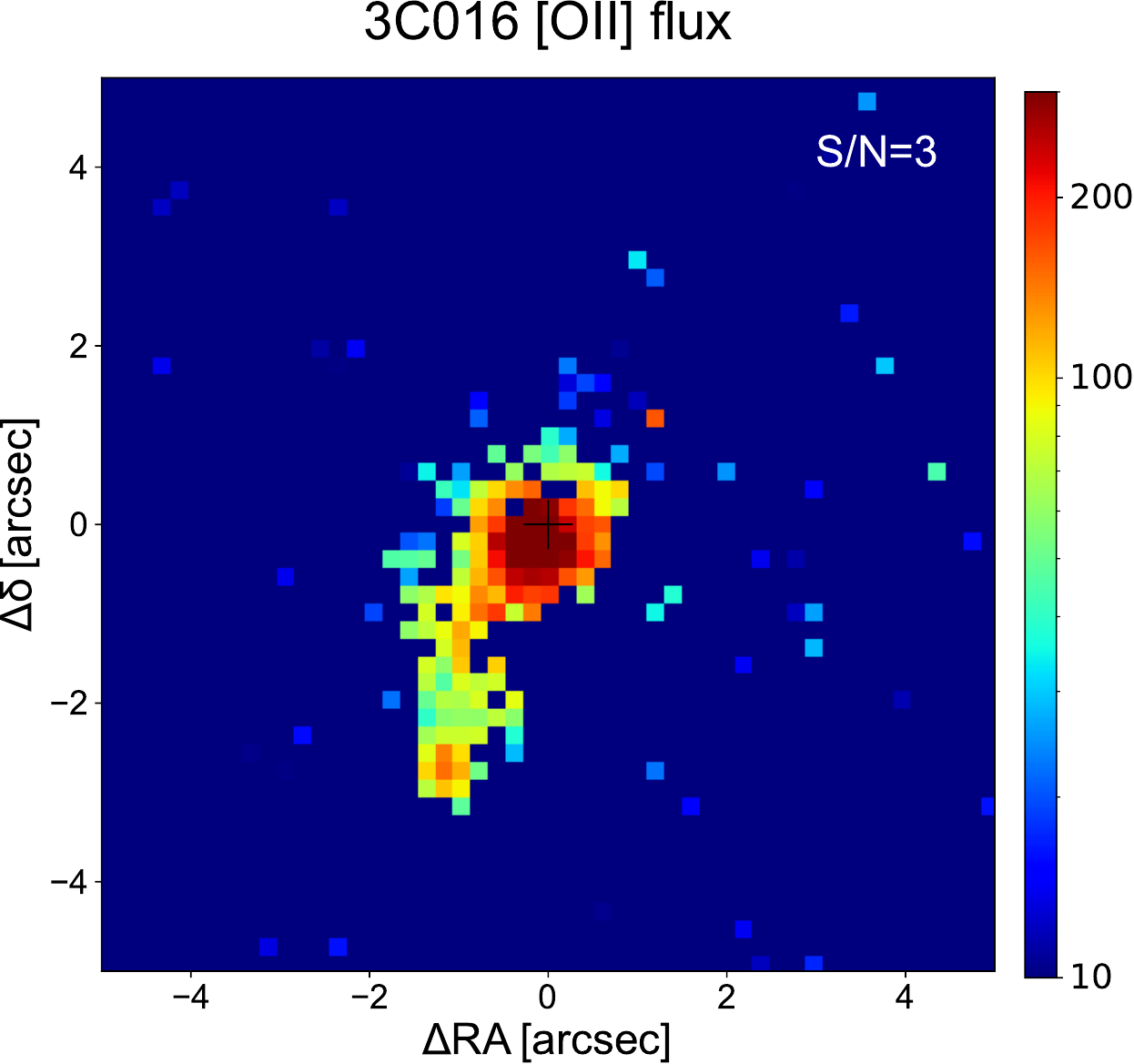}
\includegraphics[width=0.25\linewidth]{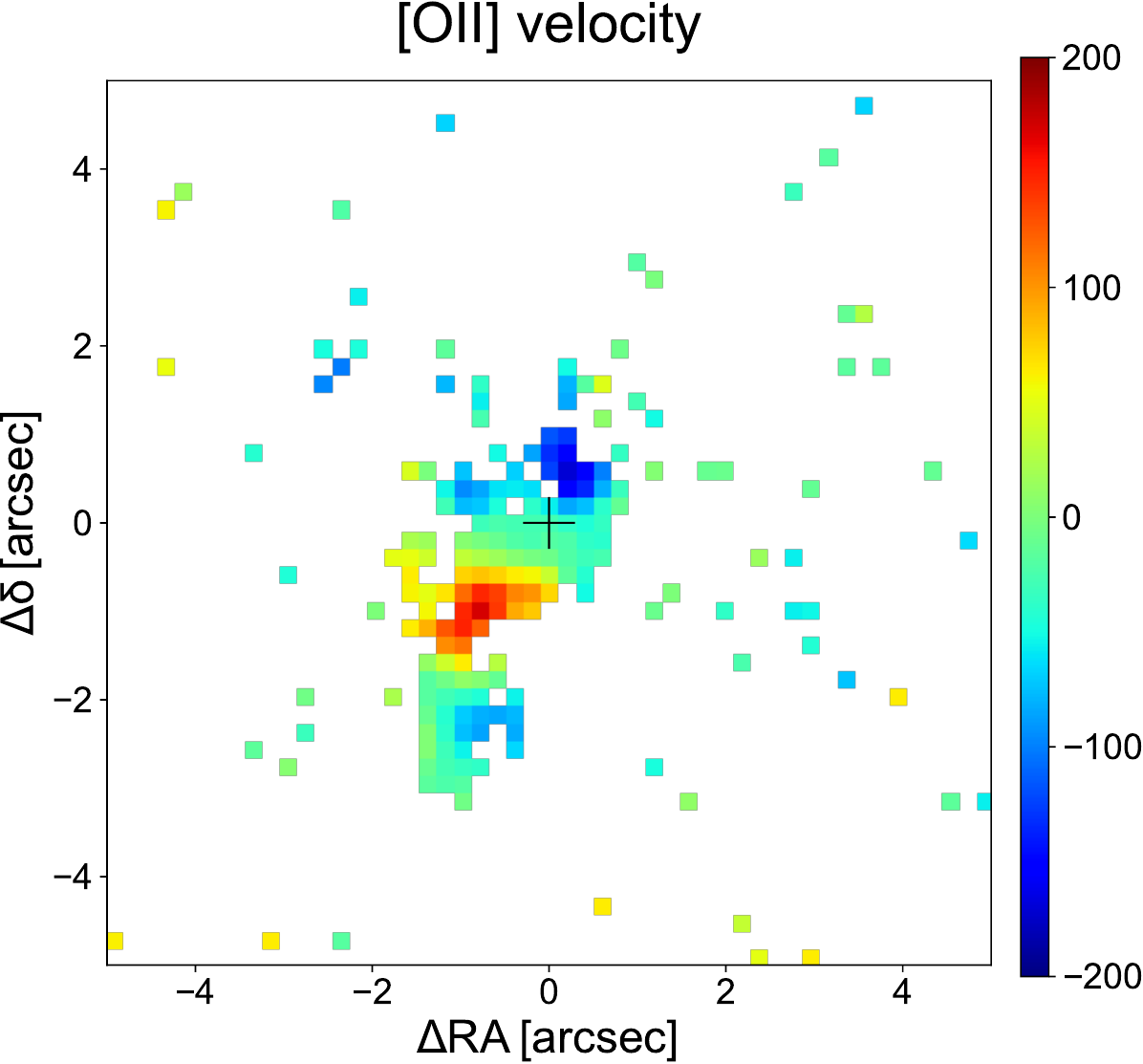}
\includegraphics[width=0.245\linewidth]{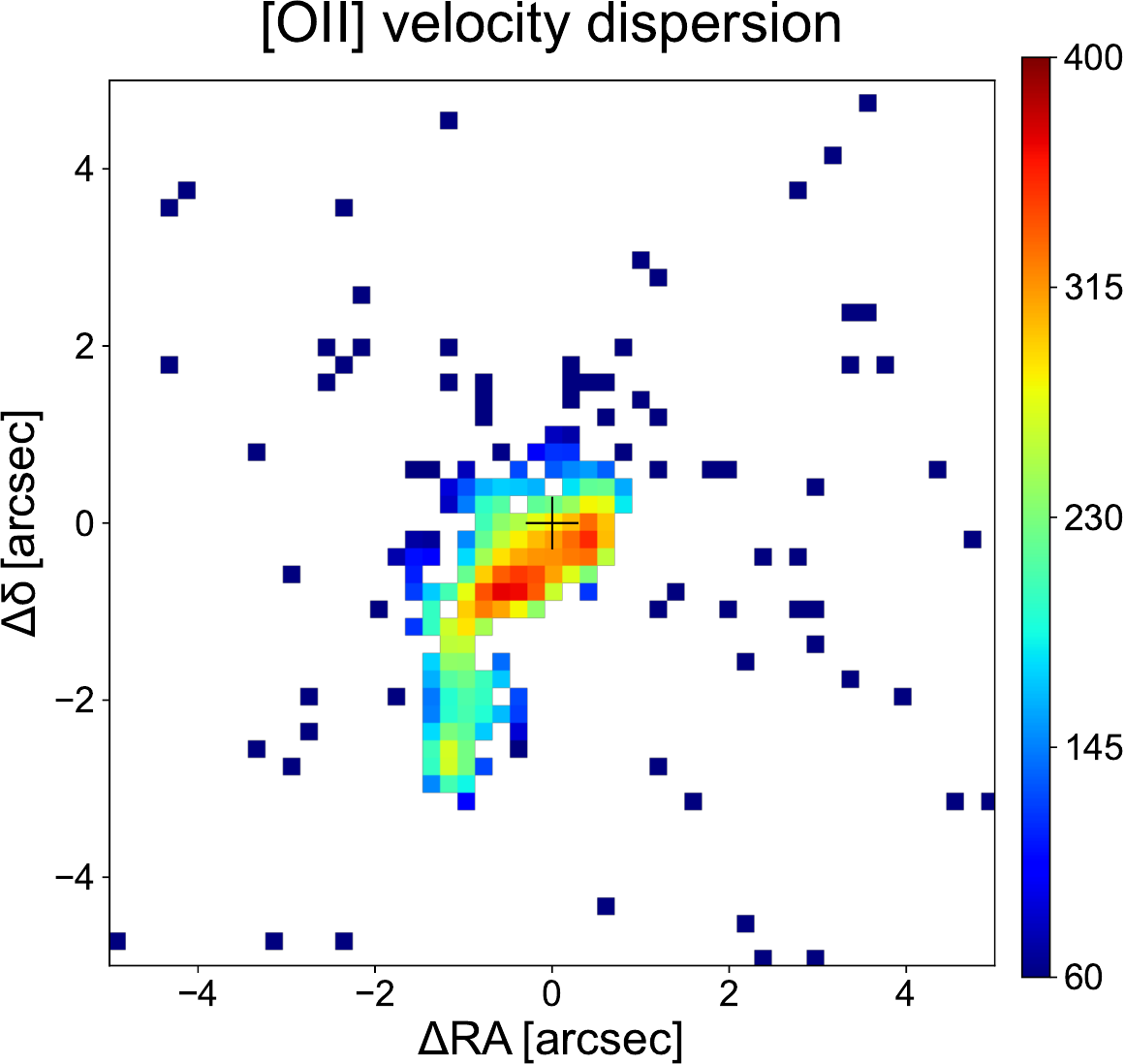}

\includegraphics[width=0.21\linewidth]{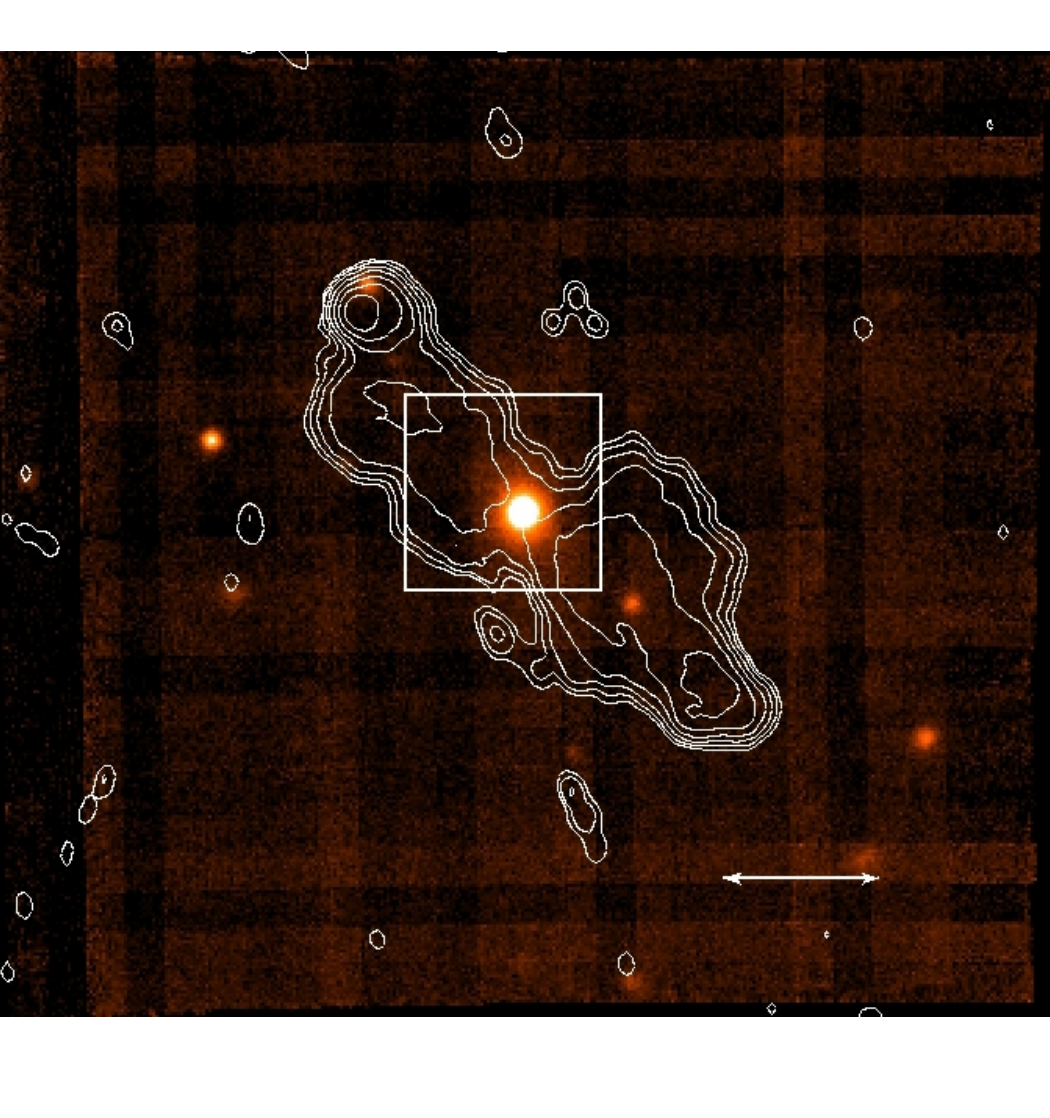}
\includegraphics[width=0.25\linewidth]{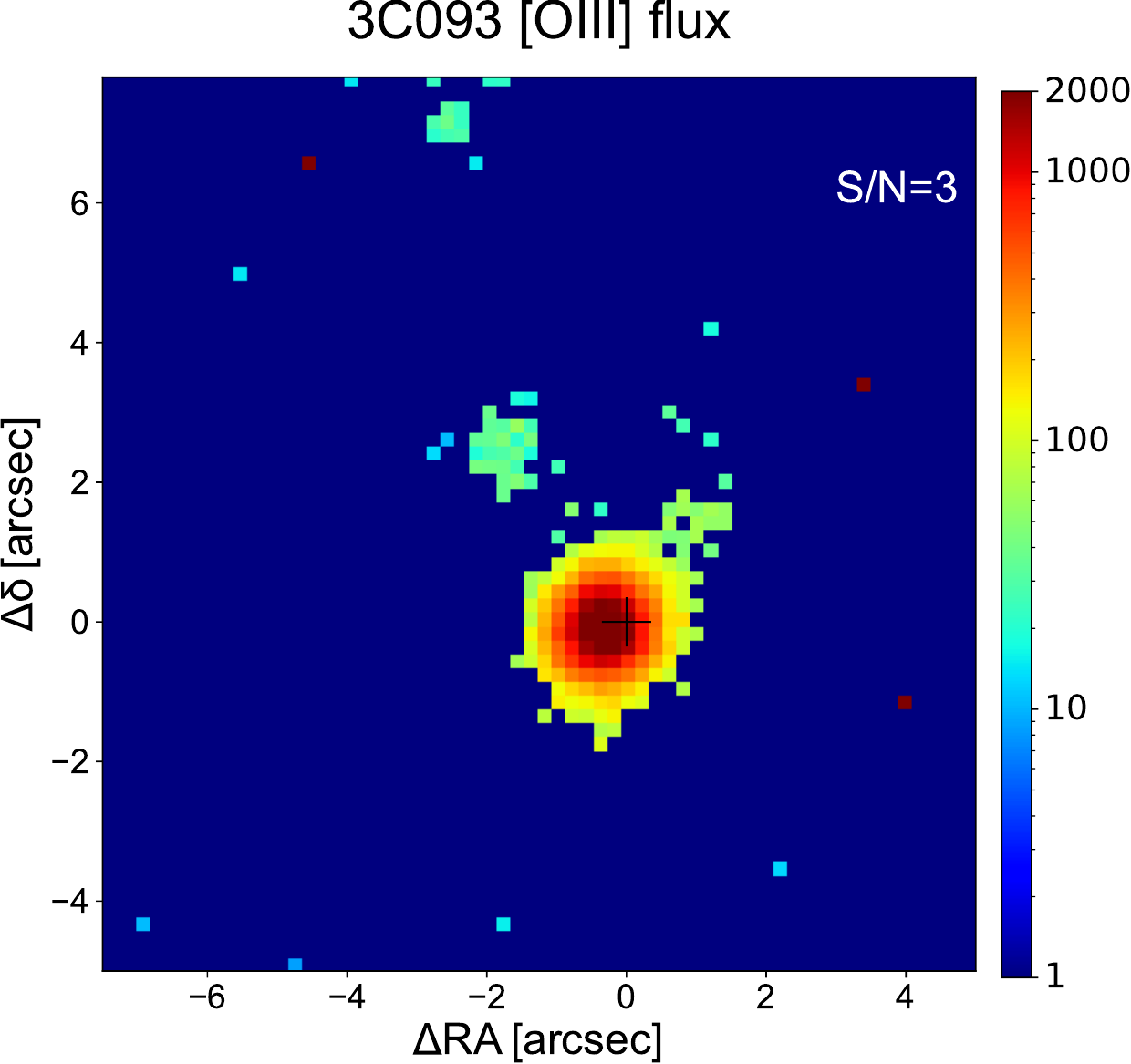}
\includegraphics[width=0.25\linewidth]{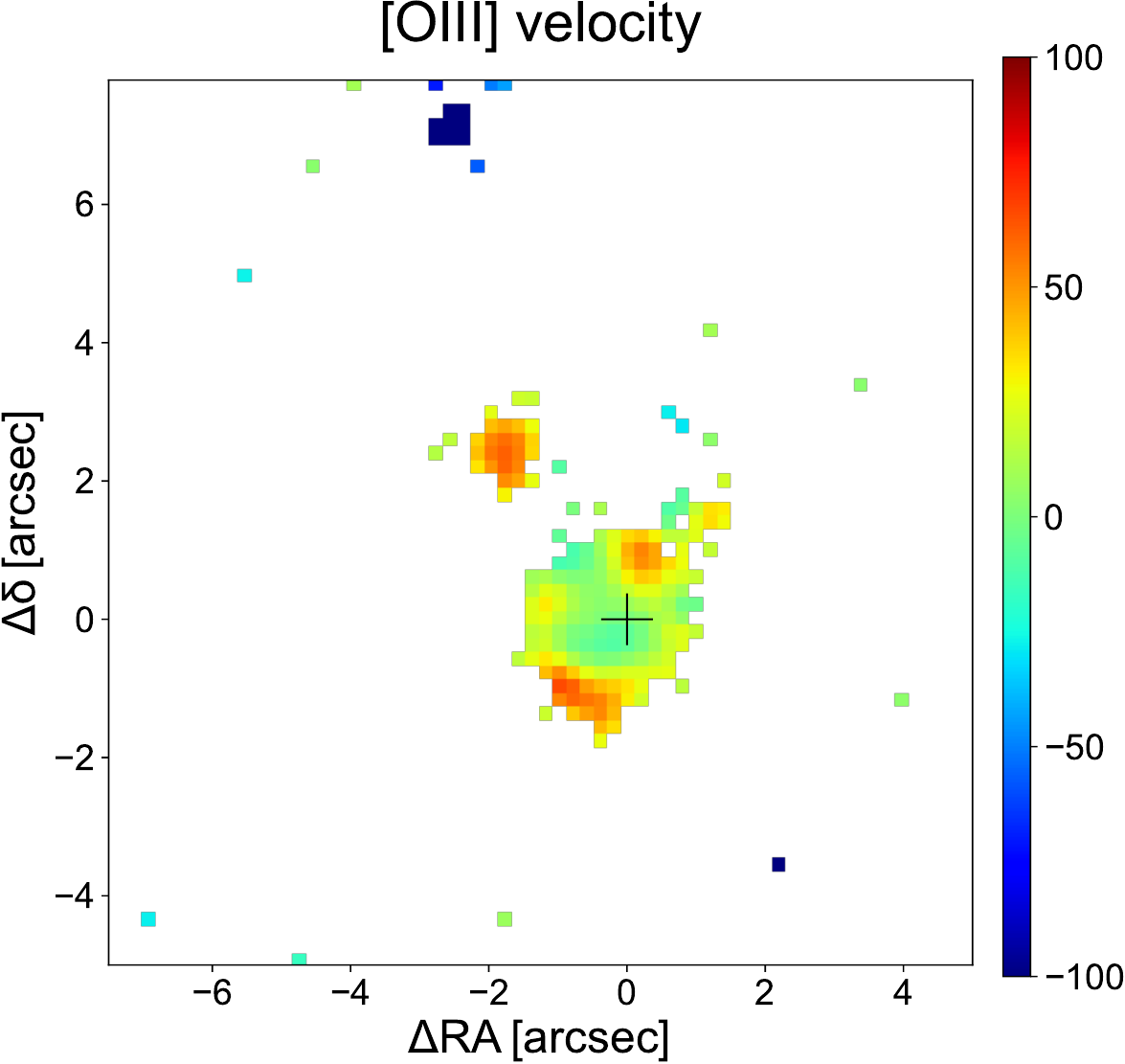}
\includegraphics[width=0.245\linewidth]{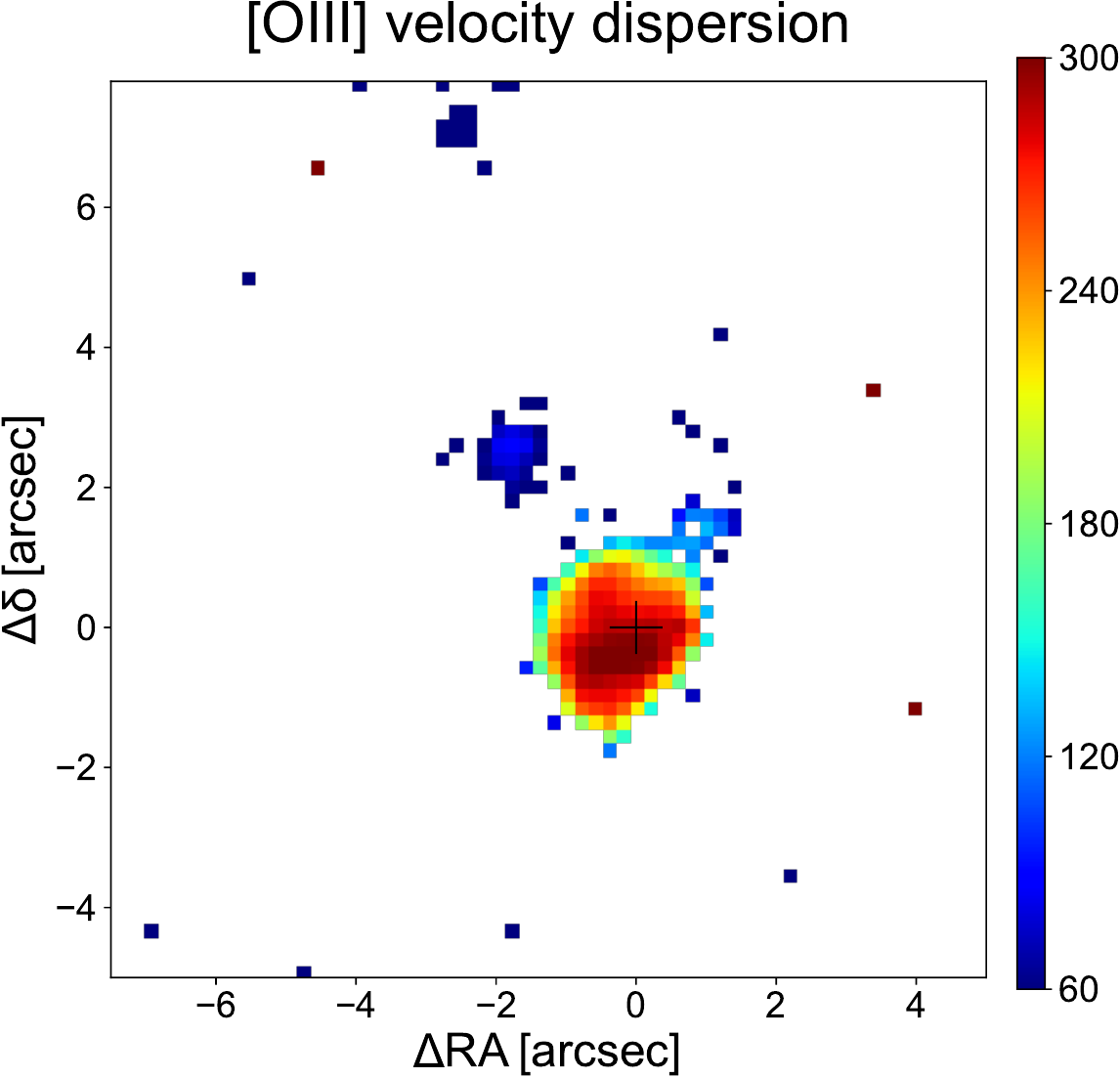}

\includegraphics[width=0.21\linewidth]{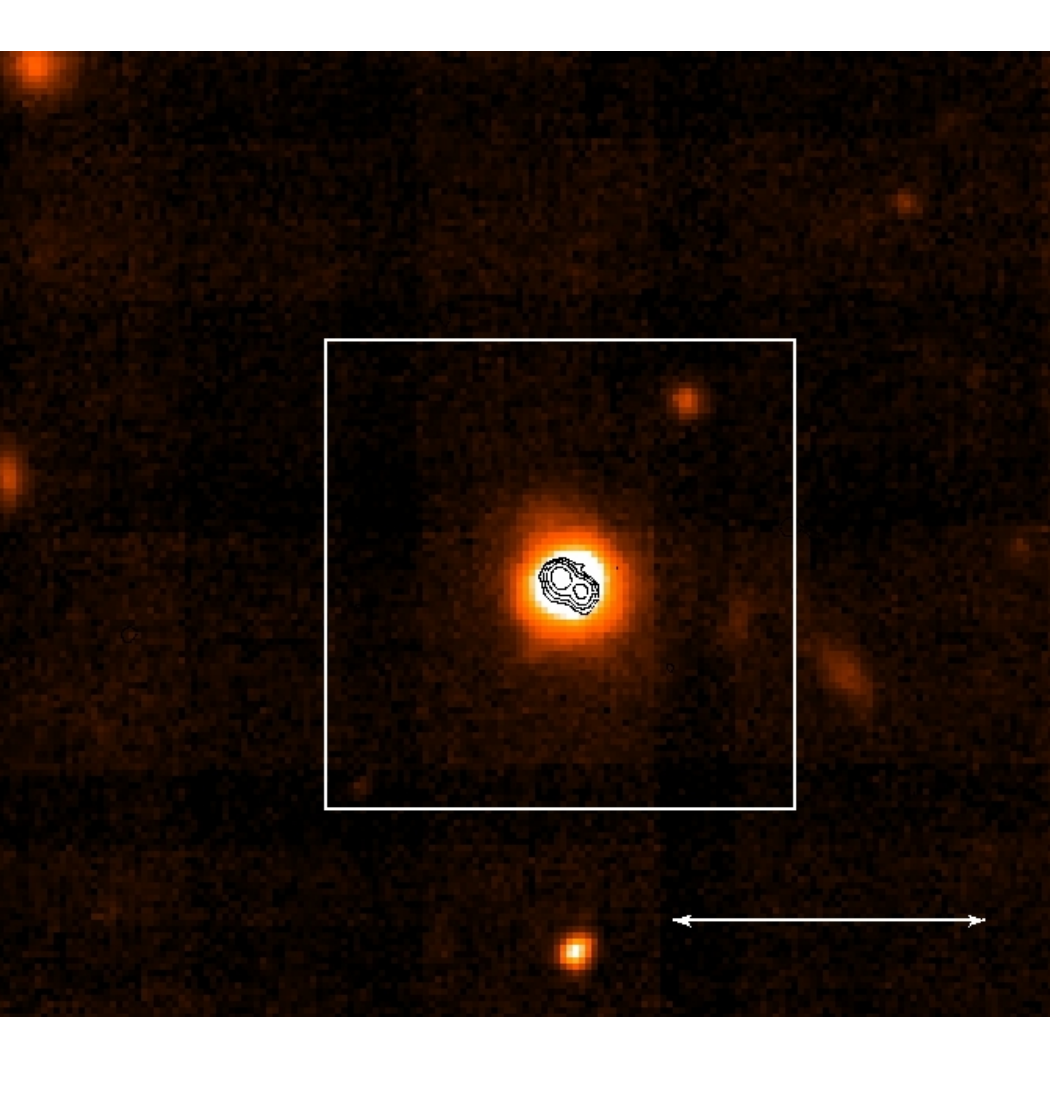}
\includegraphics[width=0.25\linewidth]{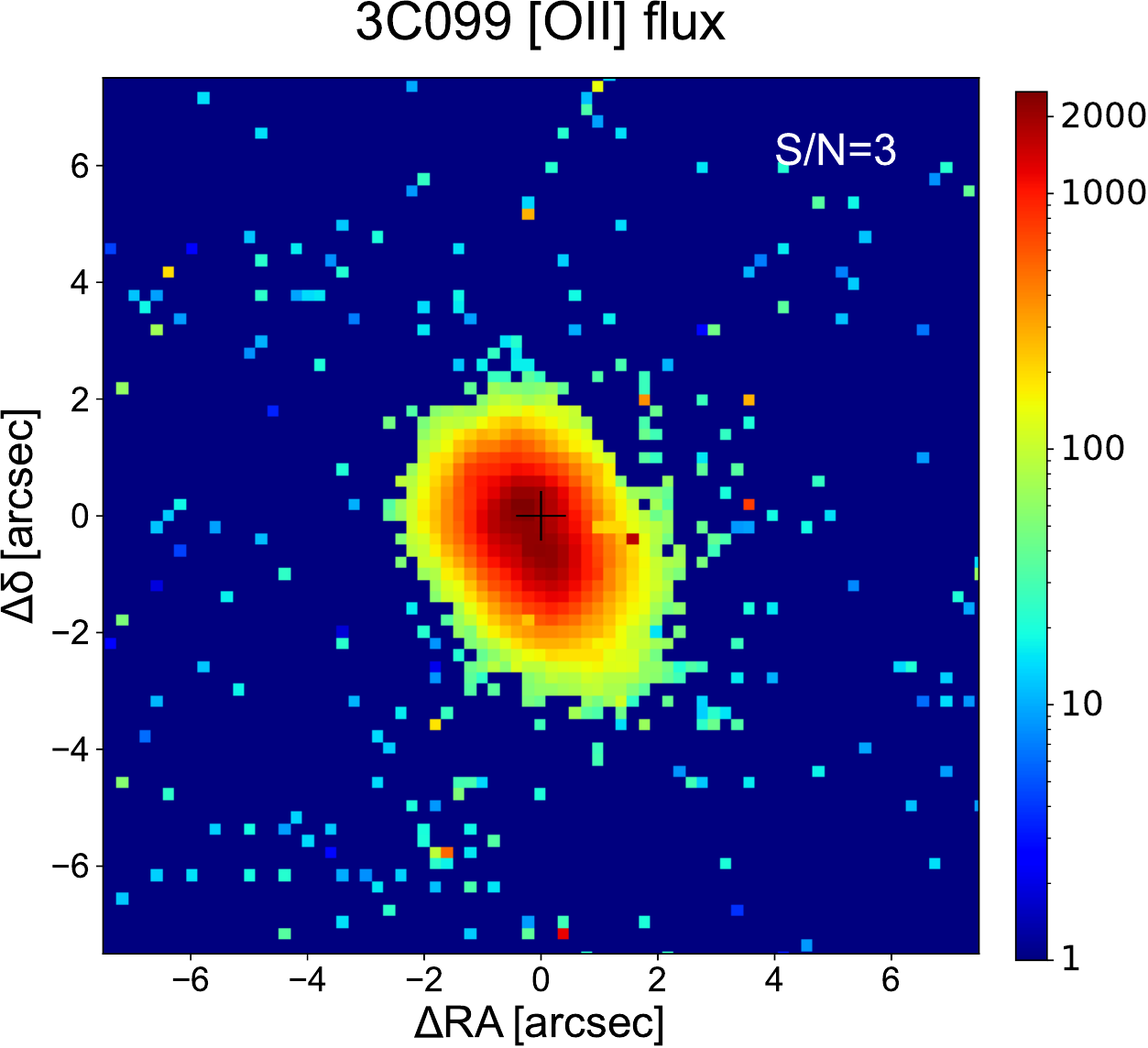}
\includegraphics[width=0.25\linewidth]{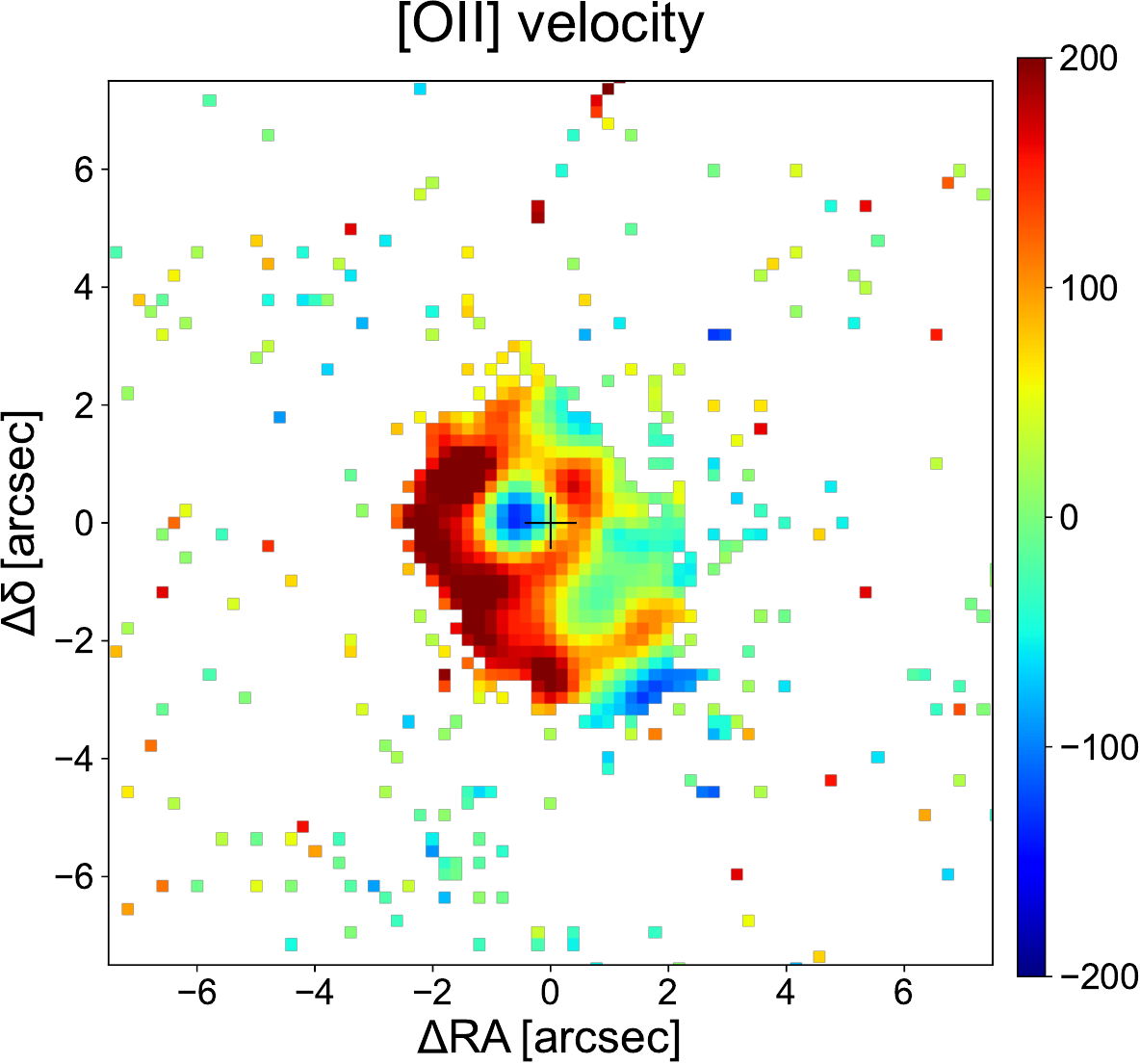}
\includegraphics[width=0.25\linewidth]{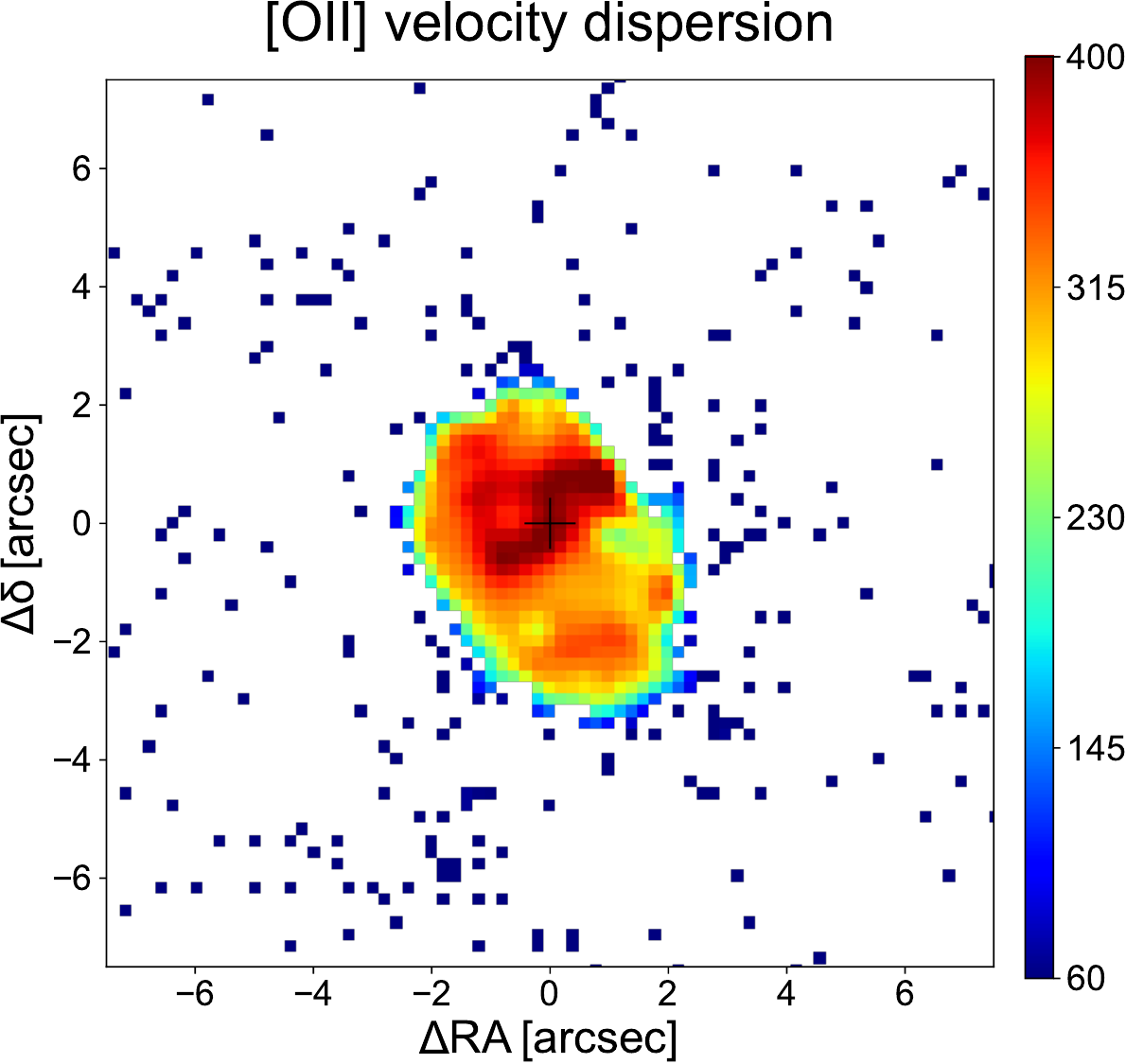}

\includegraphics[width=0.21\linewidth]{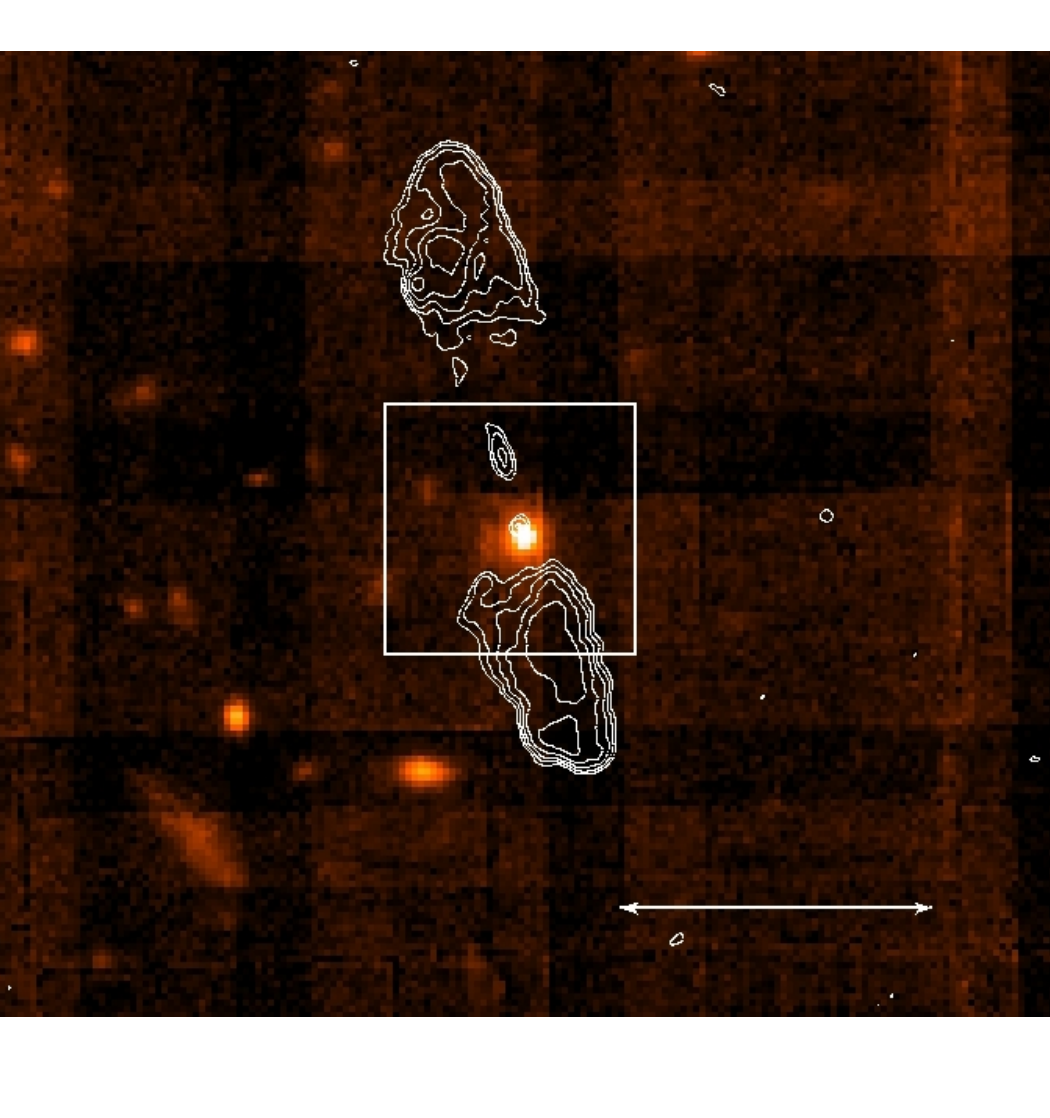}
\includegraphics[width=0.25\linewidth]{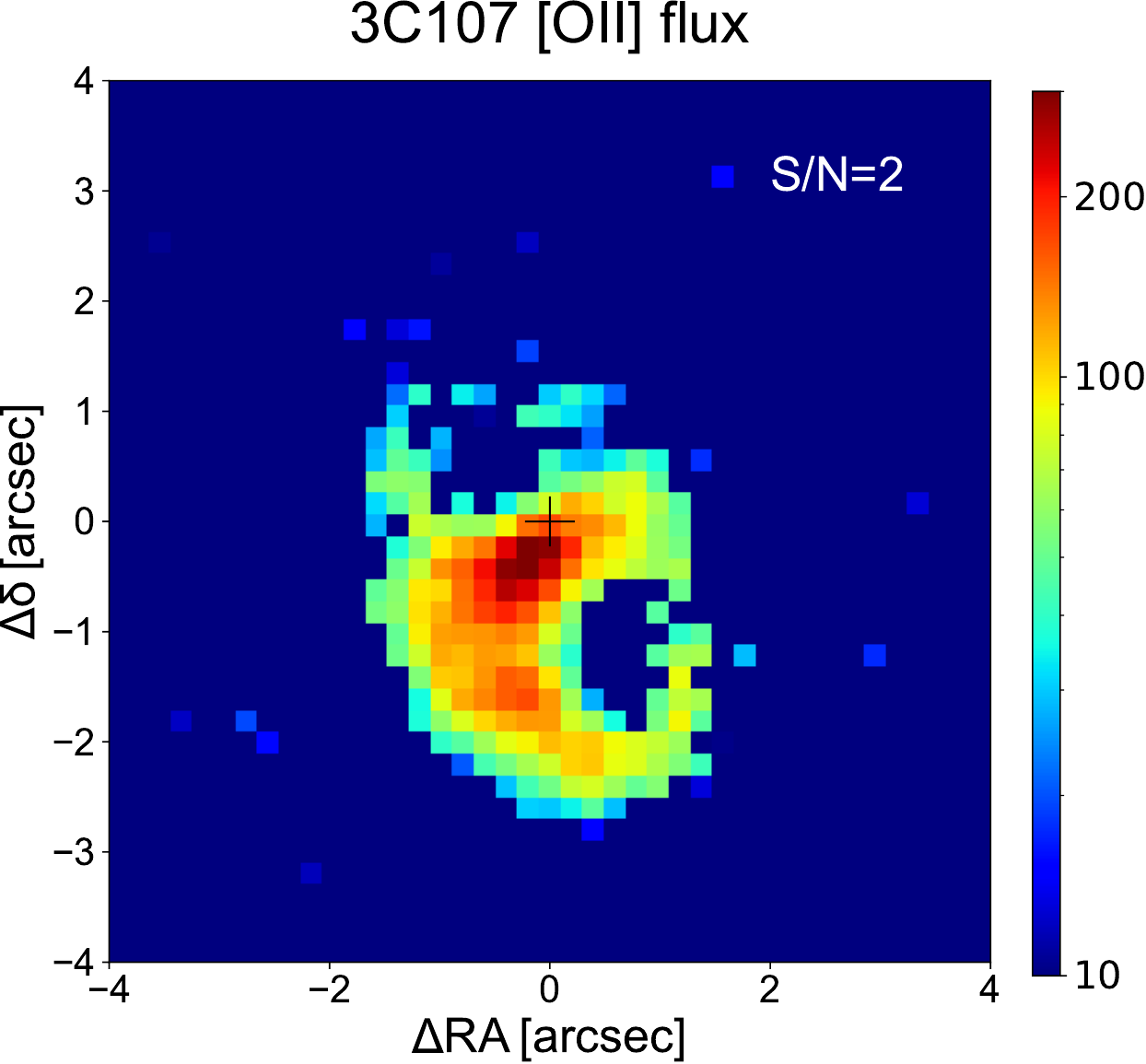}
\includegraphics[width=0.25\linewidth]{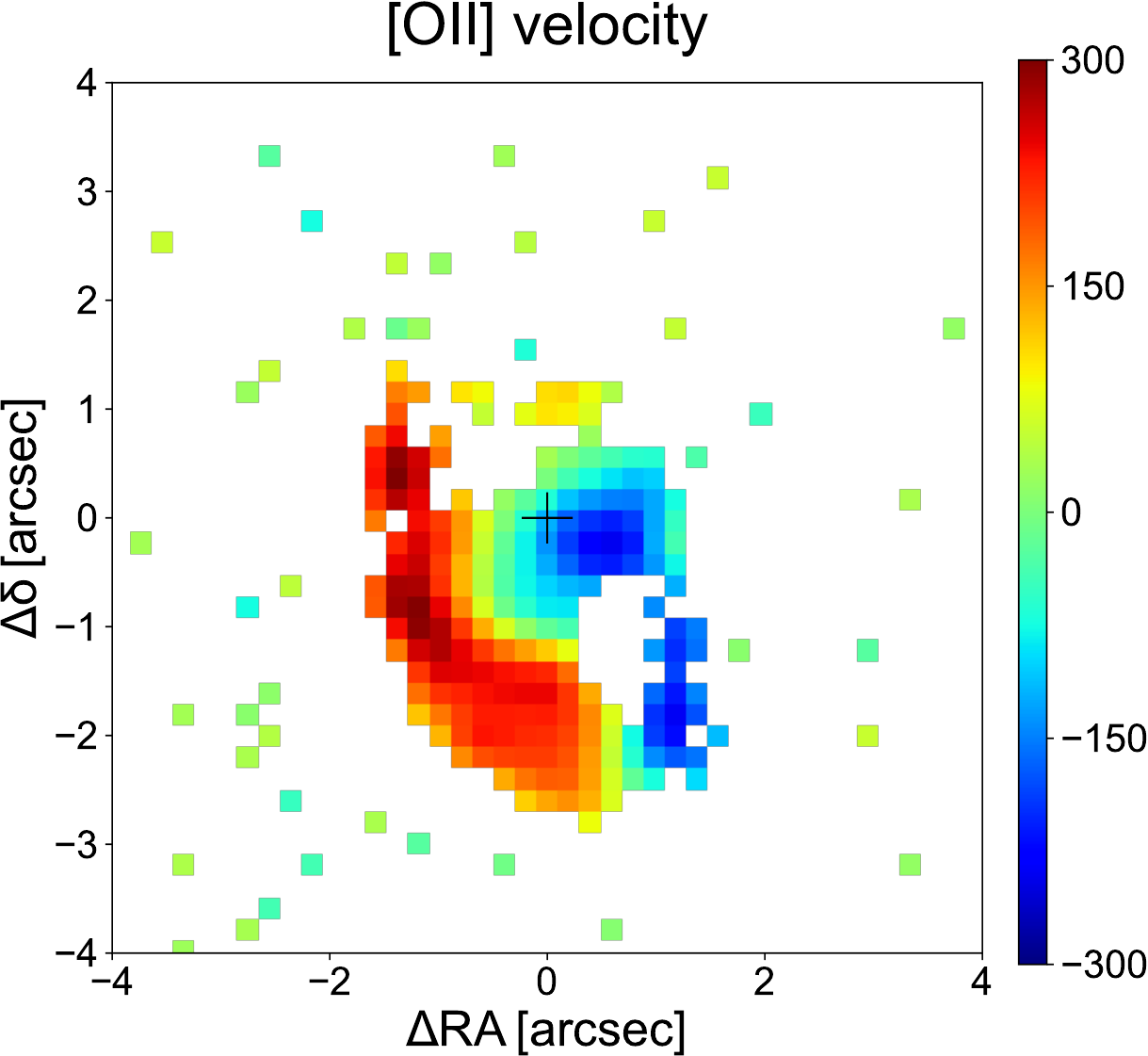}
\includegraphics[width=0.25\linewidth]{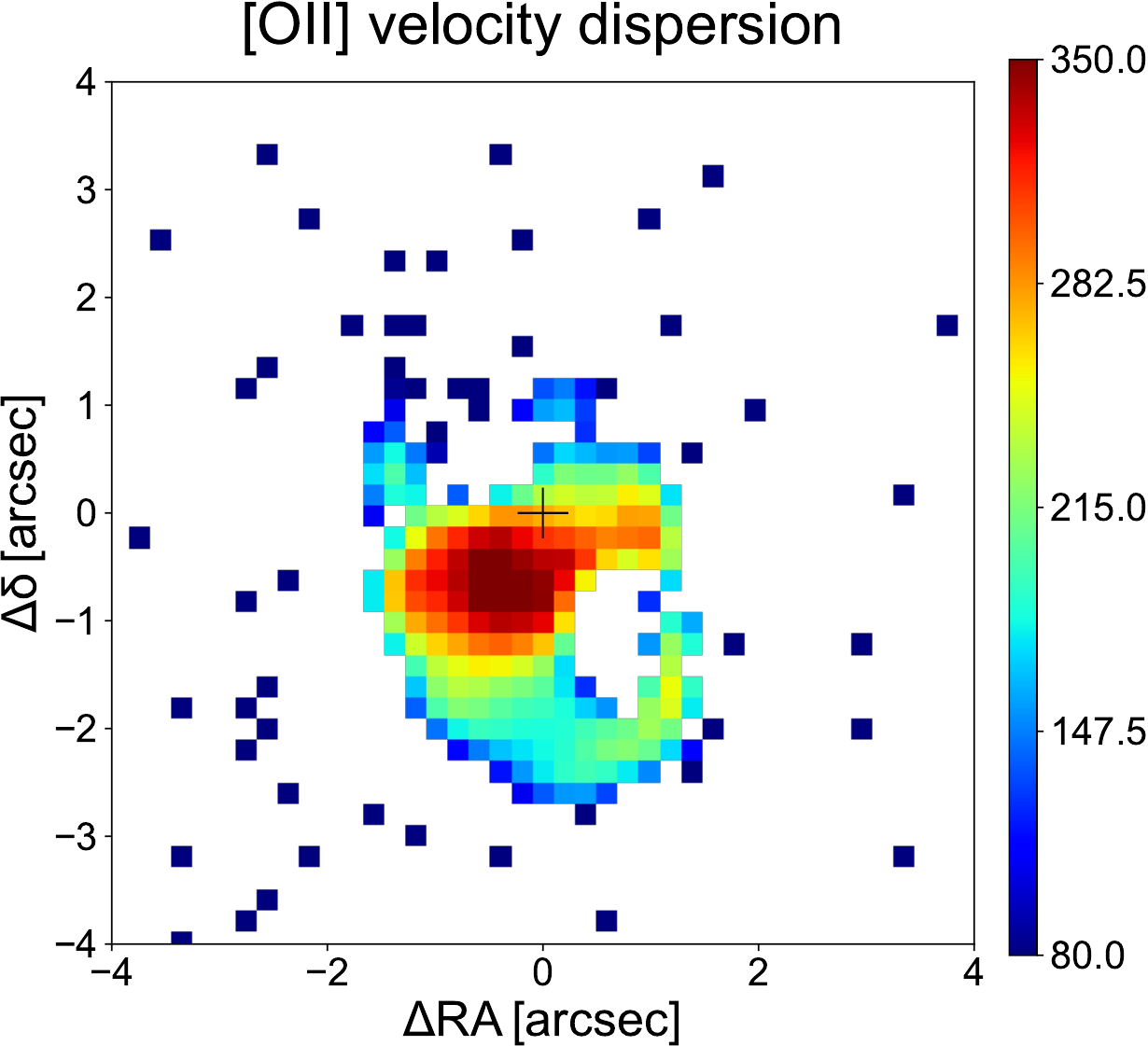}

\includegraphics[width=0.21\linewidth]{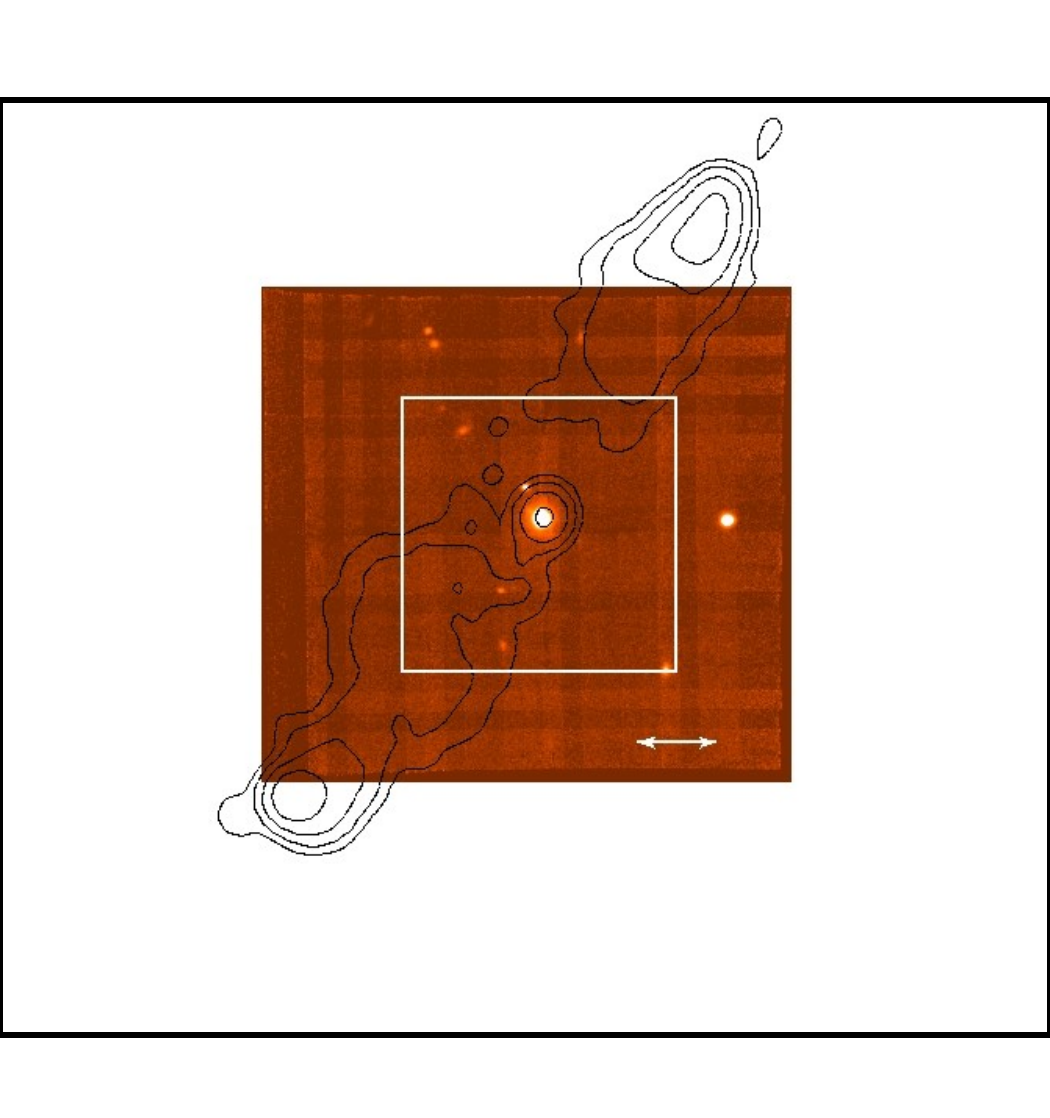}
\includegraphics[width=0.25\linewidth]{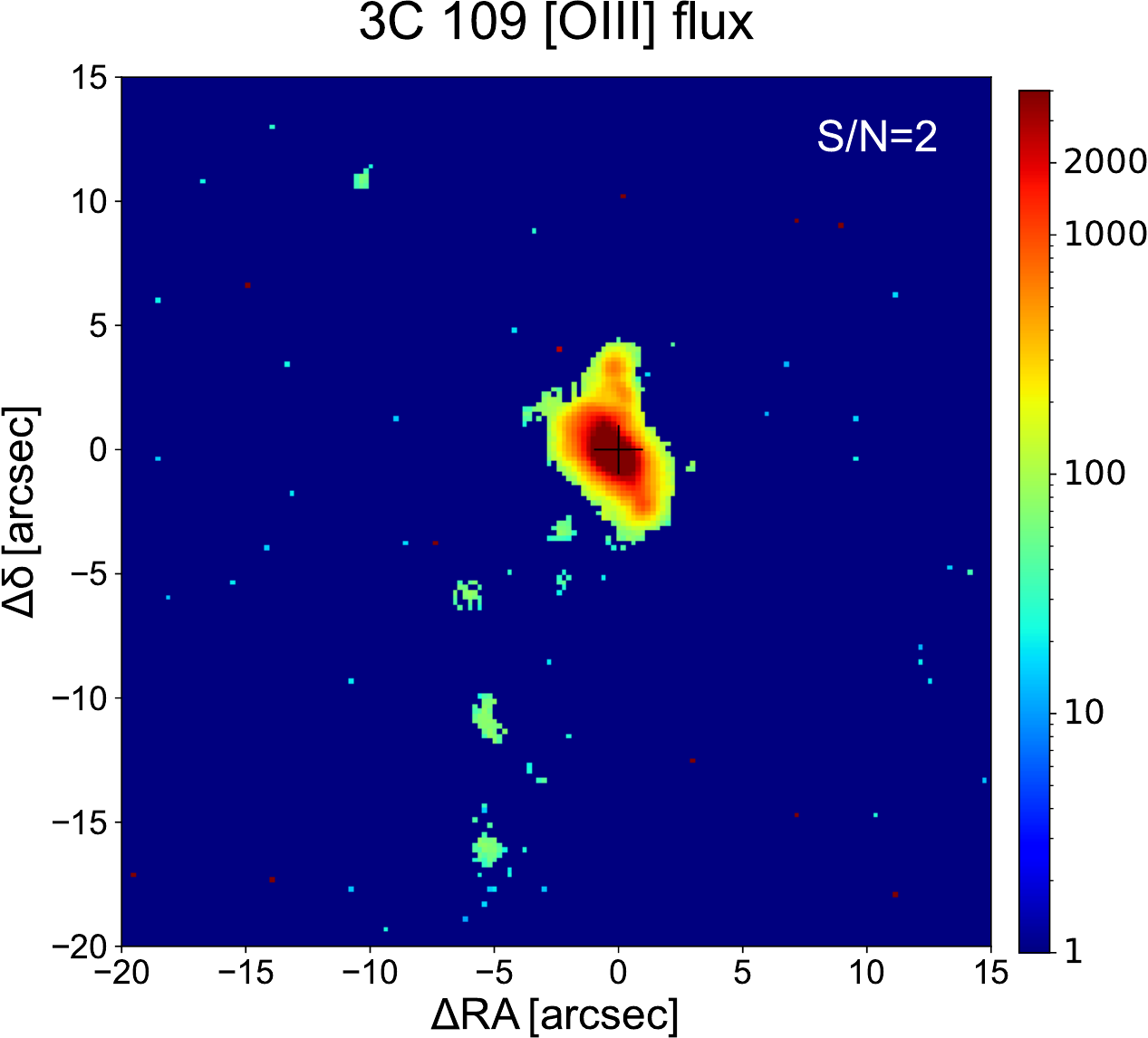}
\includegraphics[width=0.25\linewidth]{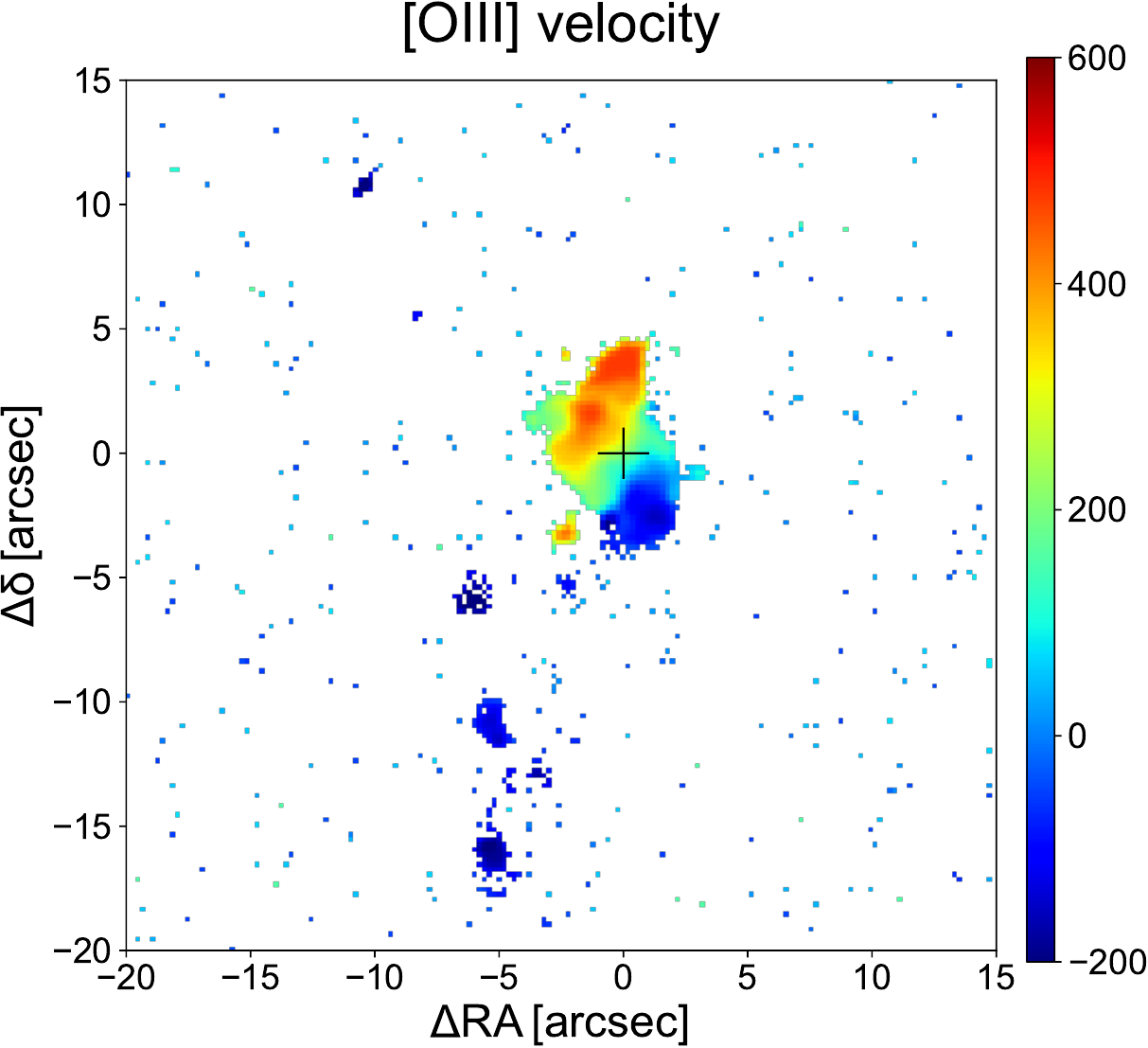}
\includegraphics[width=0.25\linewidth]{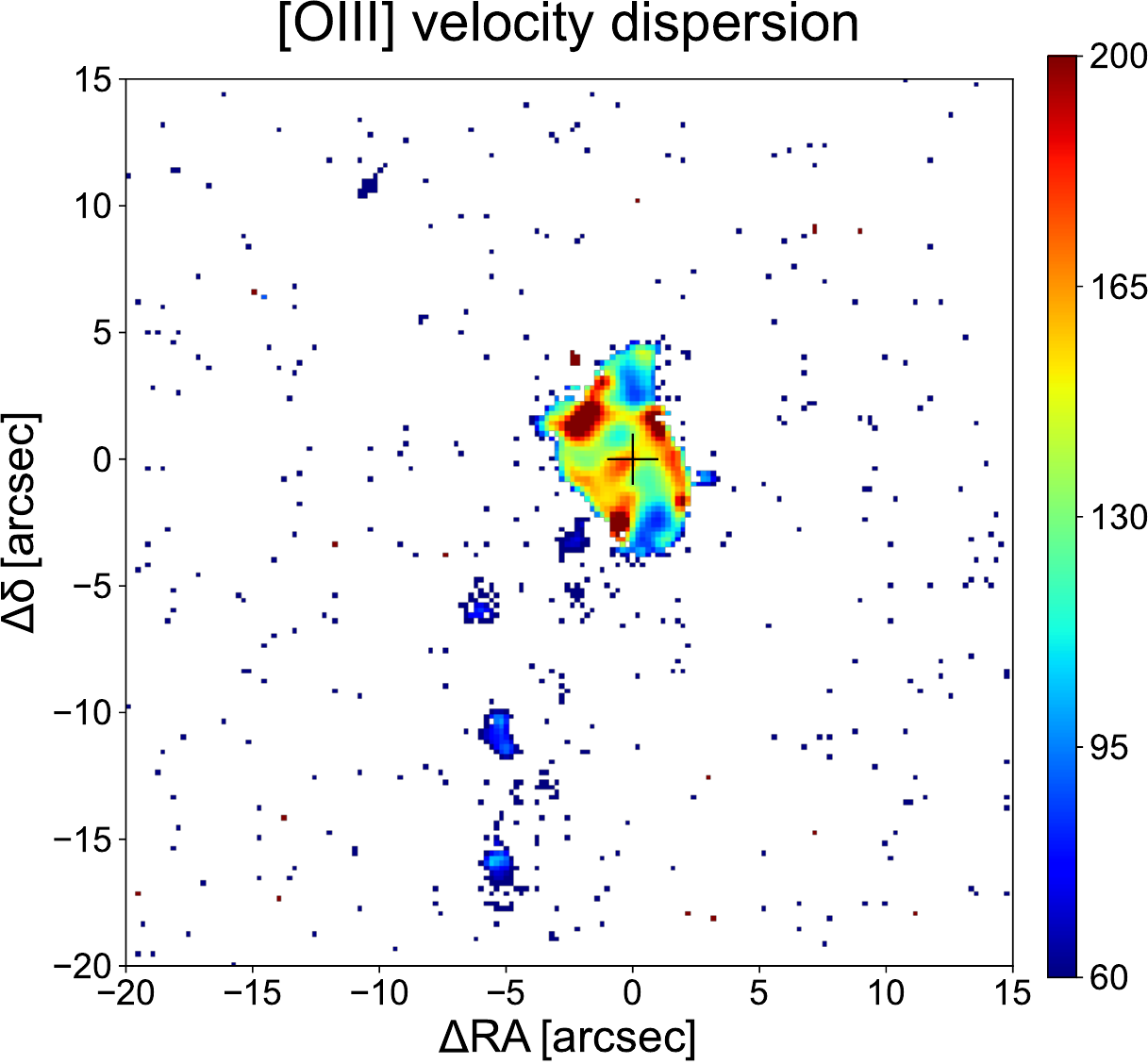}

\caption{Left: continuum image over the whole 1\arcmin$\times$1\arcmin\ MUSE field ov view,  extracted between 4500 and 4800 \AA\ in the rest frame of the radio galaxy, with 
  superposed the radio isophotes. {\bf North is on top, East to the left.} The white box corresponds to the
  size of the figures on the right where the emission line image is shown. {\bf The color scale is in units of 10$^{-18}$ erg s$^{-1}$
cm$^{-2}$ arcsec$^{-2}$
.} In the two rightmost panel we report the gas velocity and gas velocity dispersion, {\bf both in \kms\ units. In all images we marked with a cross the location of the continuum peak, likely to be the position of the AGN.}}
    \label{fig1}
\end{figure*}

\addtocounter{figure}{-1}
\begin{figure*}
\centering
\includegraphics[width=0.20\linewidth]{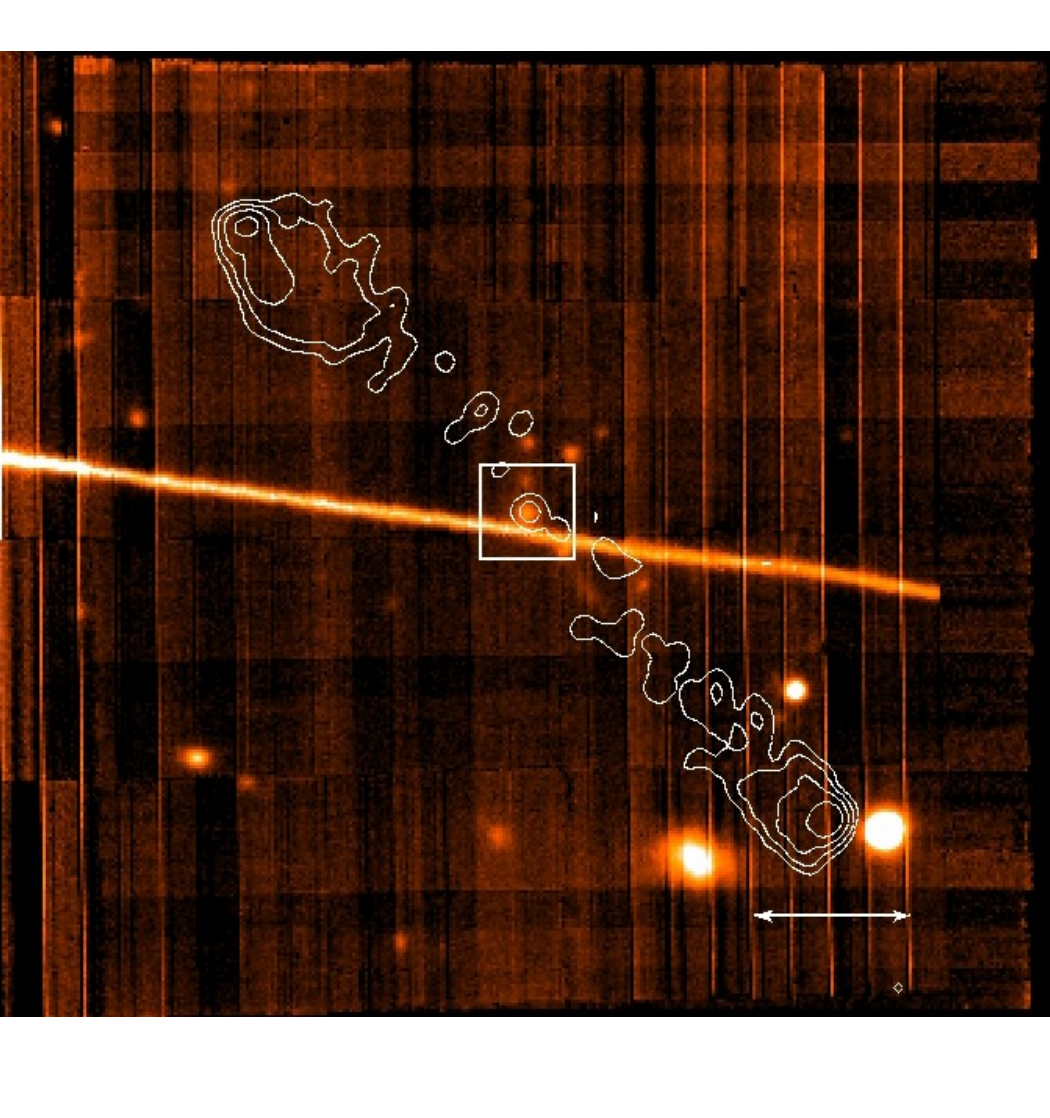}
\includegraphics[width=0.25\linewidth]{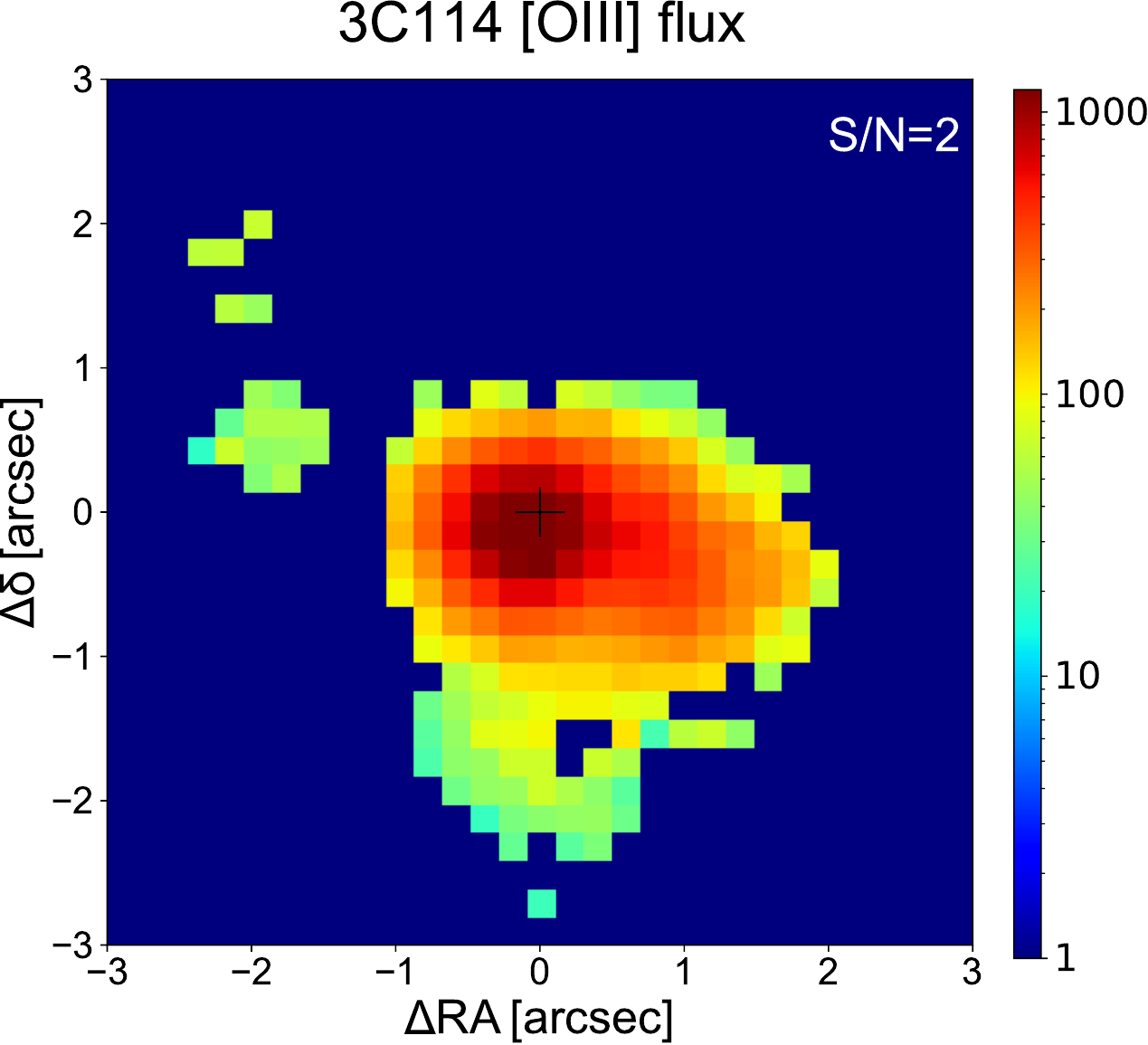}
\includegraphics[width=0.25\linewidth]{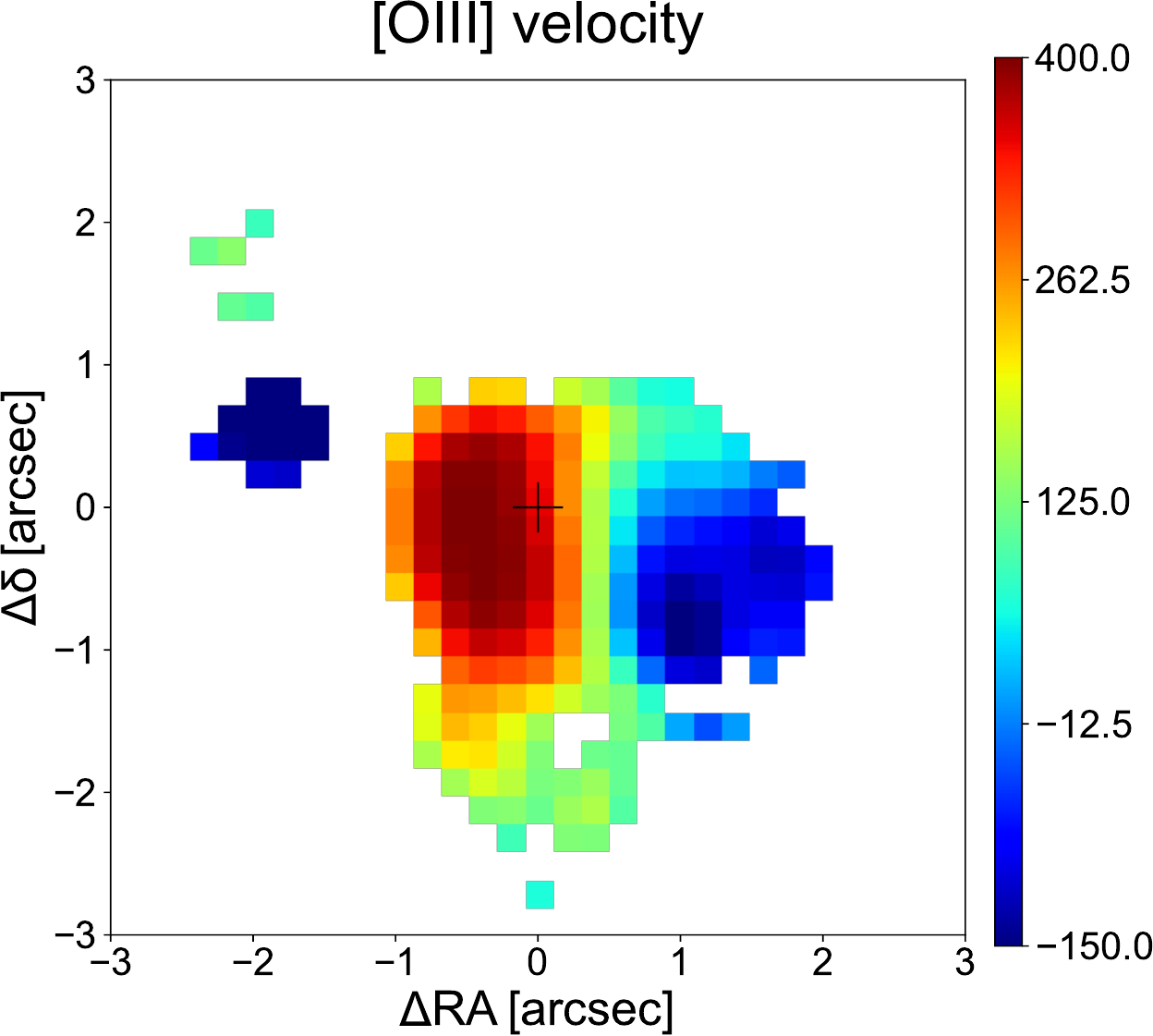}
\includegraphics[width=0.25\linewidth]{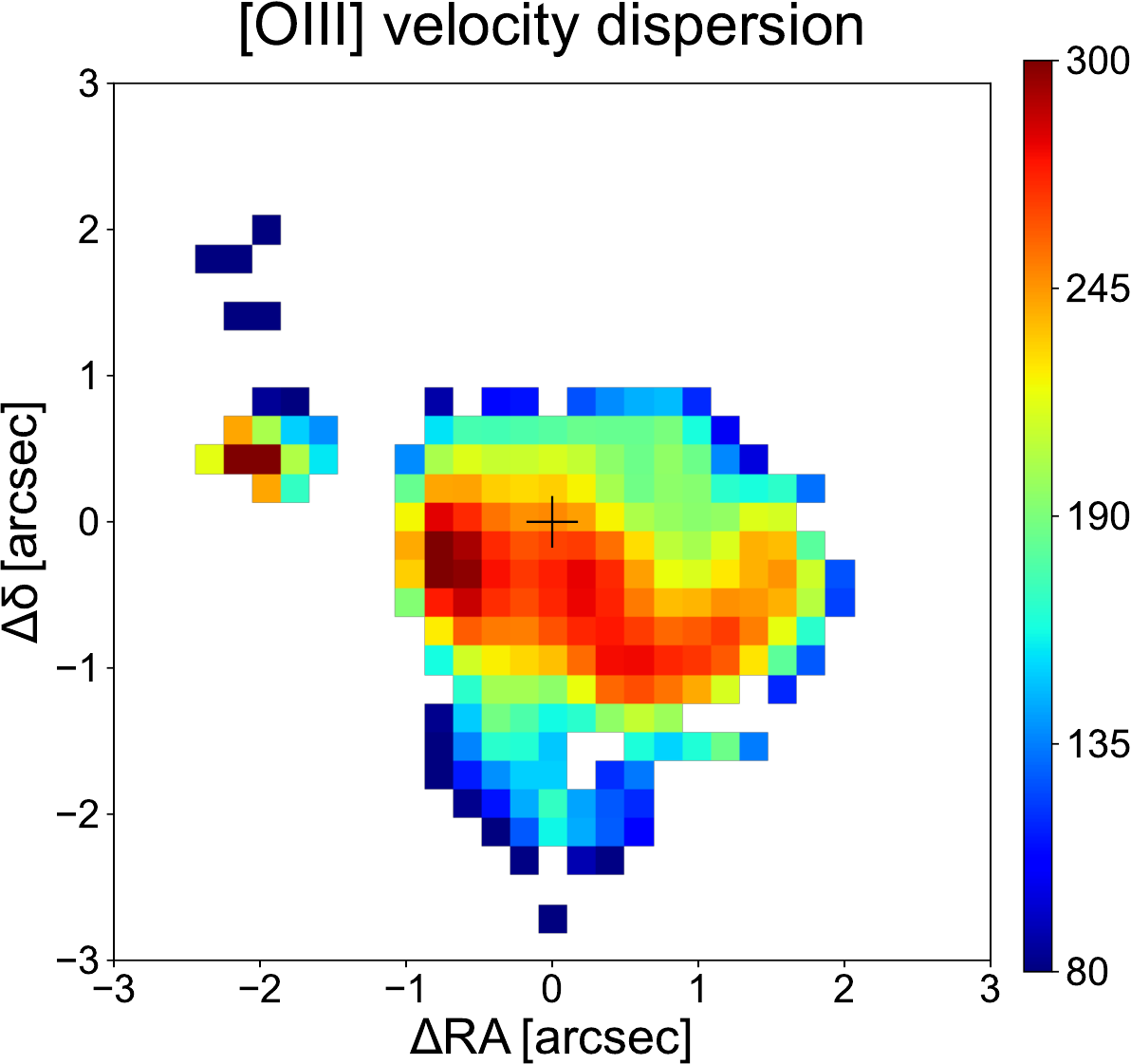}

\includegraphics[width=0.205\linewidth]{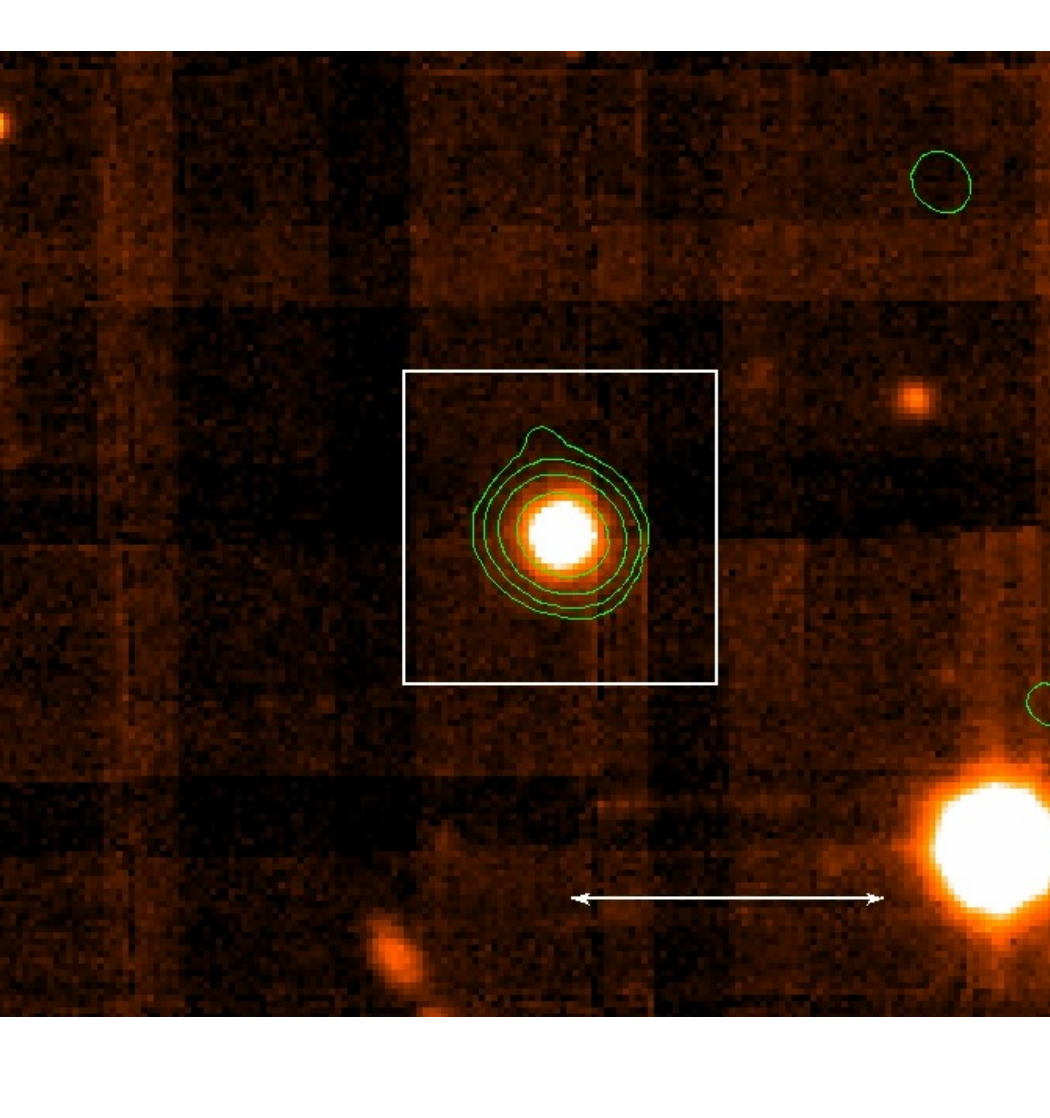}
\includegraphics[width=0.255\linewidth]{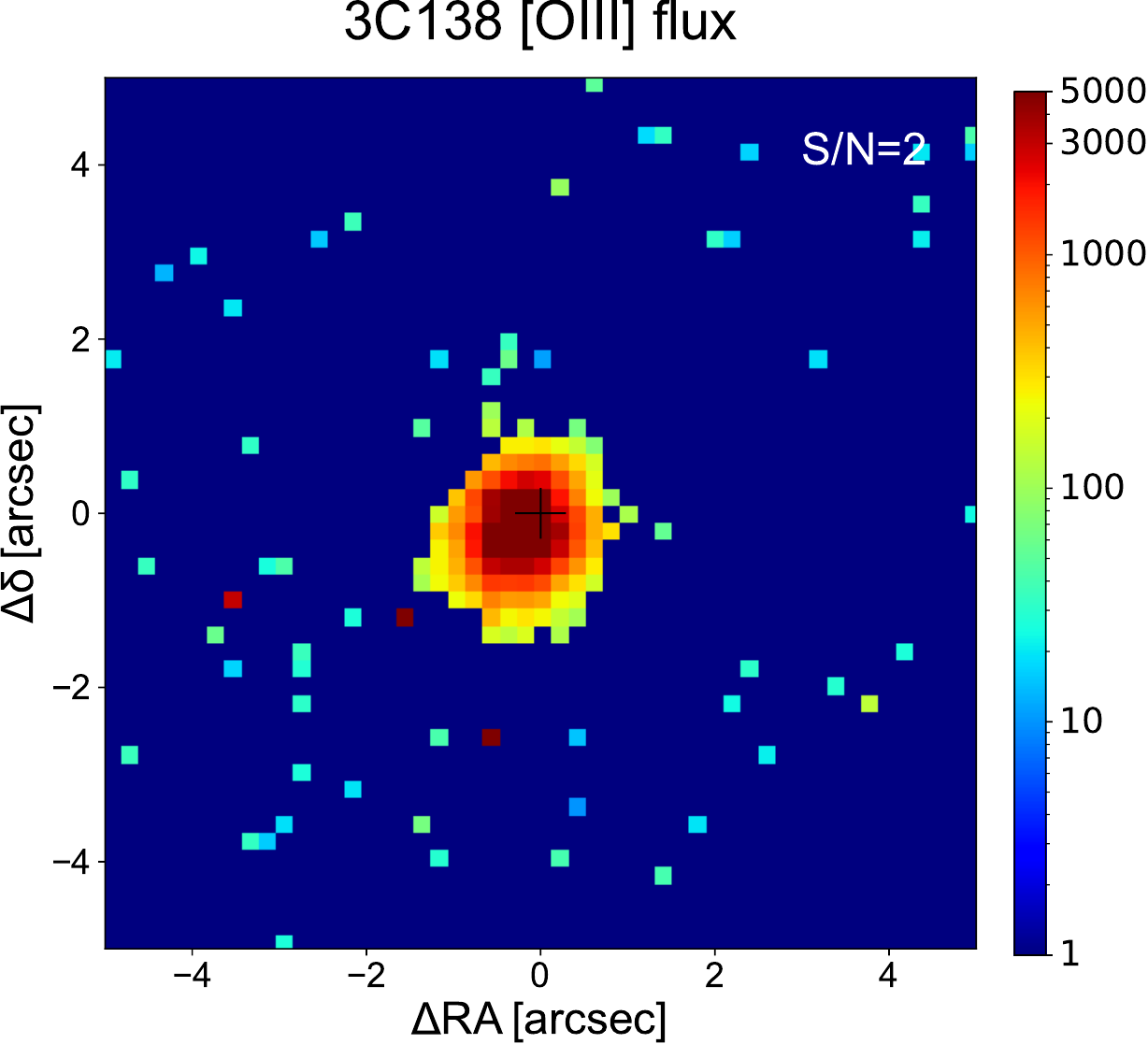}
\includegraphics[width=0.25\linewidth]{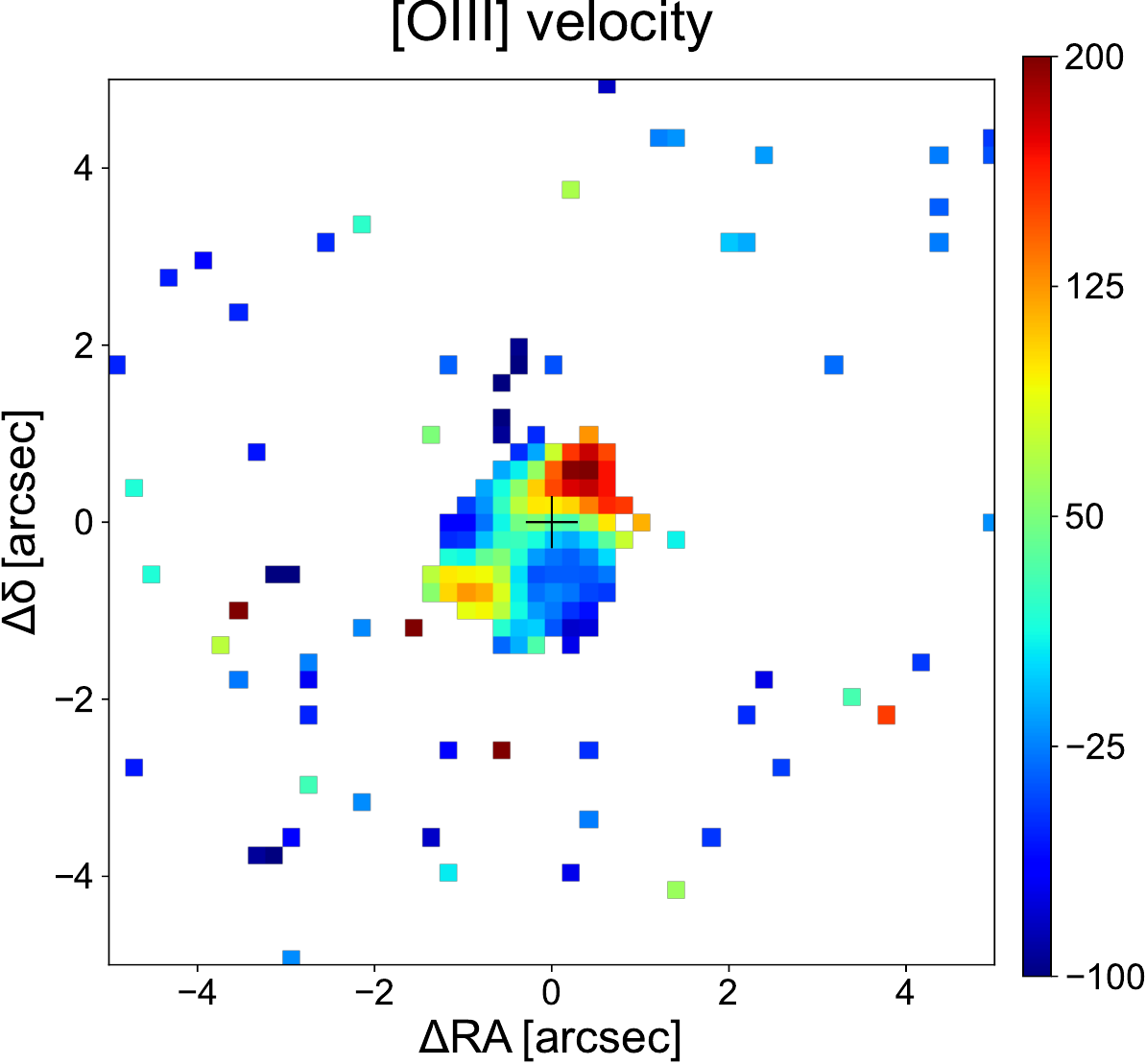}
\includegraphics[width=0.25\linewidth]{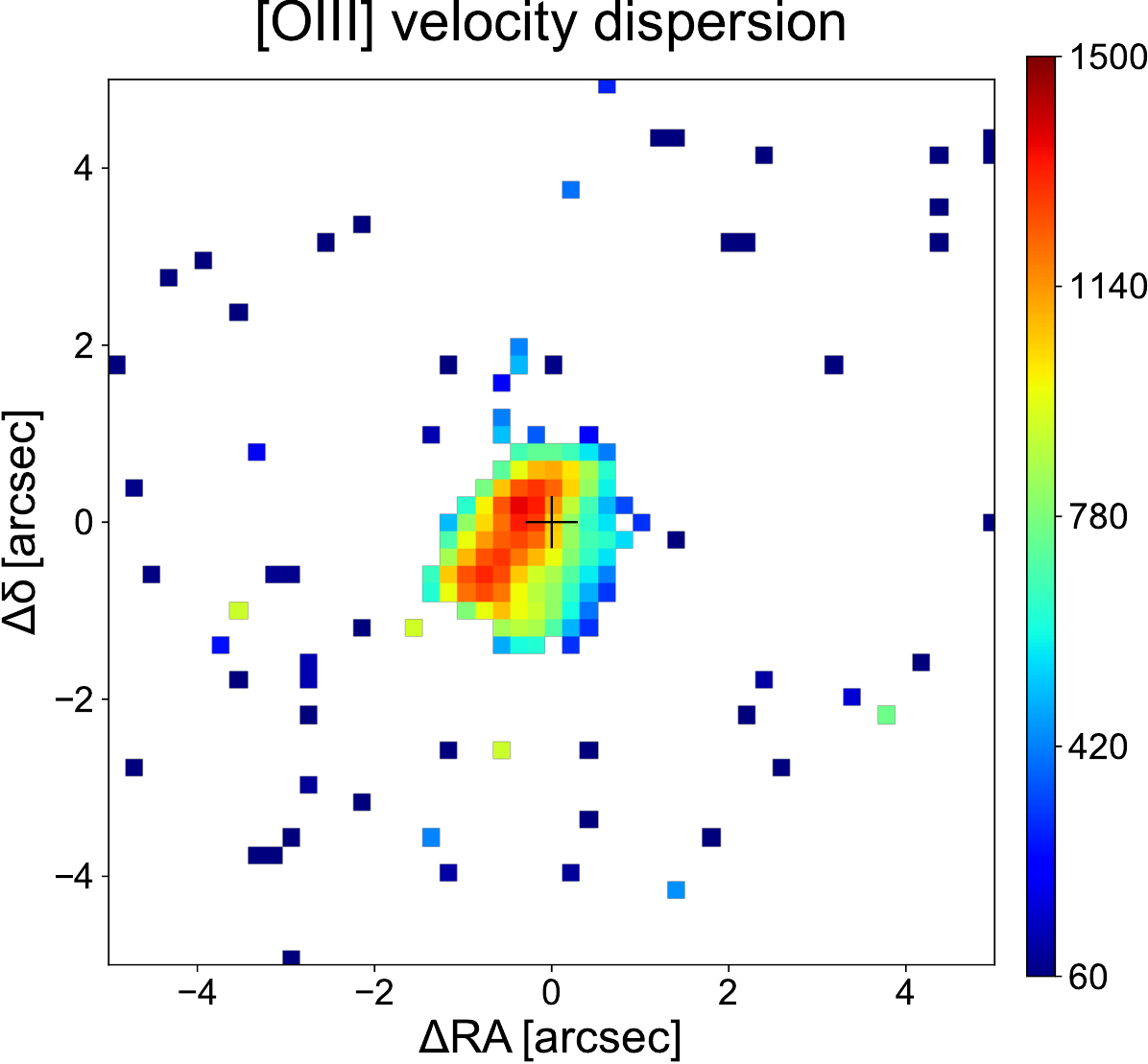}

\includegraphics[width=0.20\linewidth]{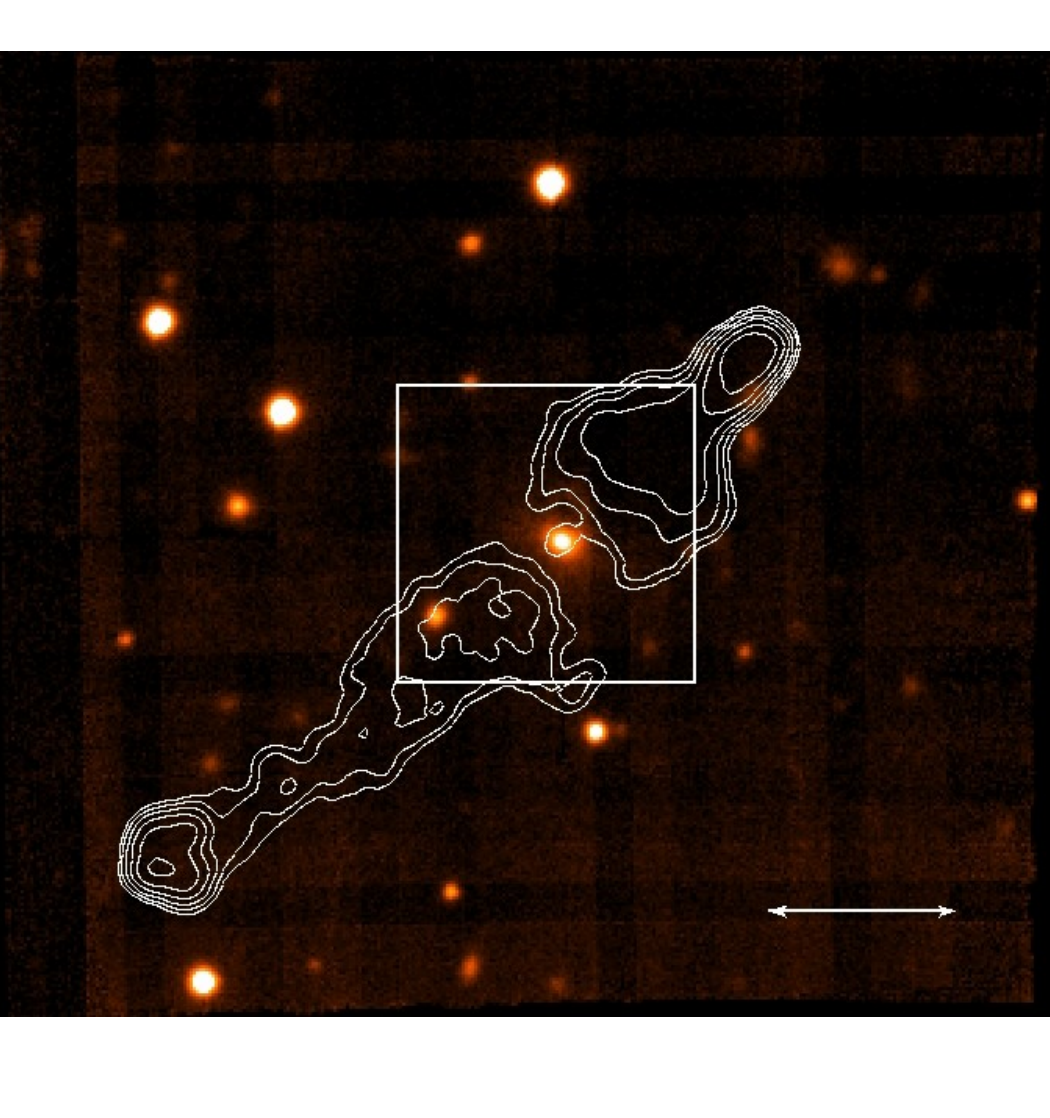}
\includegraphics[width=0.25\linewidth]{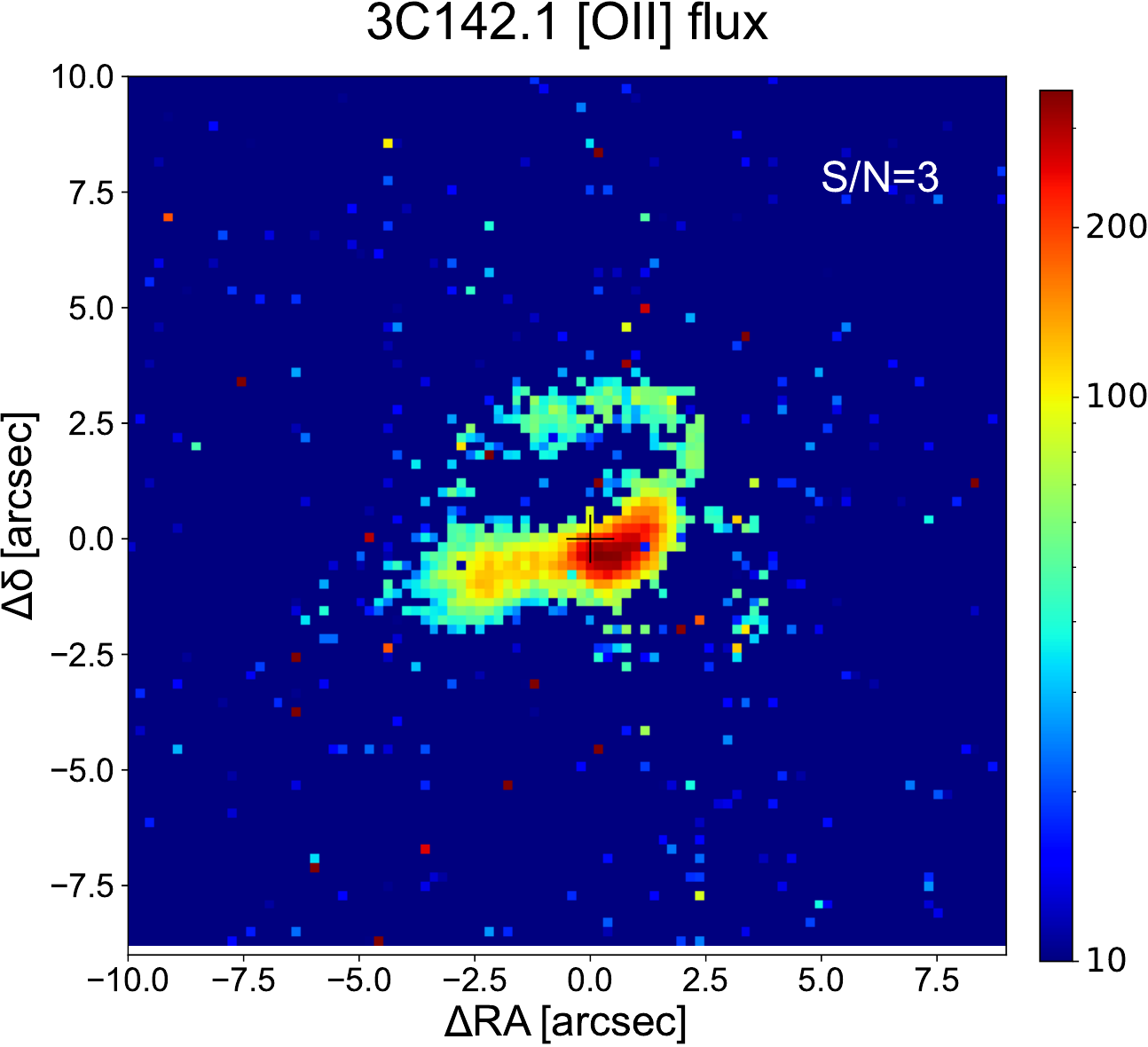}
\includegraphics[width=0.25\linewidth]{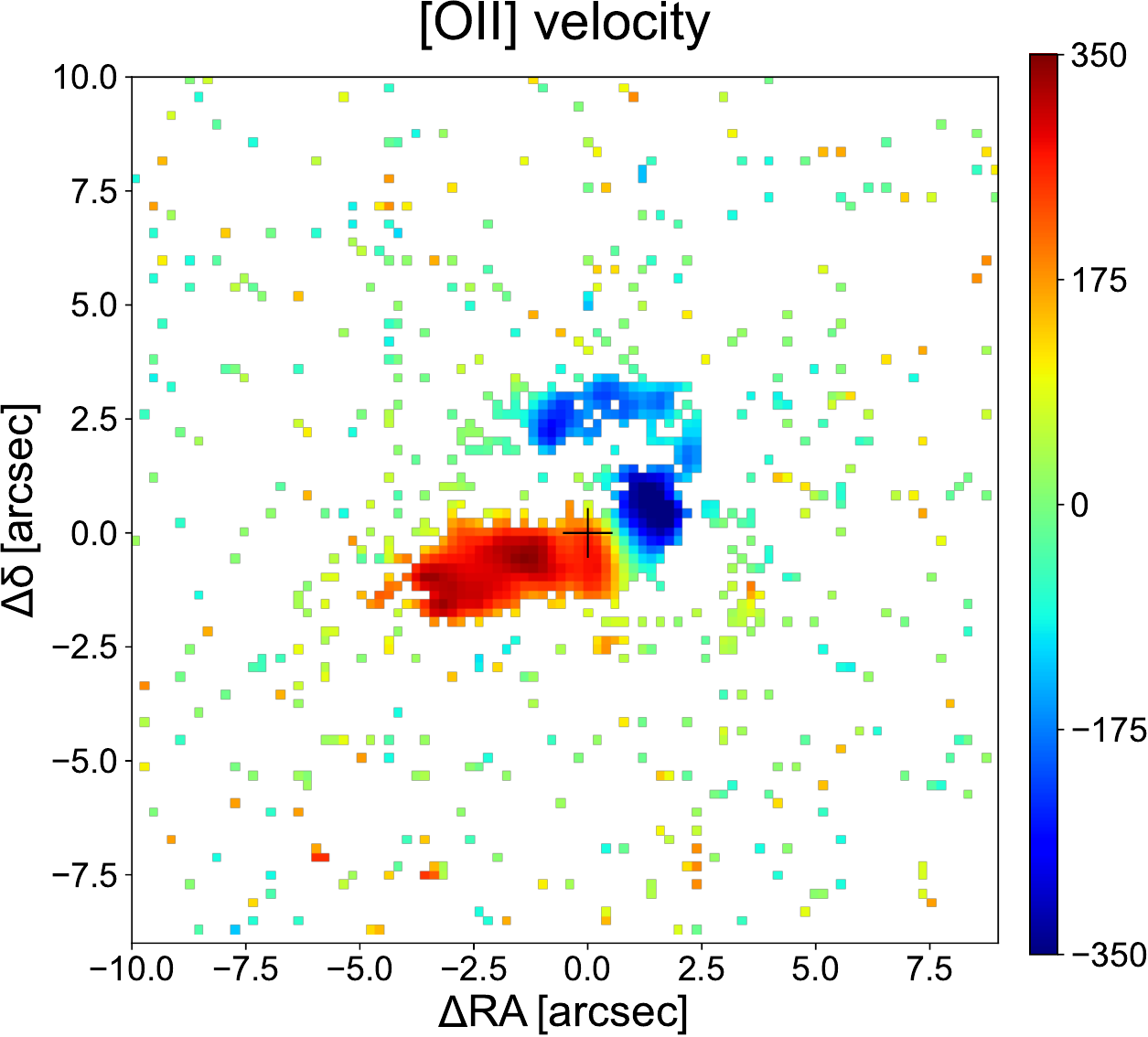}
\includegraphics[width=0.25\linewidth]{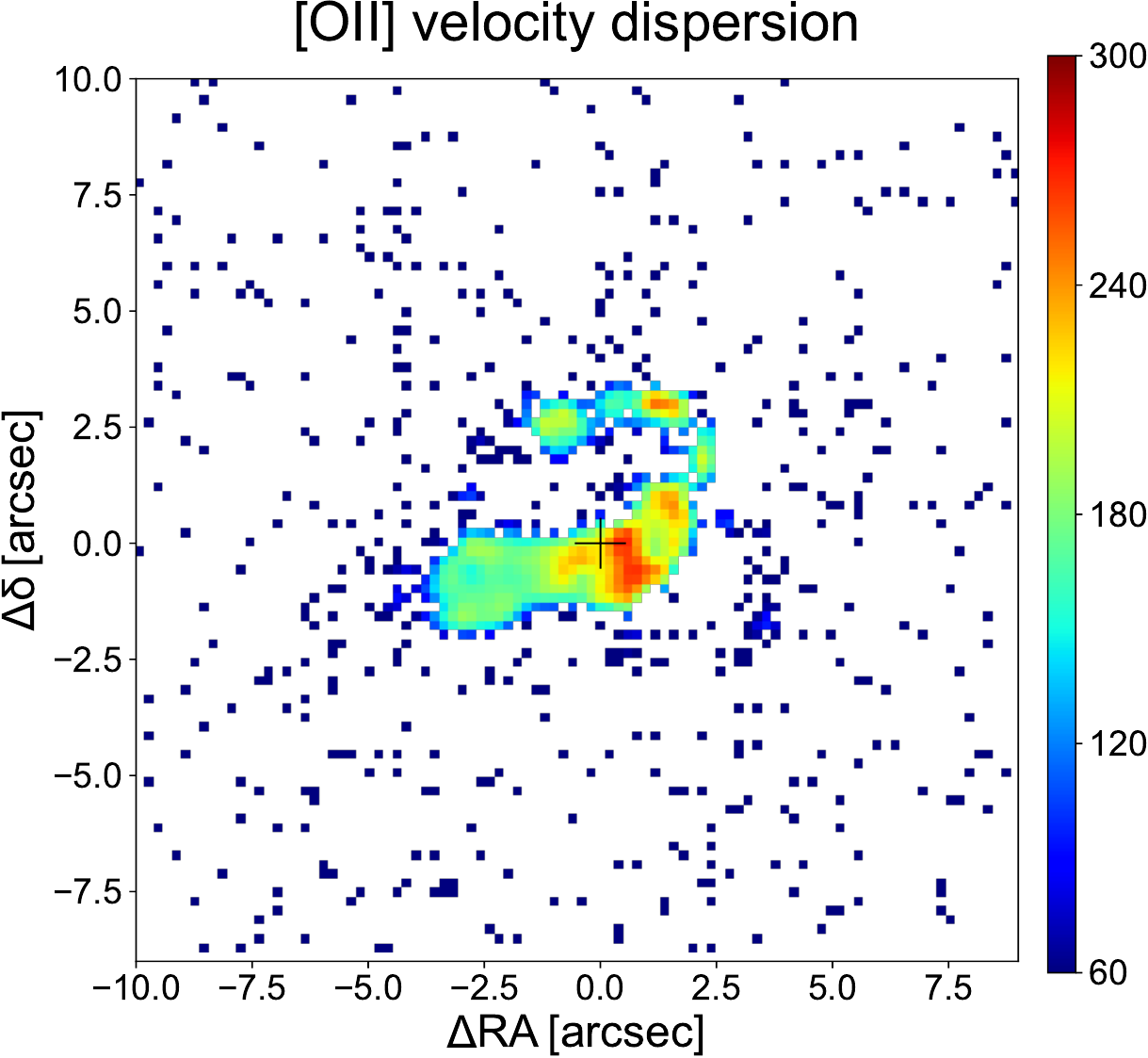}

\includegraphics[width=0.205\linewidth]{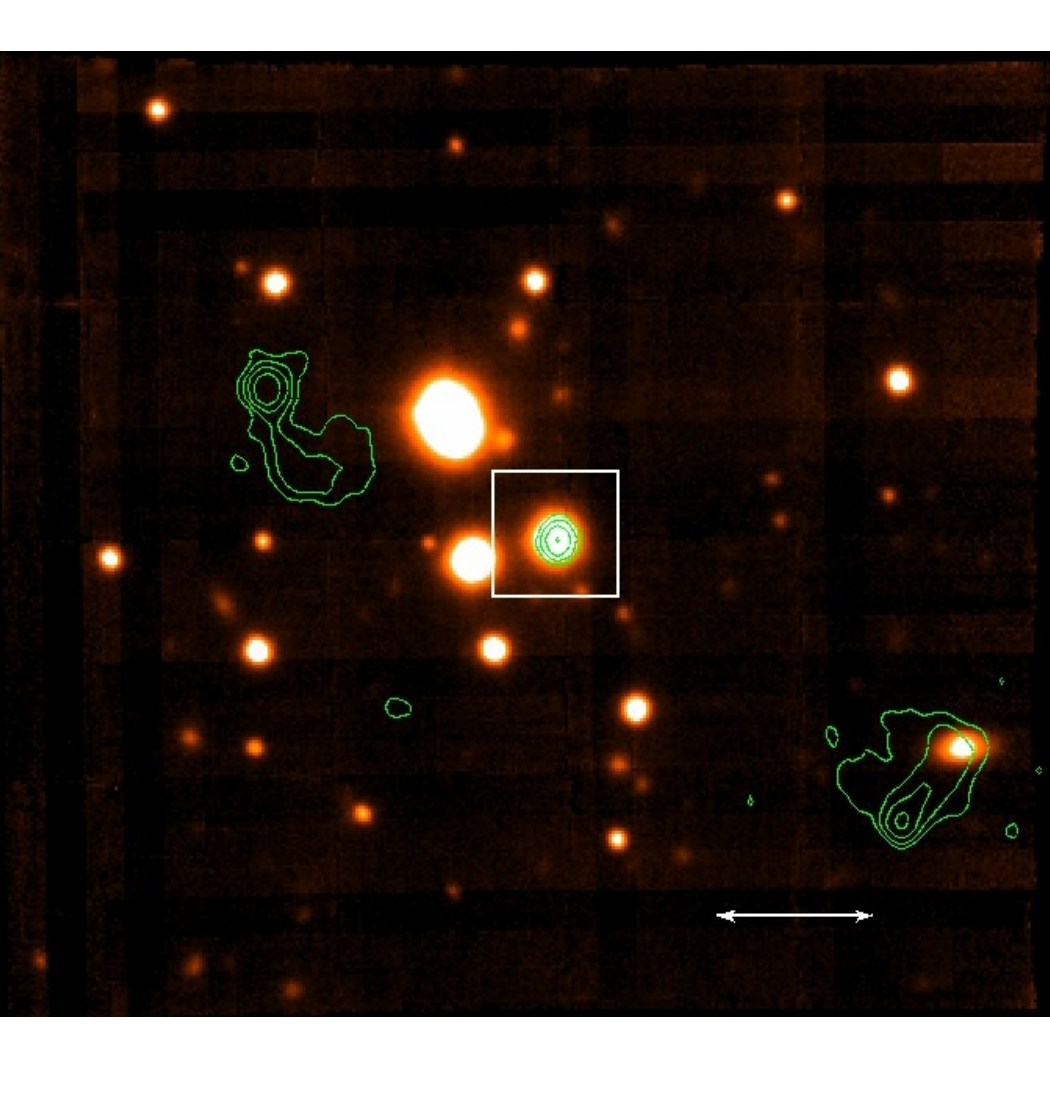}
\includegraphics[width=0.27\linewidth]{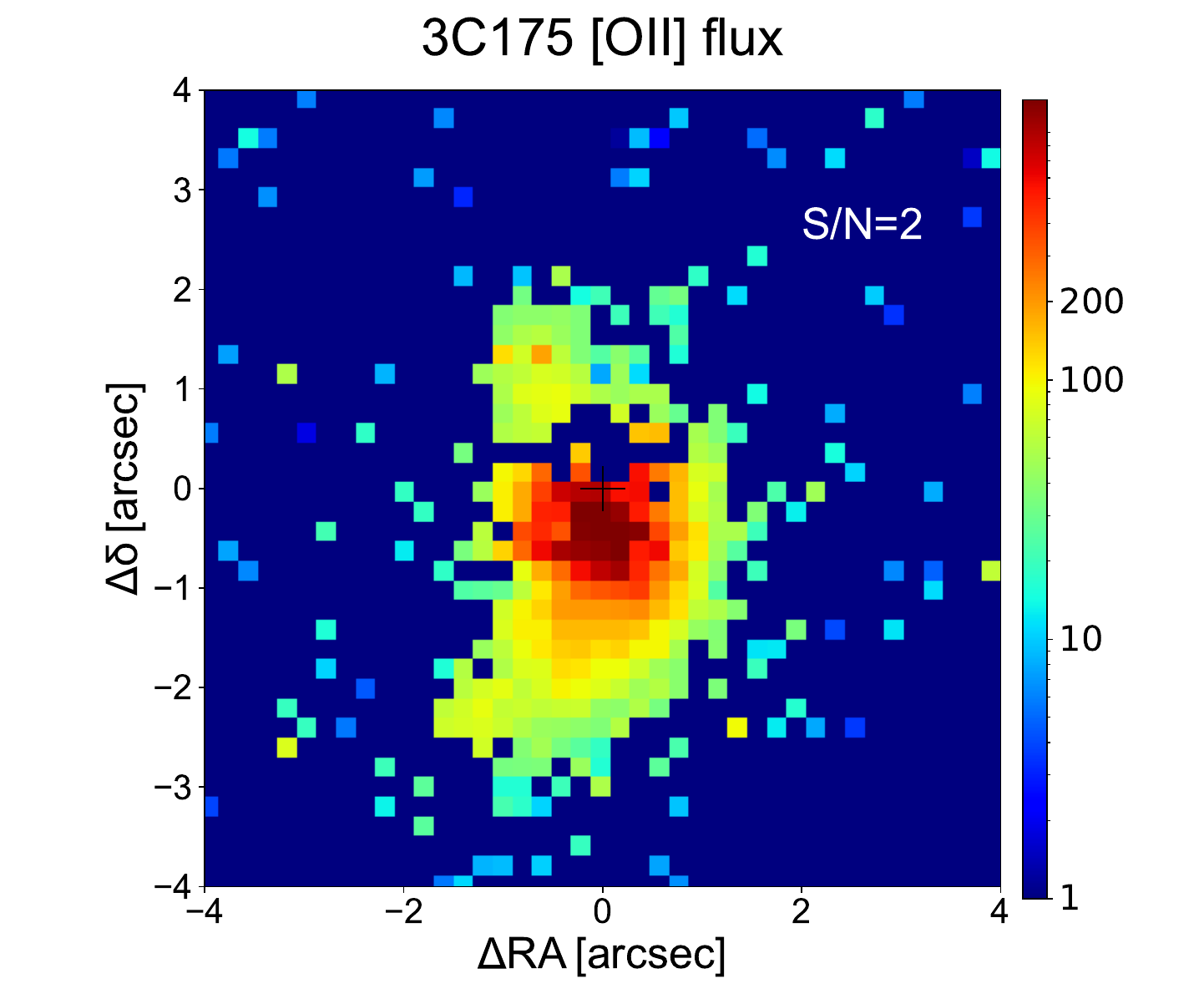}
\includegraphics[width=0.25\linewidth]{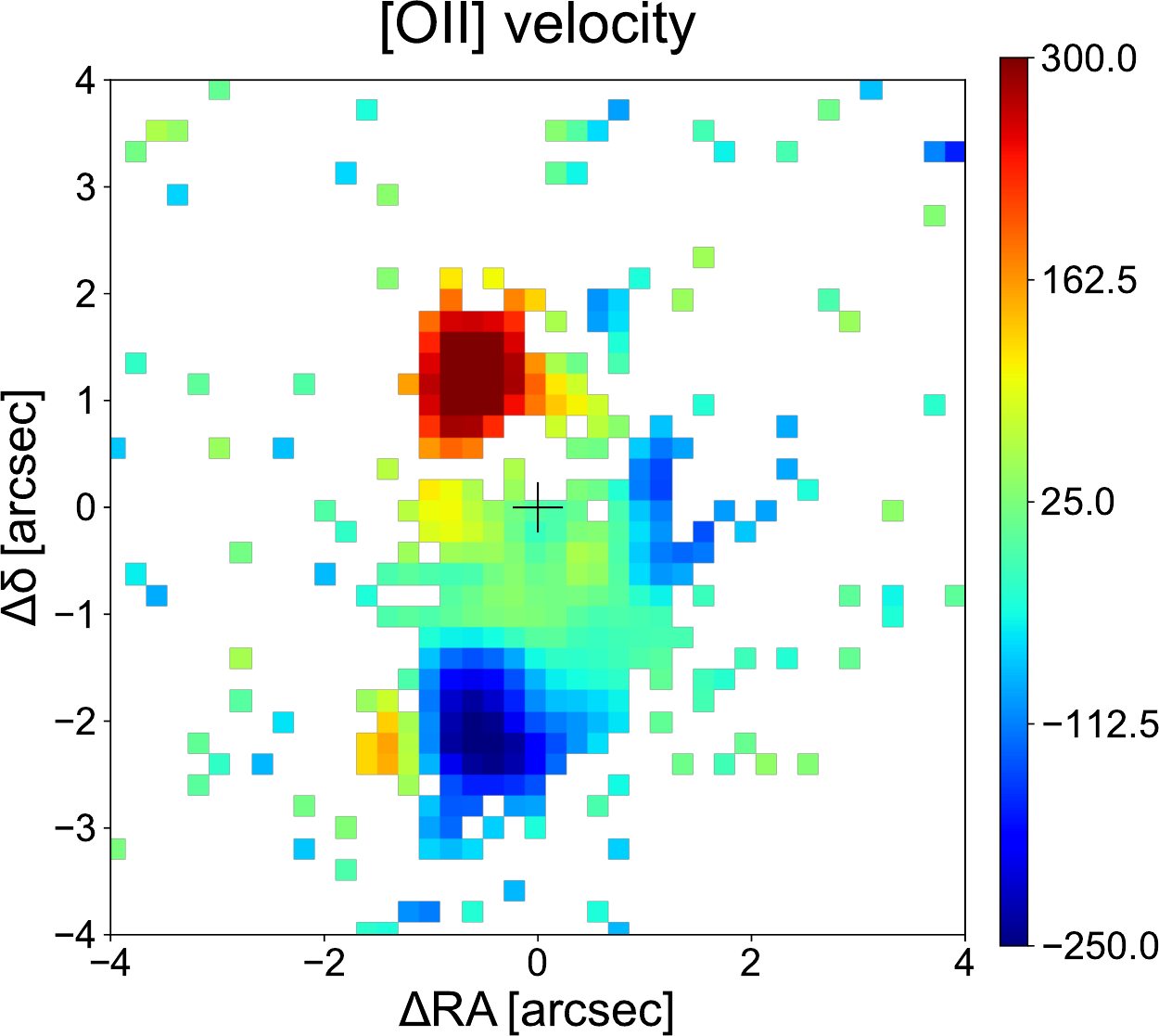}
\includegraphics[width=0.25\linewidth]{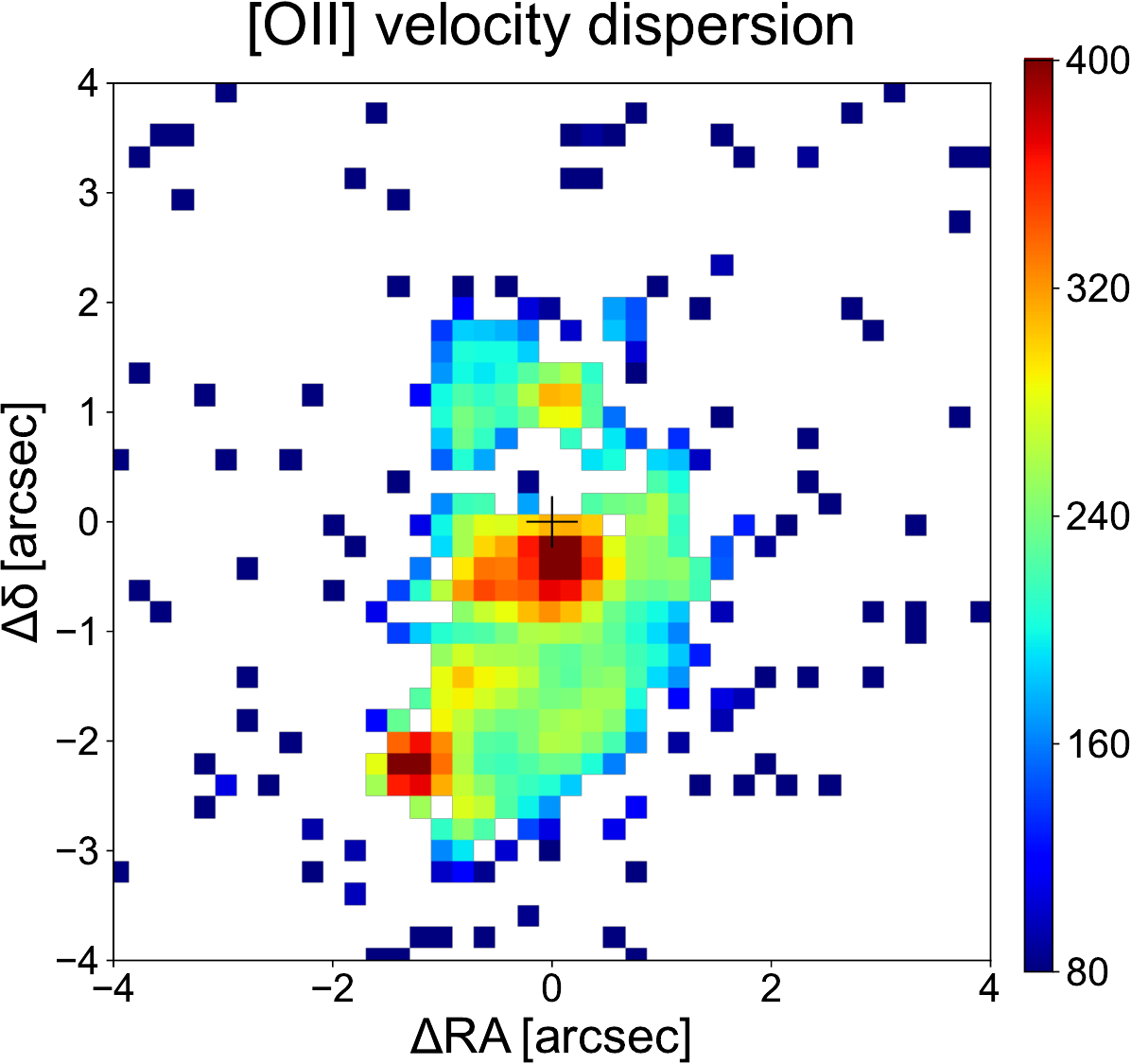}

\includegraphics[width=0.21\linewidth]{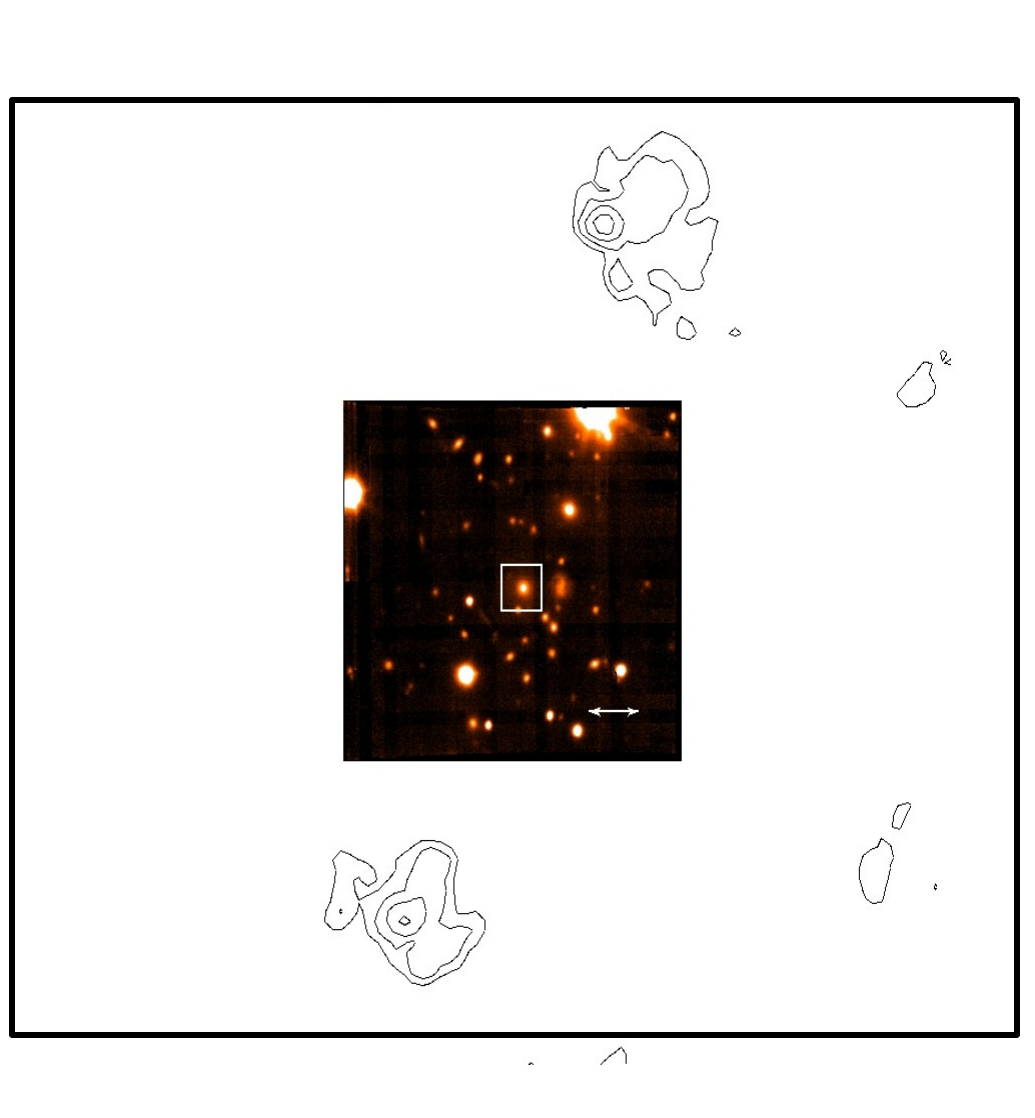}
\includegraphics[width=0.25\linewidth]{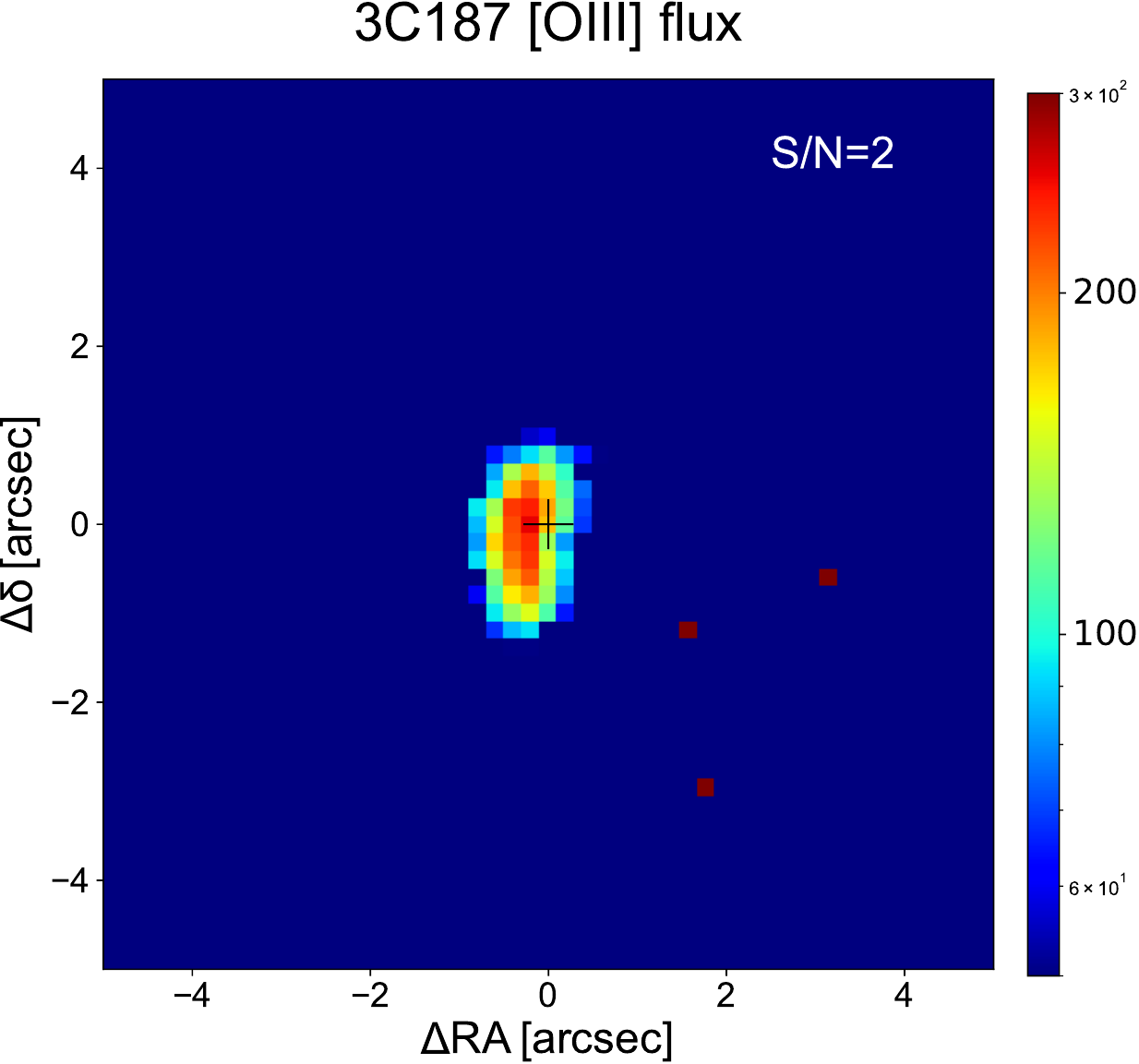}
\includegraphics[width=0.25\linewidth]{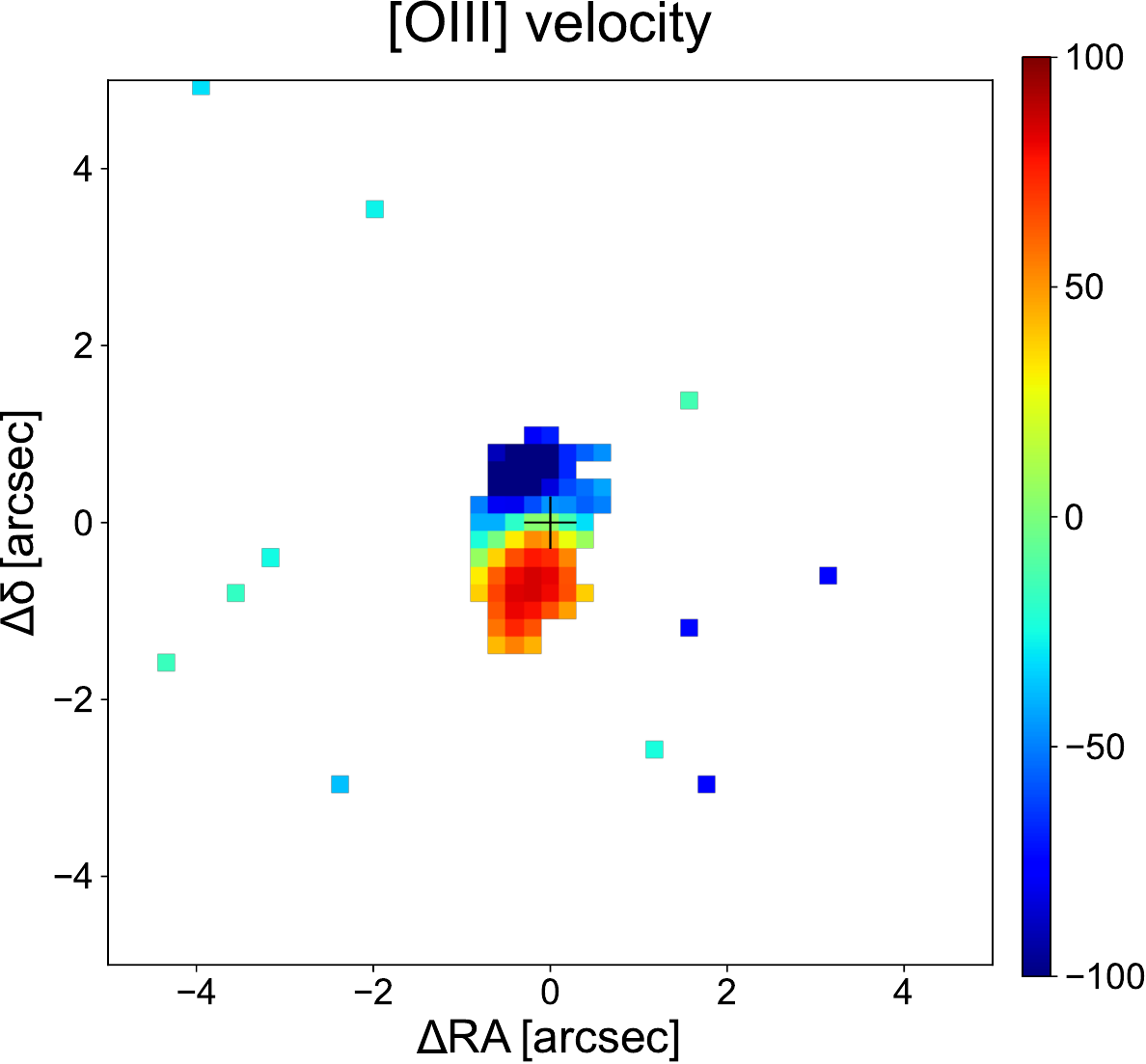}
\includegraphics[width=0.25\linewidth]{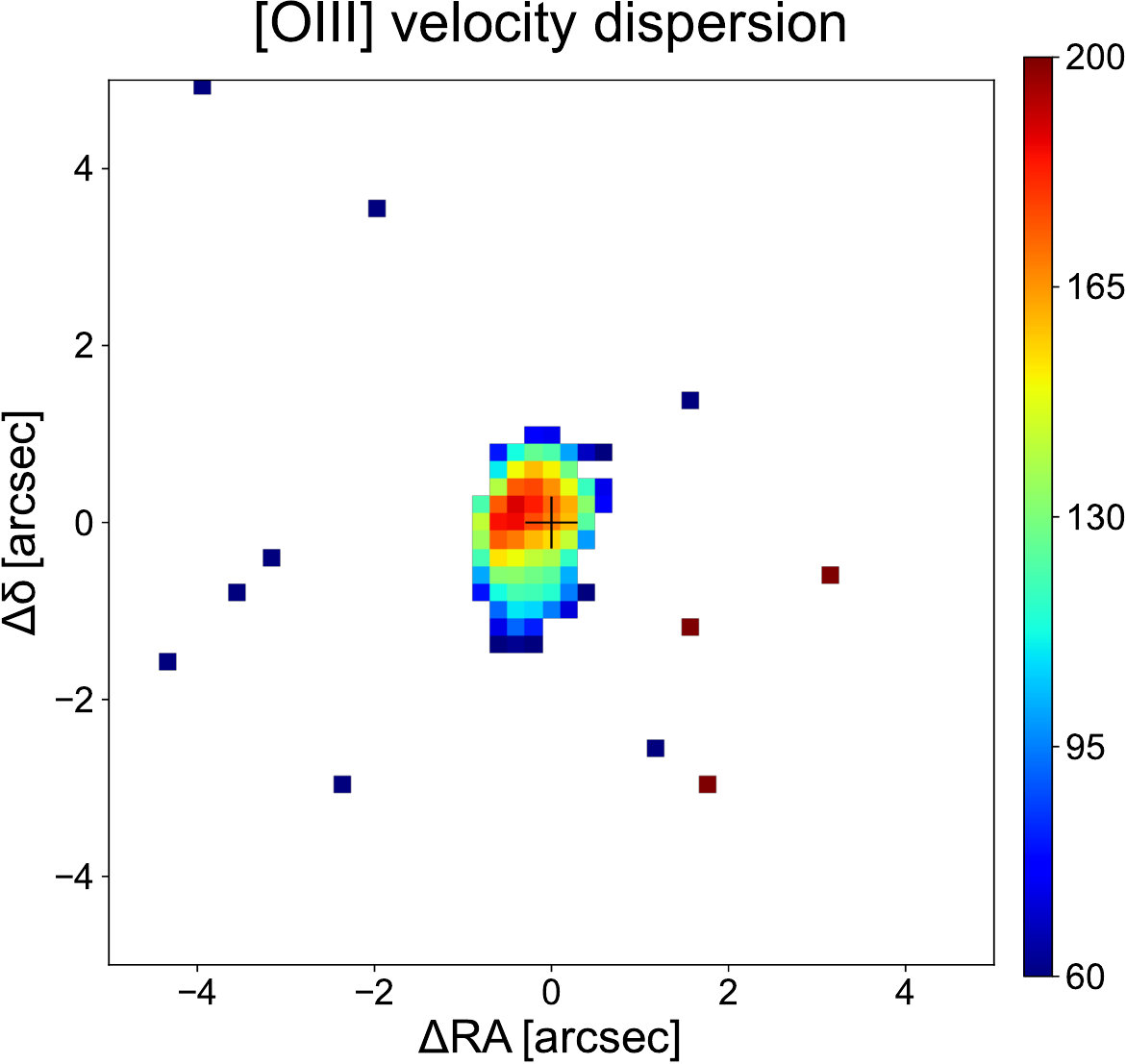}
\caption{- continued}
    \label{fig1}
\end{figure*}

\addtocounter{figure}{-1}
\begin{figure*}
\centering

\includegraphics[width=0.205\linewidth]{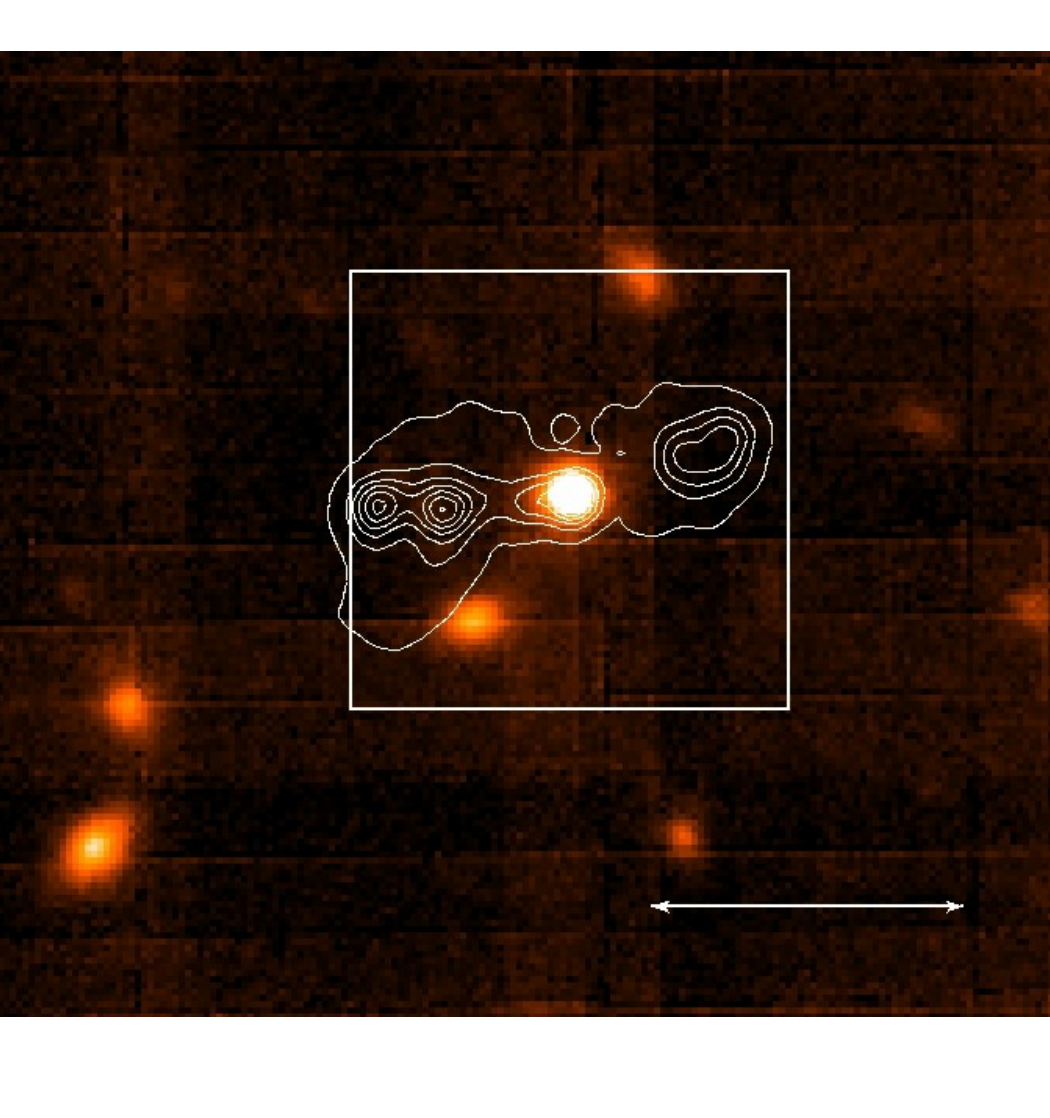}
\includegraphics[width=0.25\linewidth]{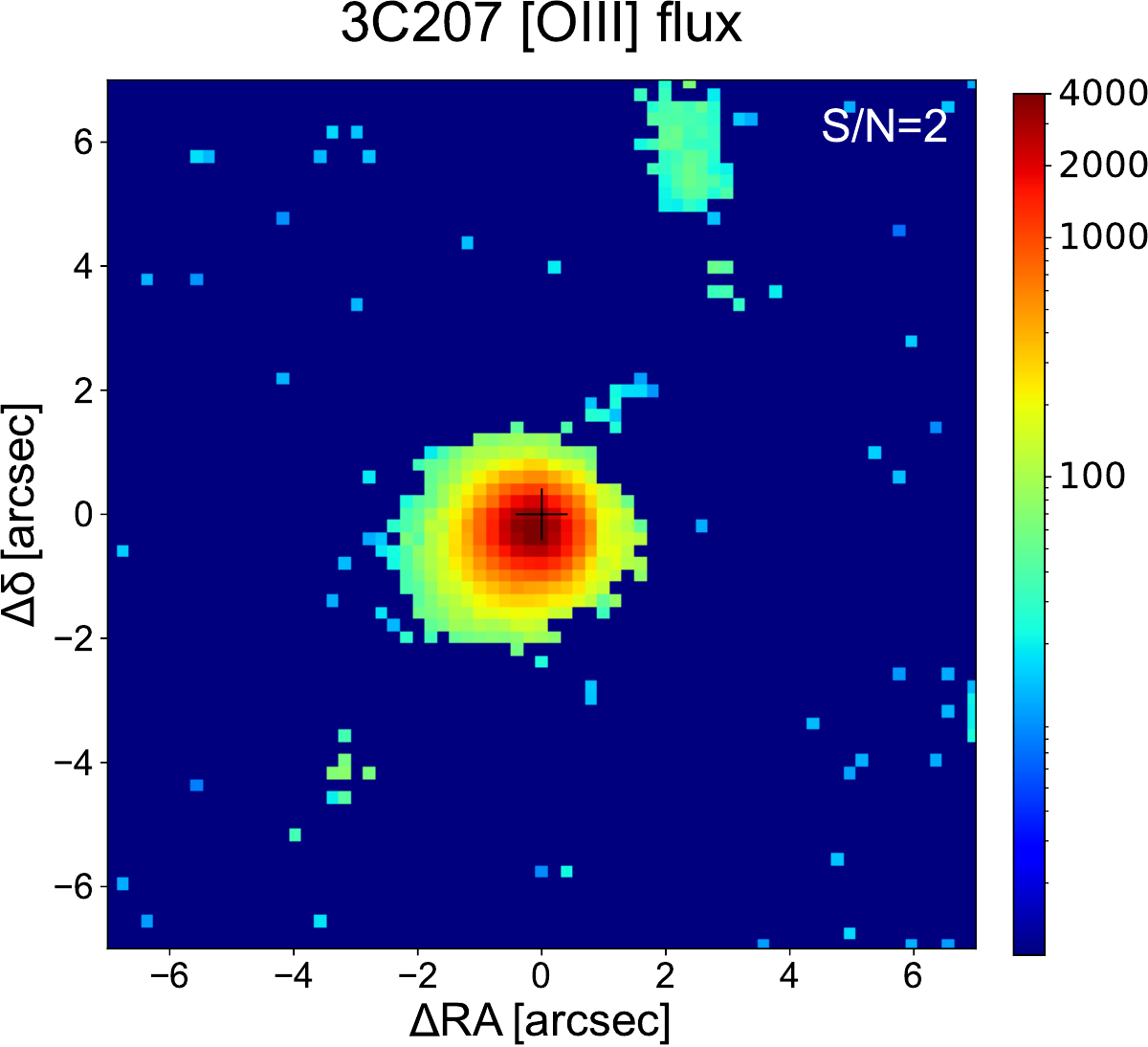}
\includegraphics[width=0.25\linewidth]{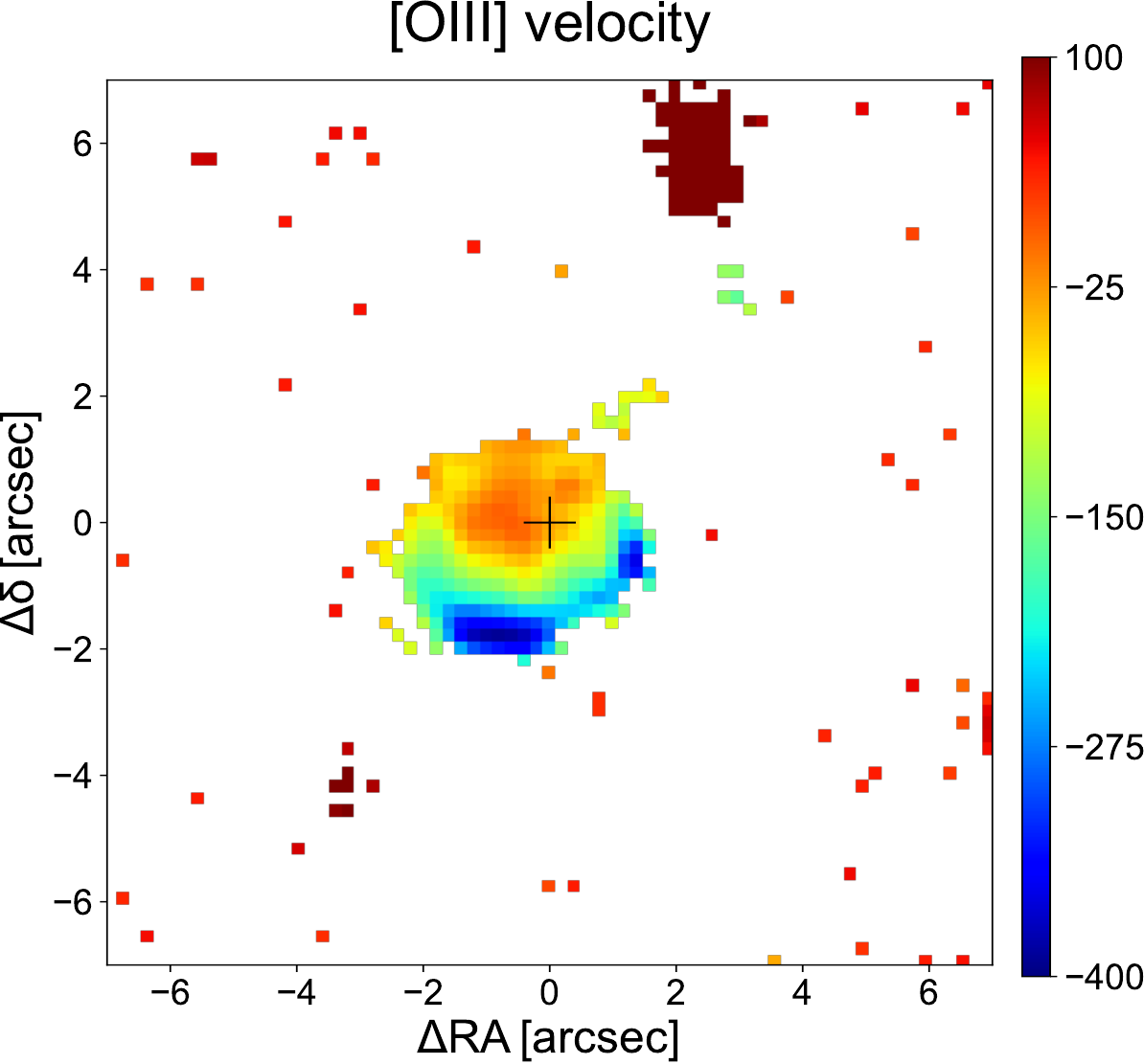}
\includegraphics[width=0.25\linewidth]{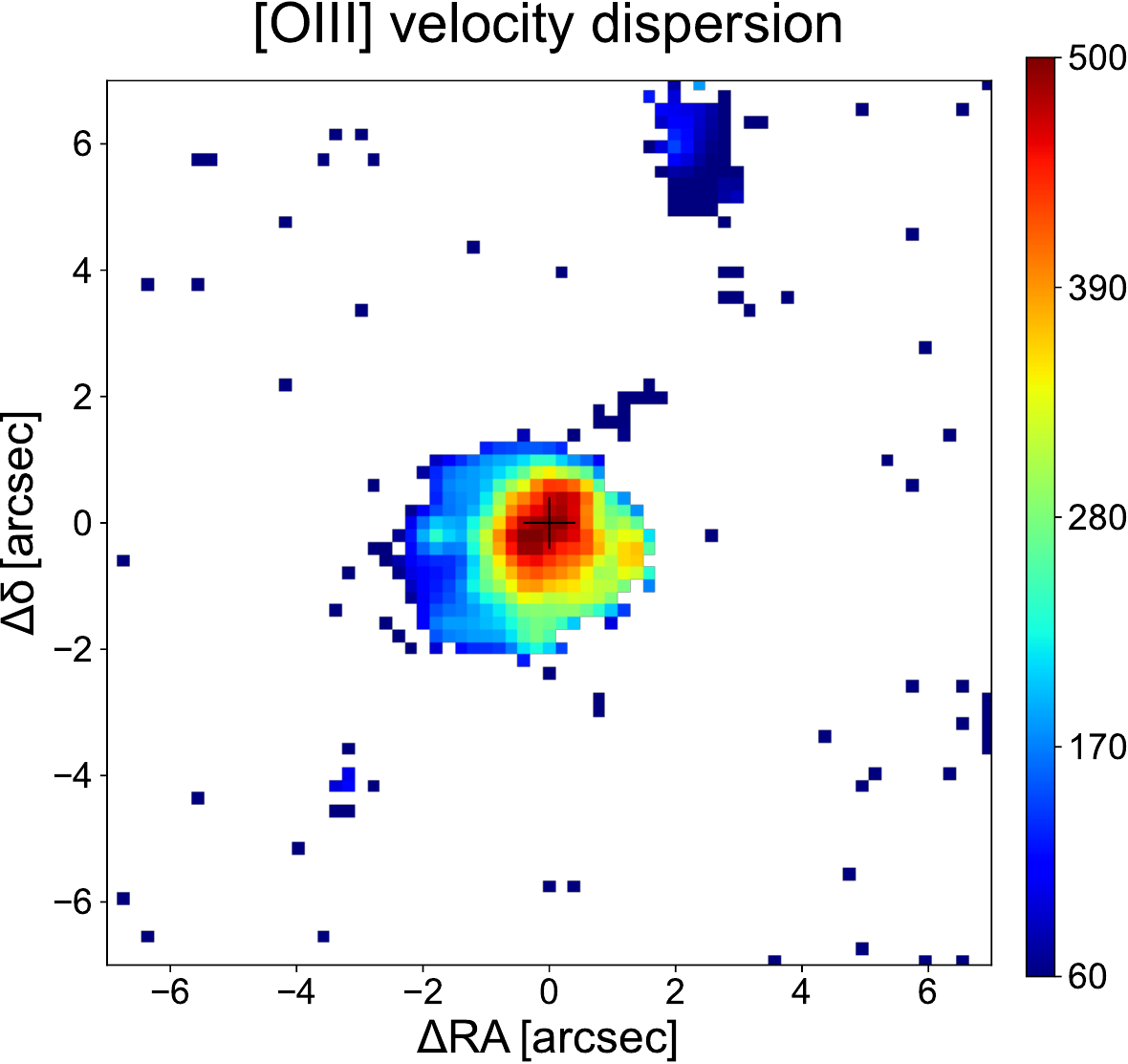}

\includegraphics[width=0.205\linewidth]{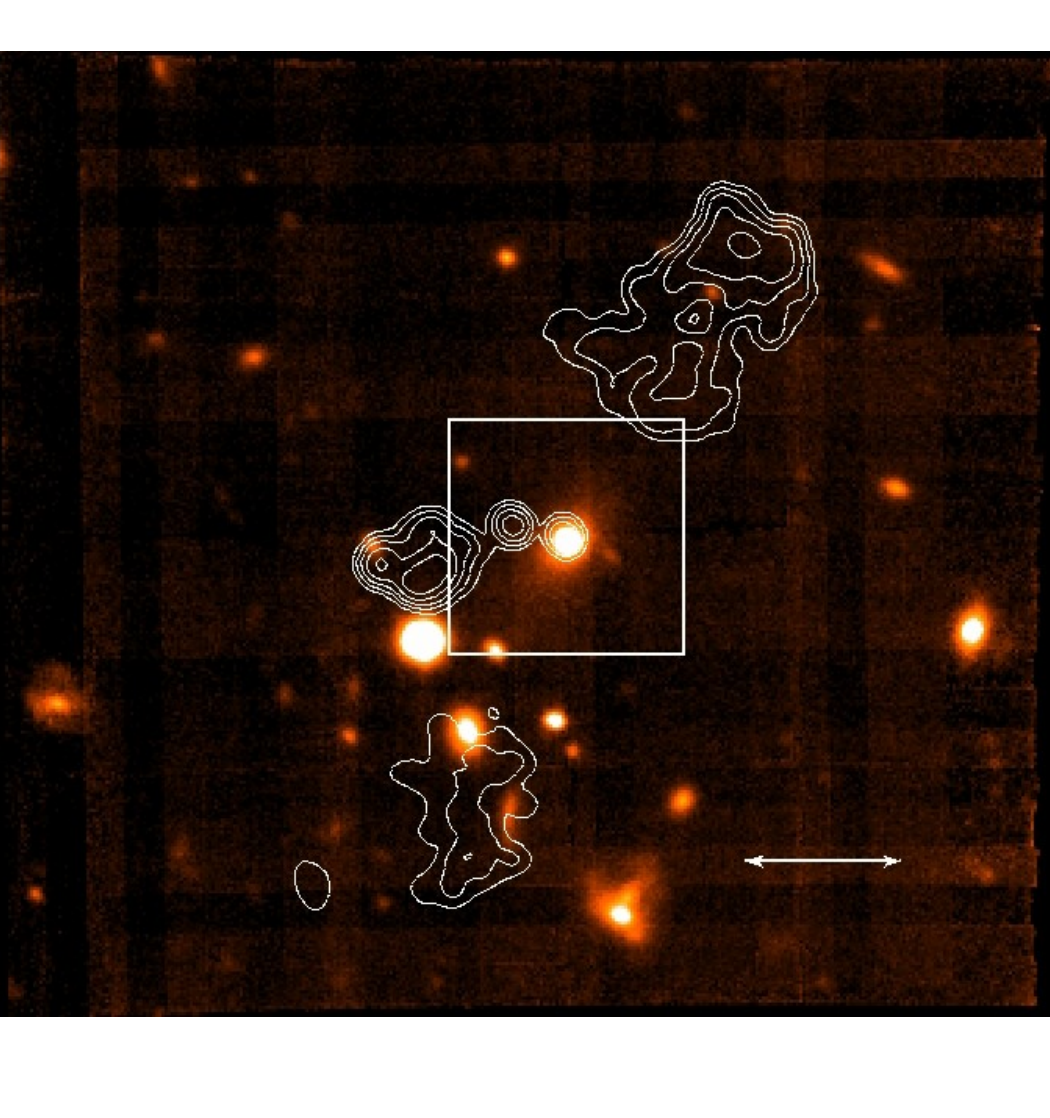}
\includegraphics[width=0.25\linewidth]{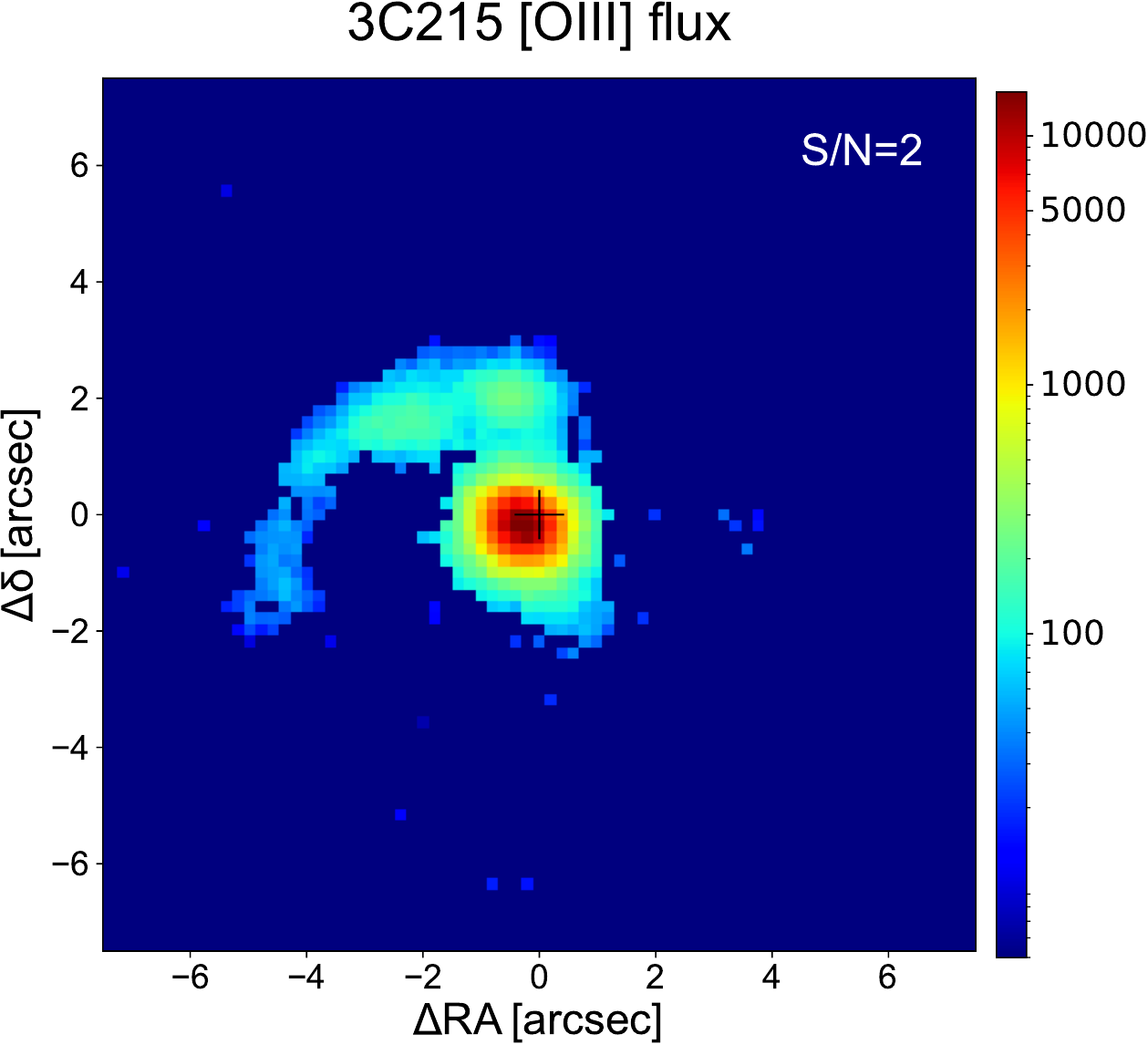}
\includegraphics[width=0.25\linewidth]{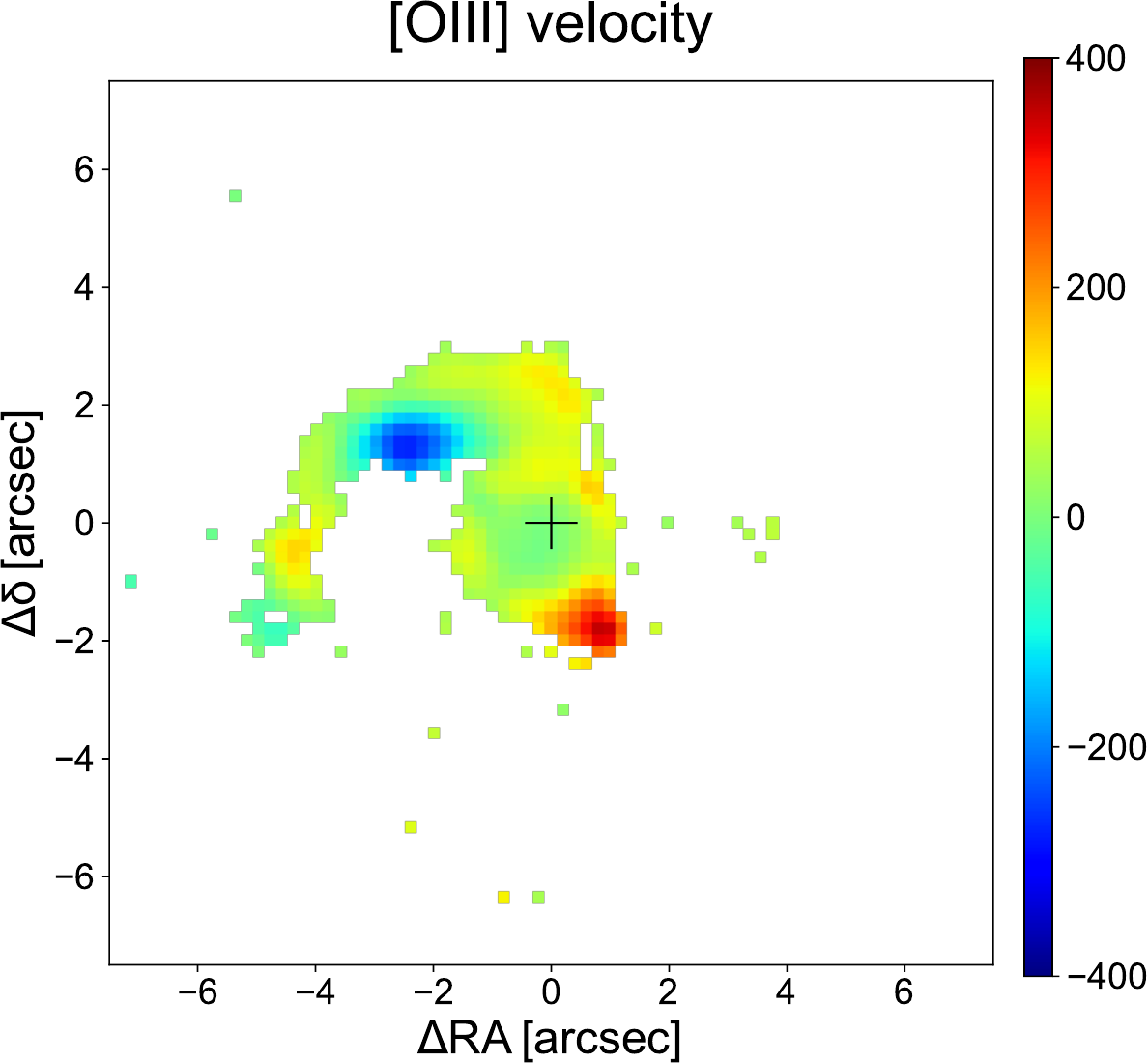}
\includegraphics[width=0.25\linewidth]{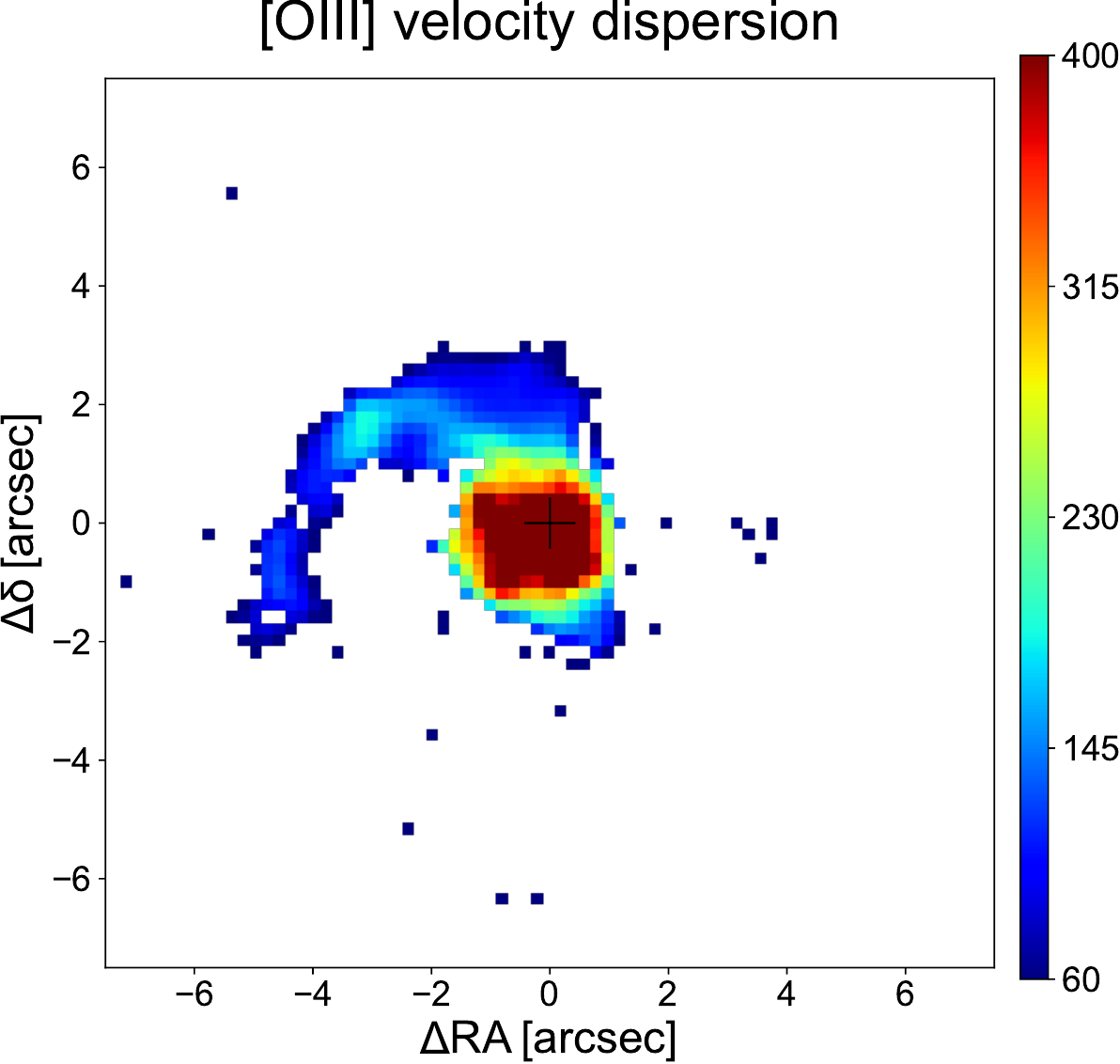}

\includegraphics[width=0.205\linewidth]{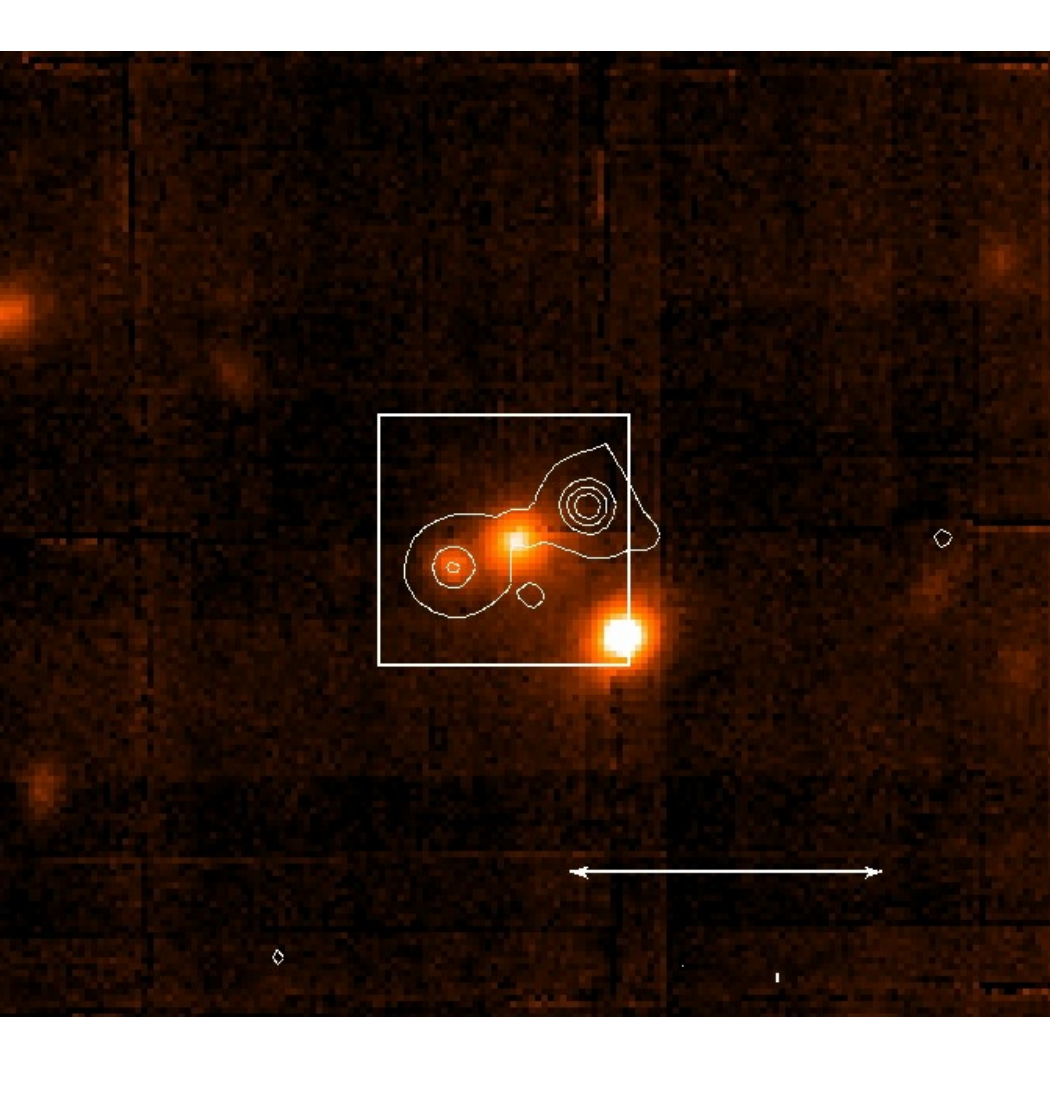}
\includegraphics[width=0.25\linewidth]{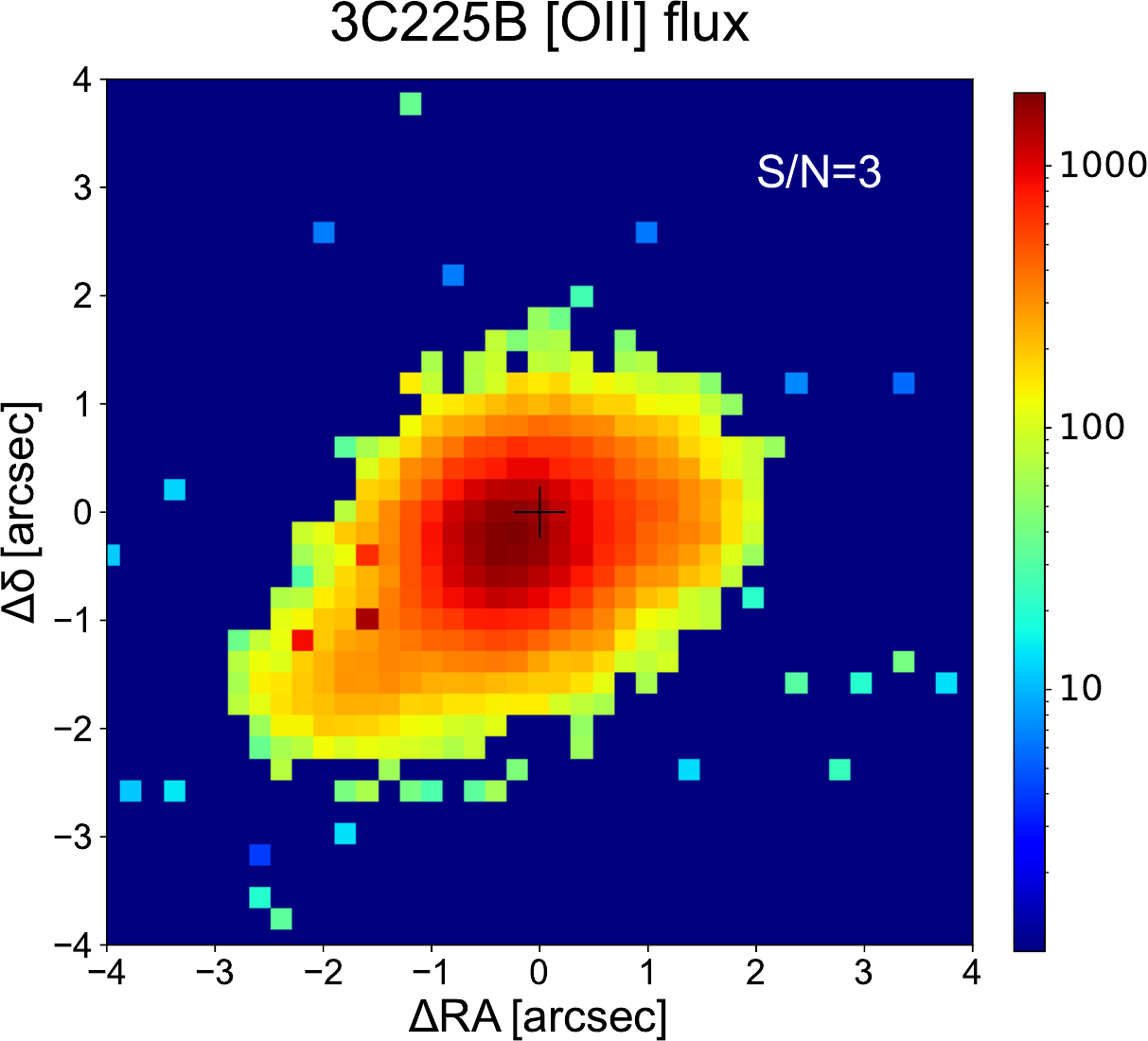}
\includegraphics[width=0.25\linewidth]{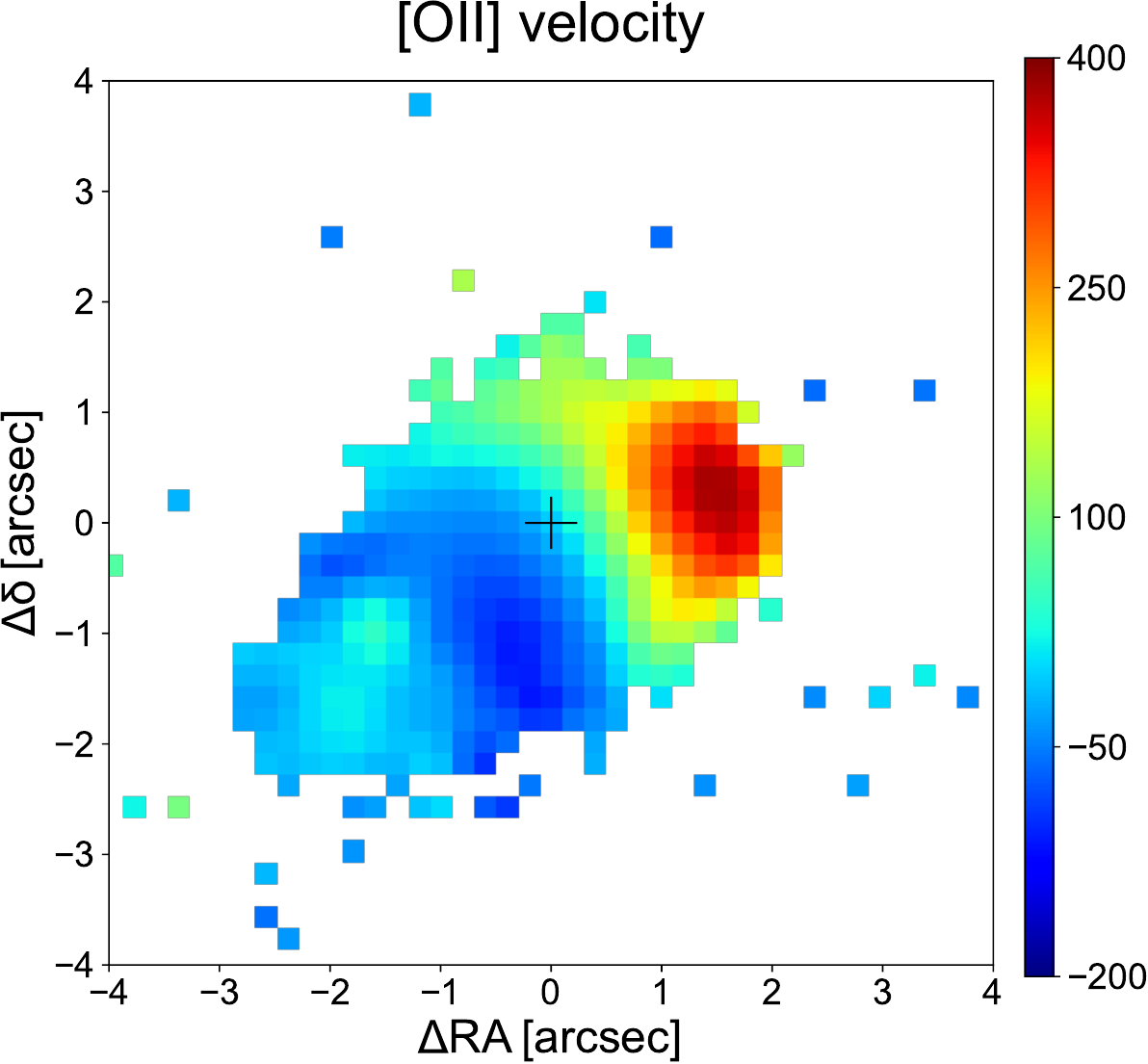}
\includegraphics[width=0.25\linewidth]{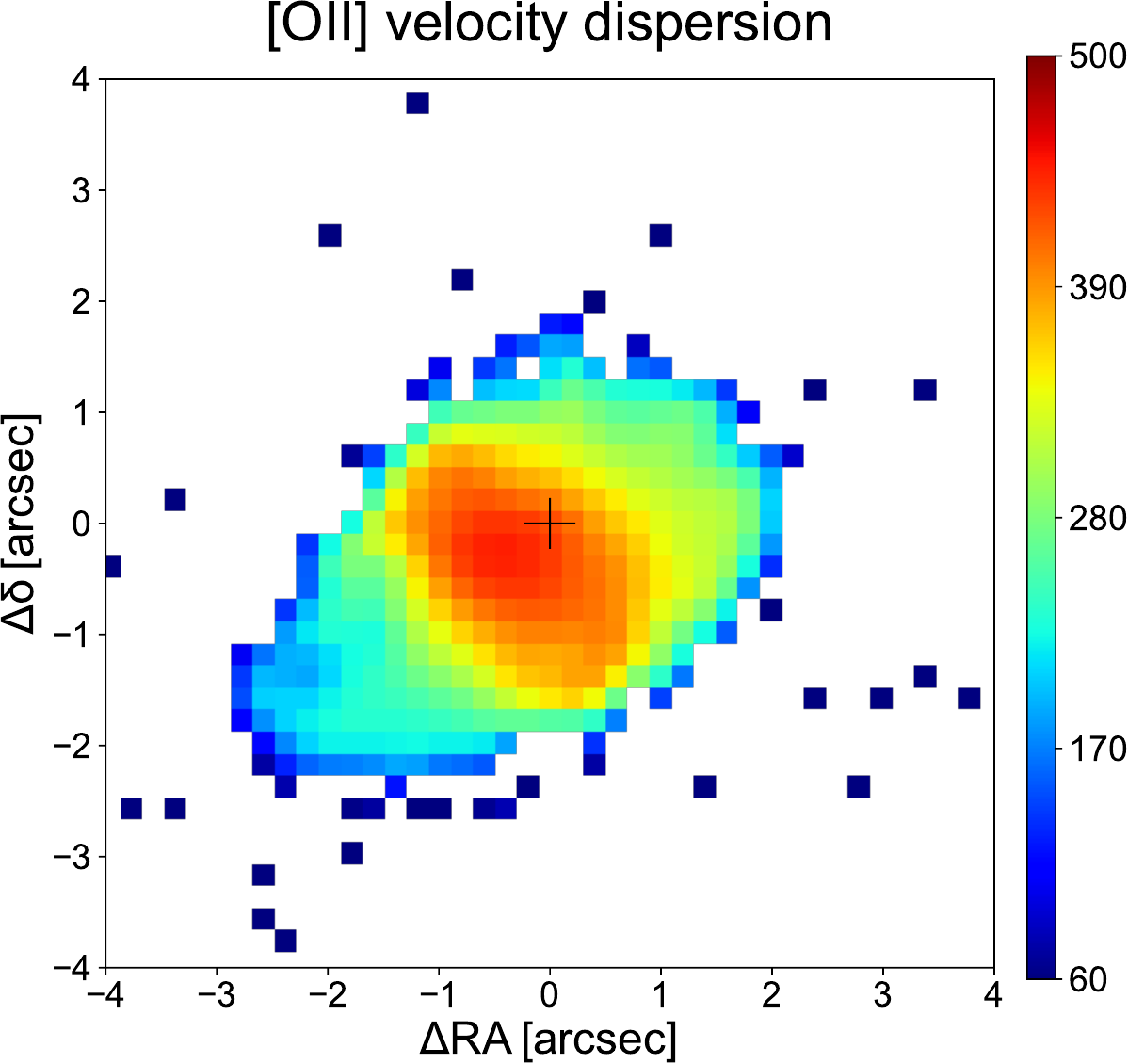}

\includegraphics[width=0.205\linewidth]{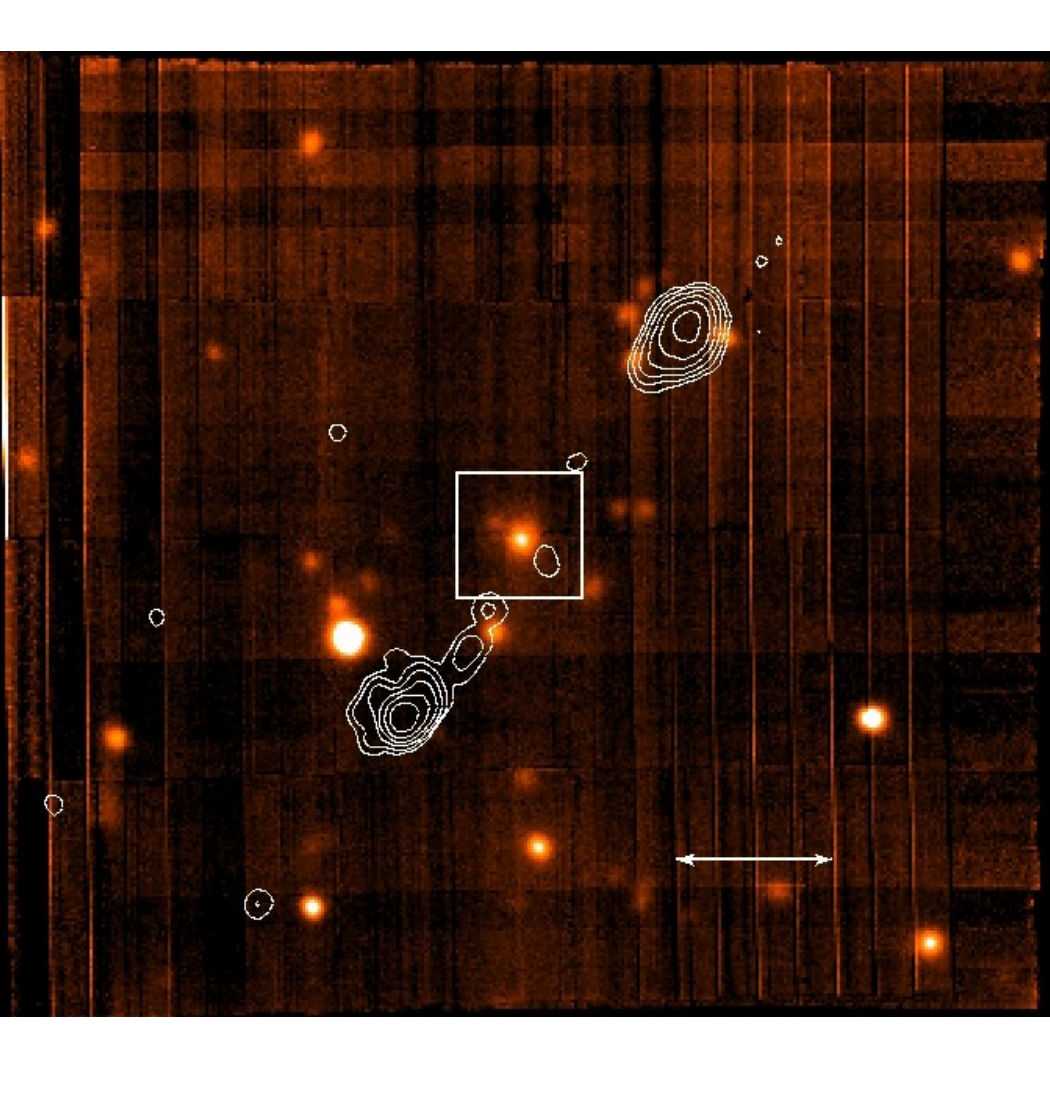}
\includegraphics[width=0.25\linewidth]{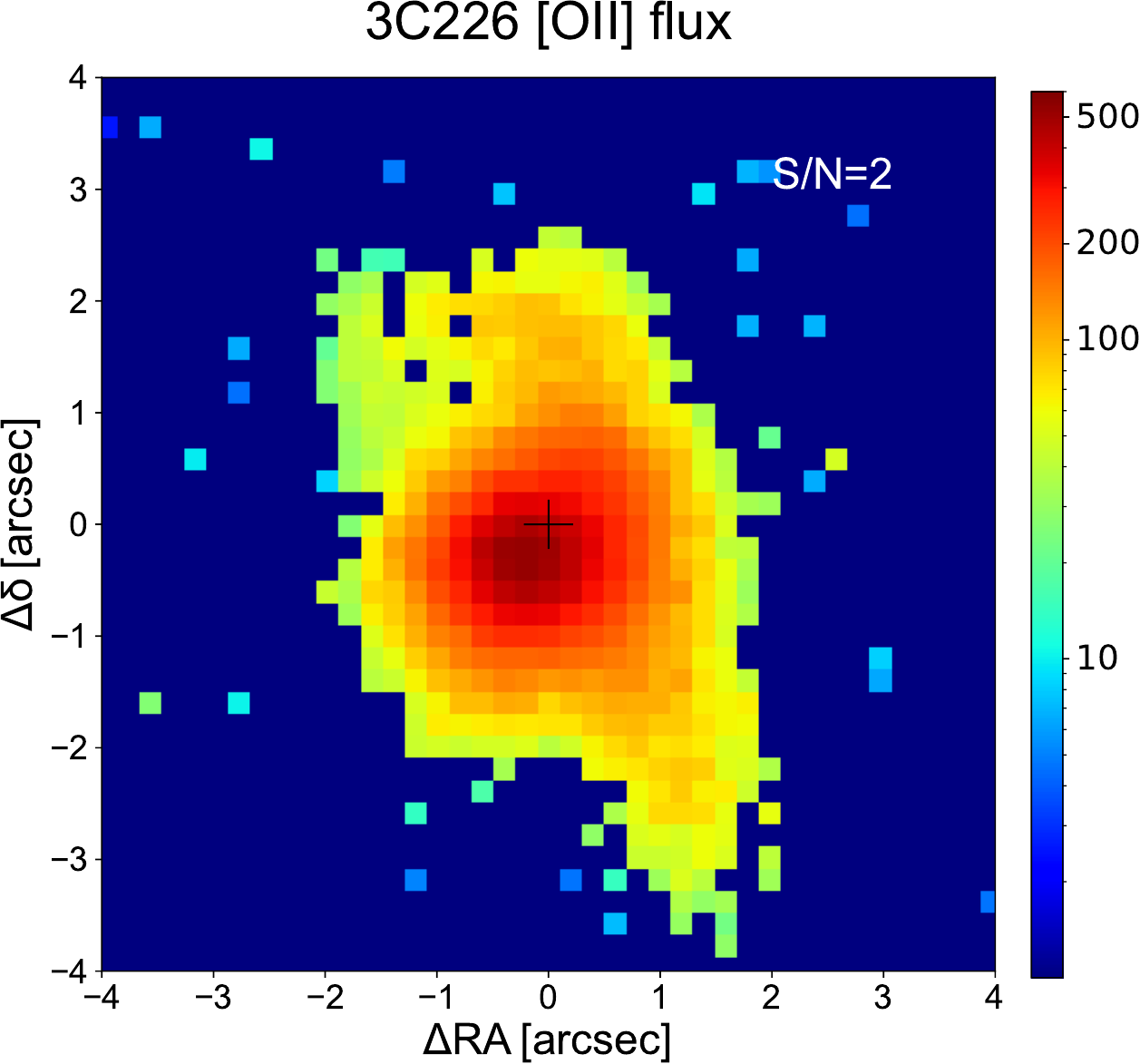}
\includegraphics[width=0.25\linewidth]{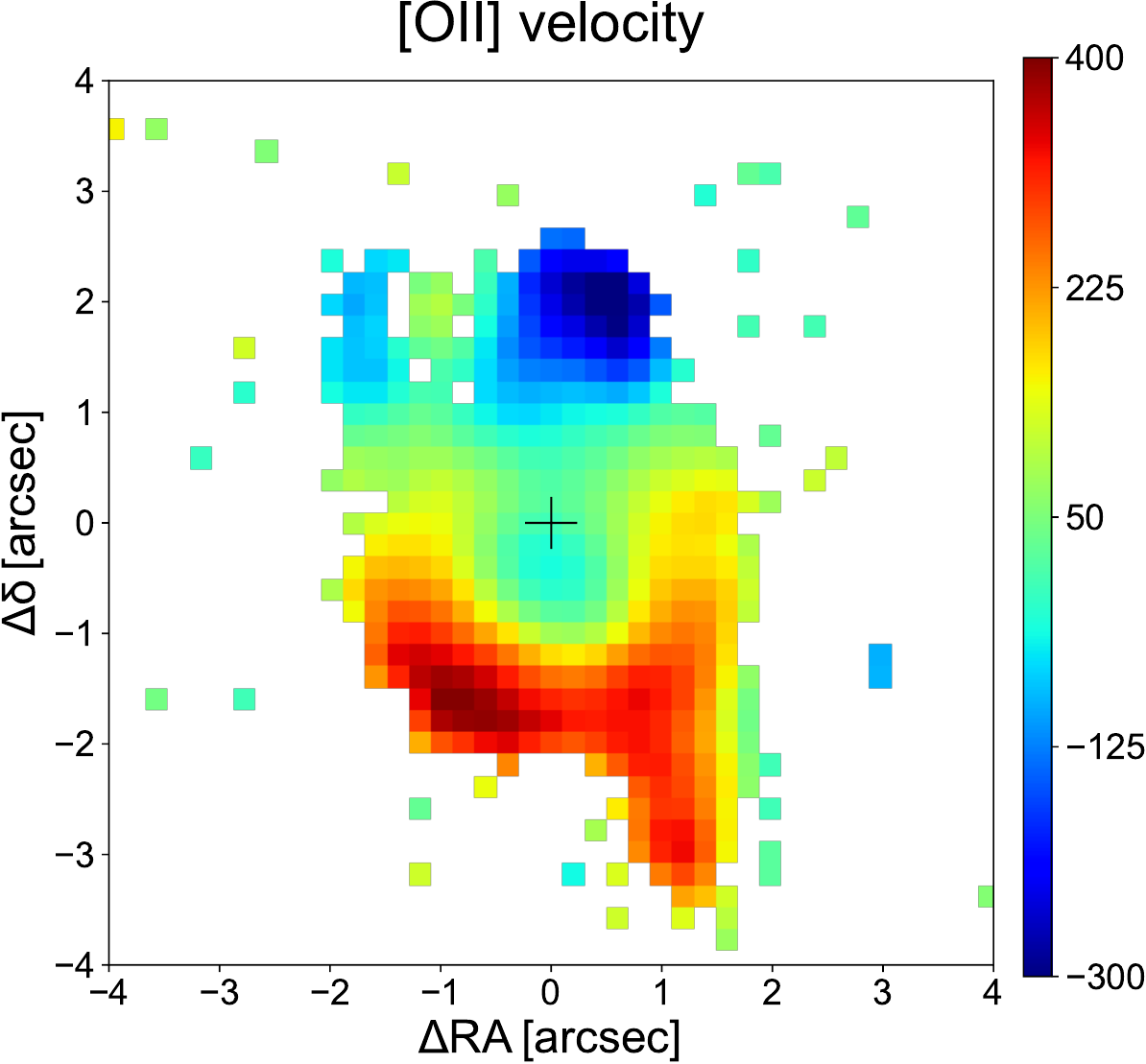}
\includegraphics[width=0.25\linewidth]{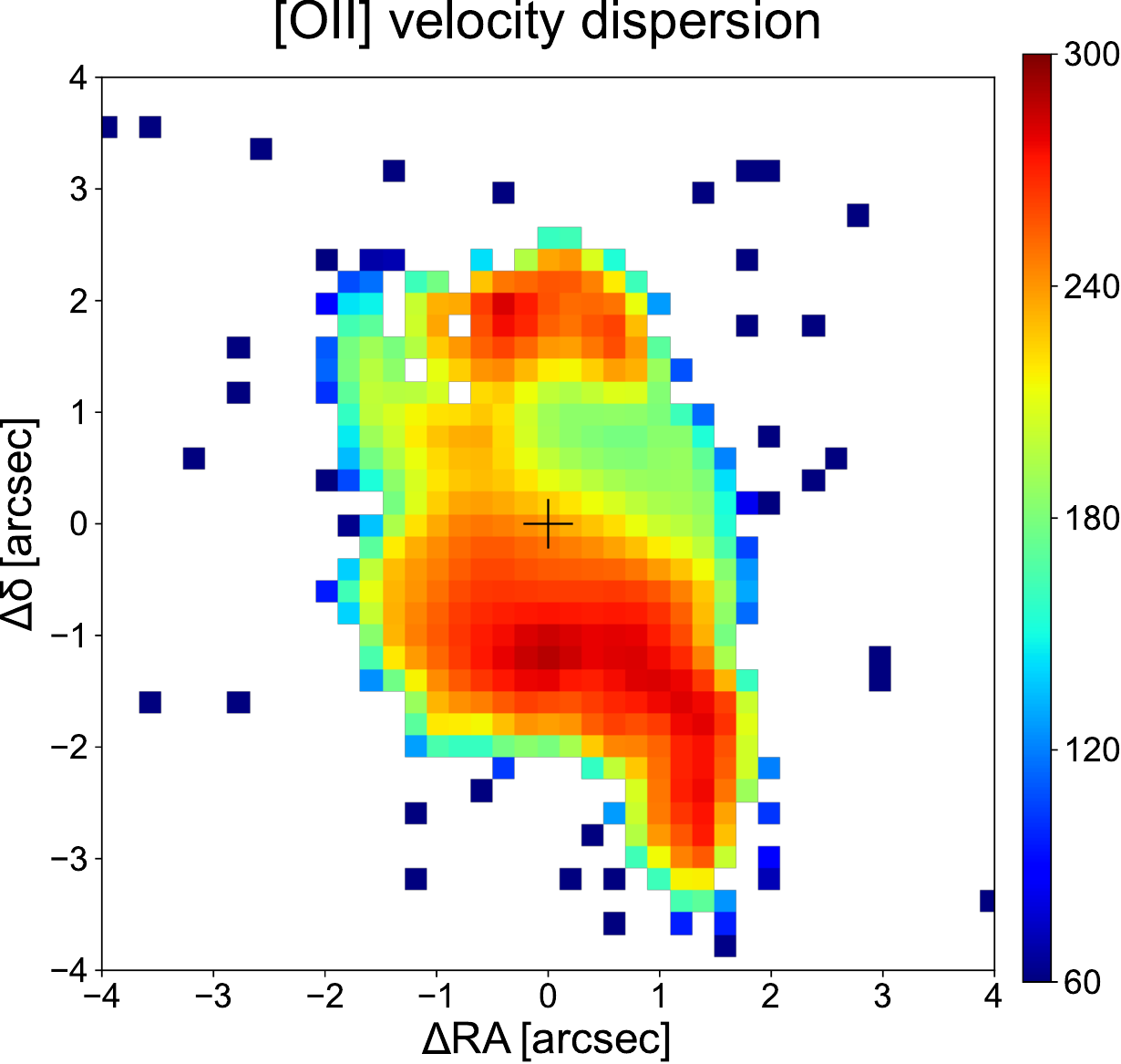}

\includegraphics[width=0.205\linewidth]{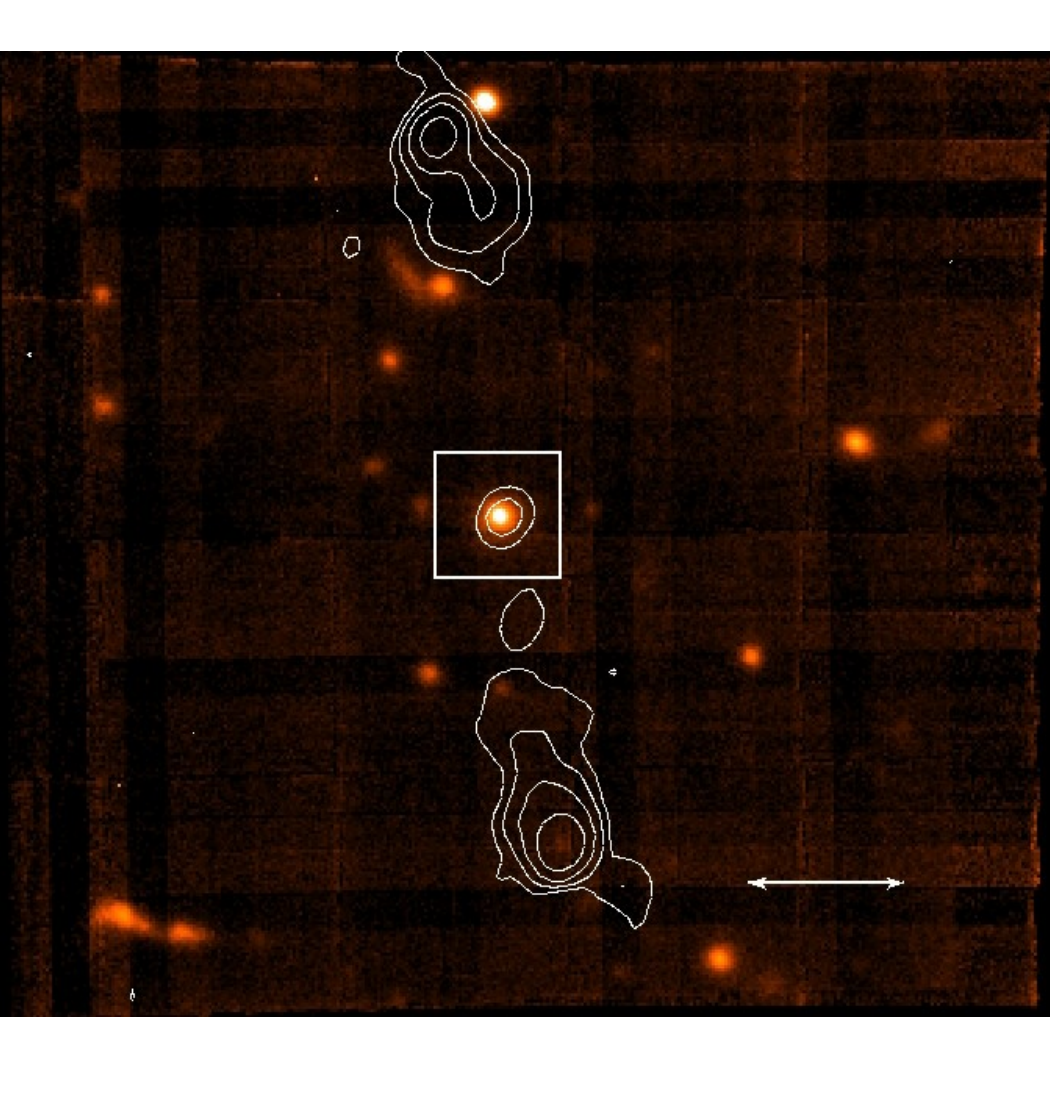}
\includegraphics[width=0.25\linewidth]{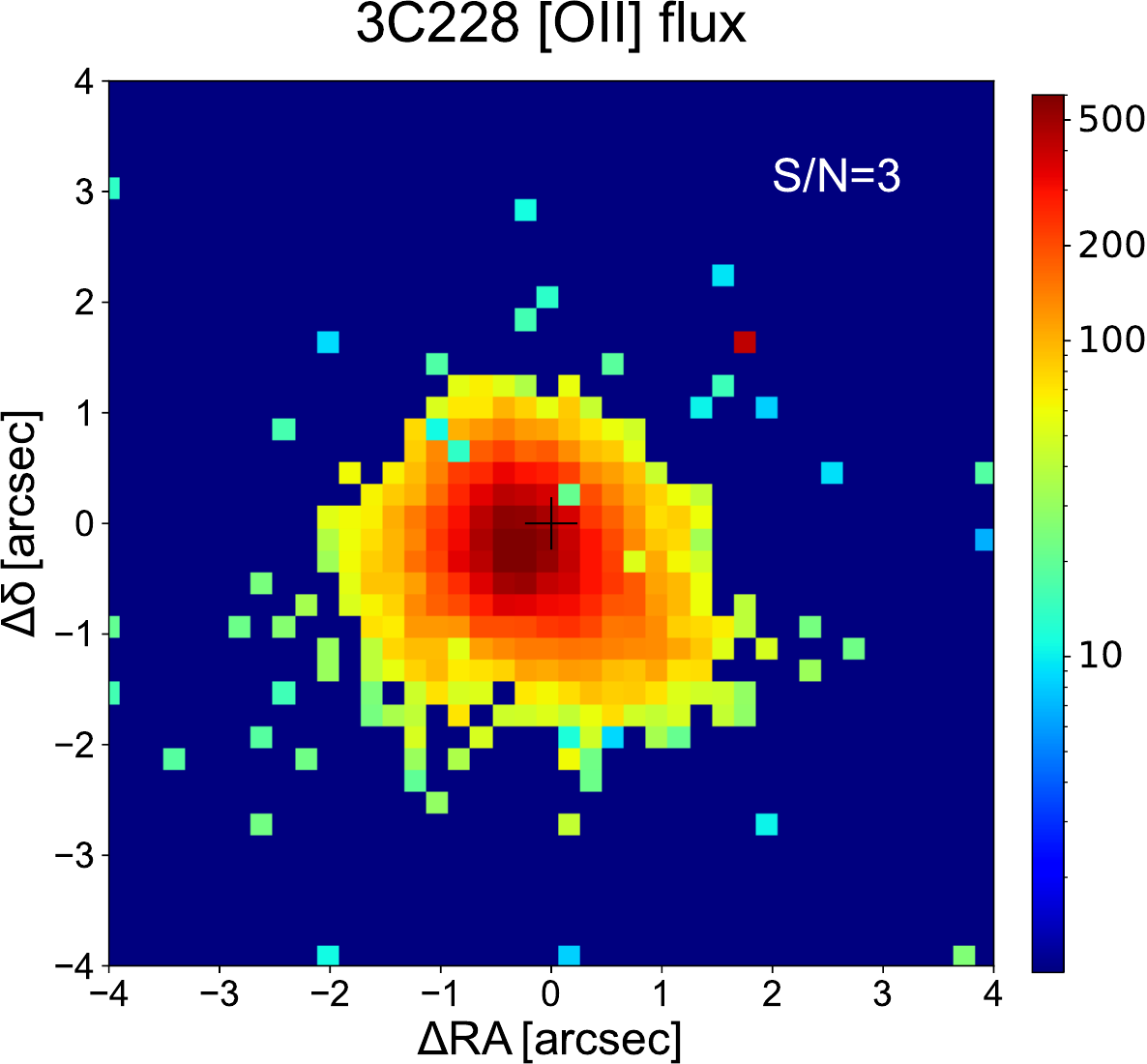}
\includegraphics[width=0.25\linewidth]{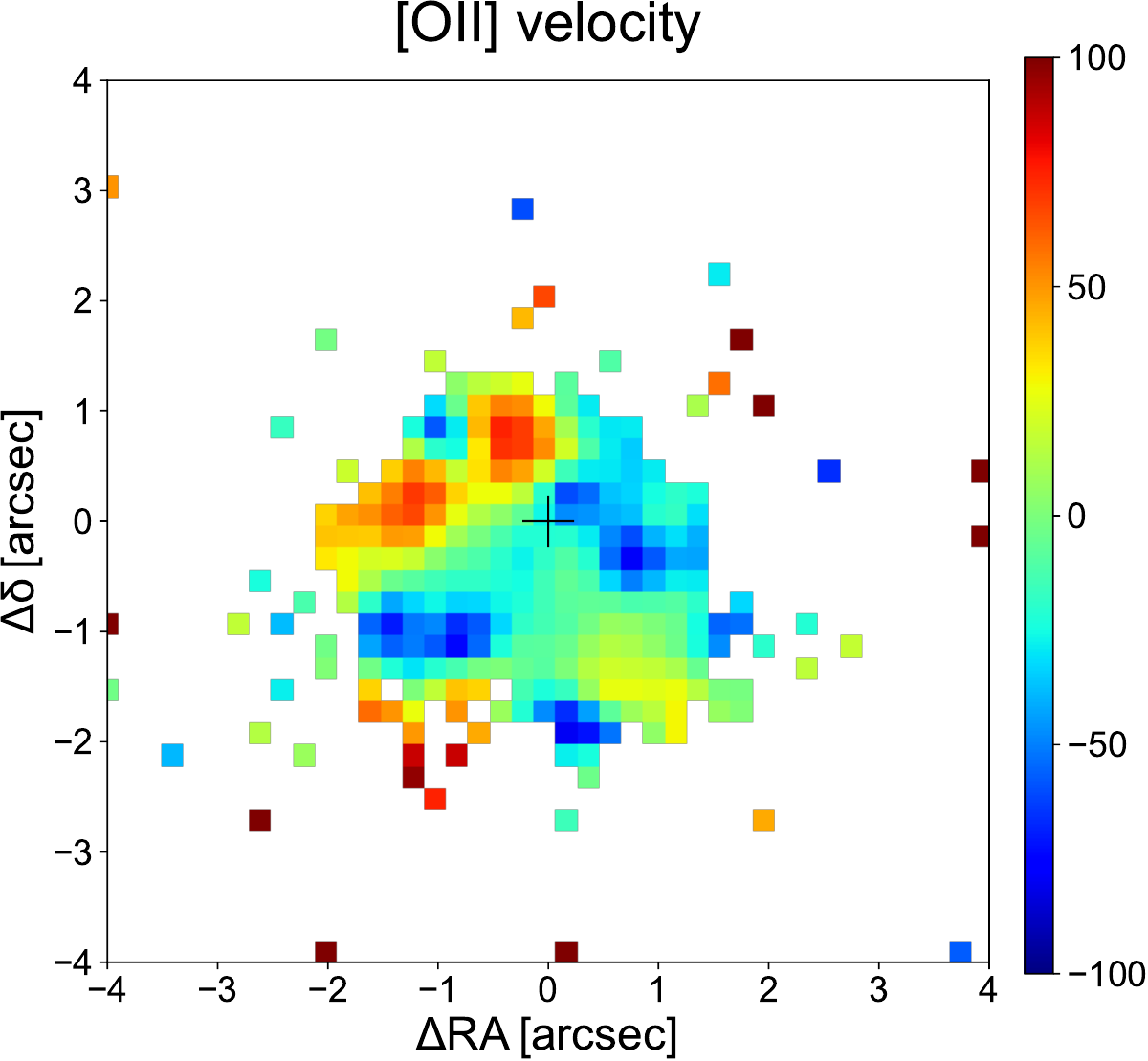}
\includegraphics[width=0.25\linewidth]{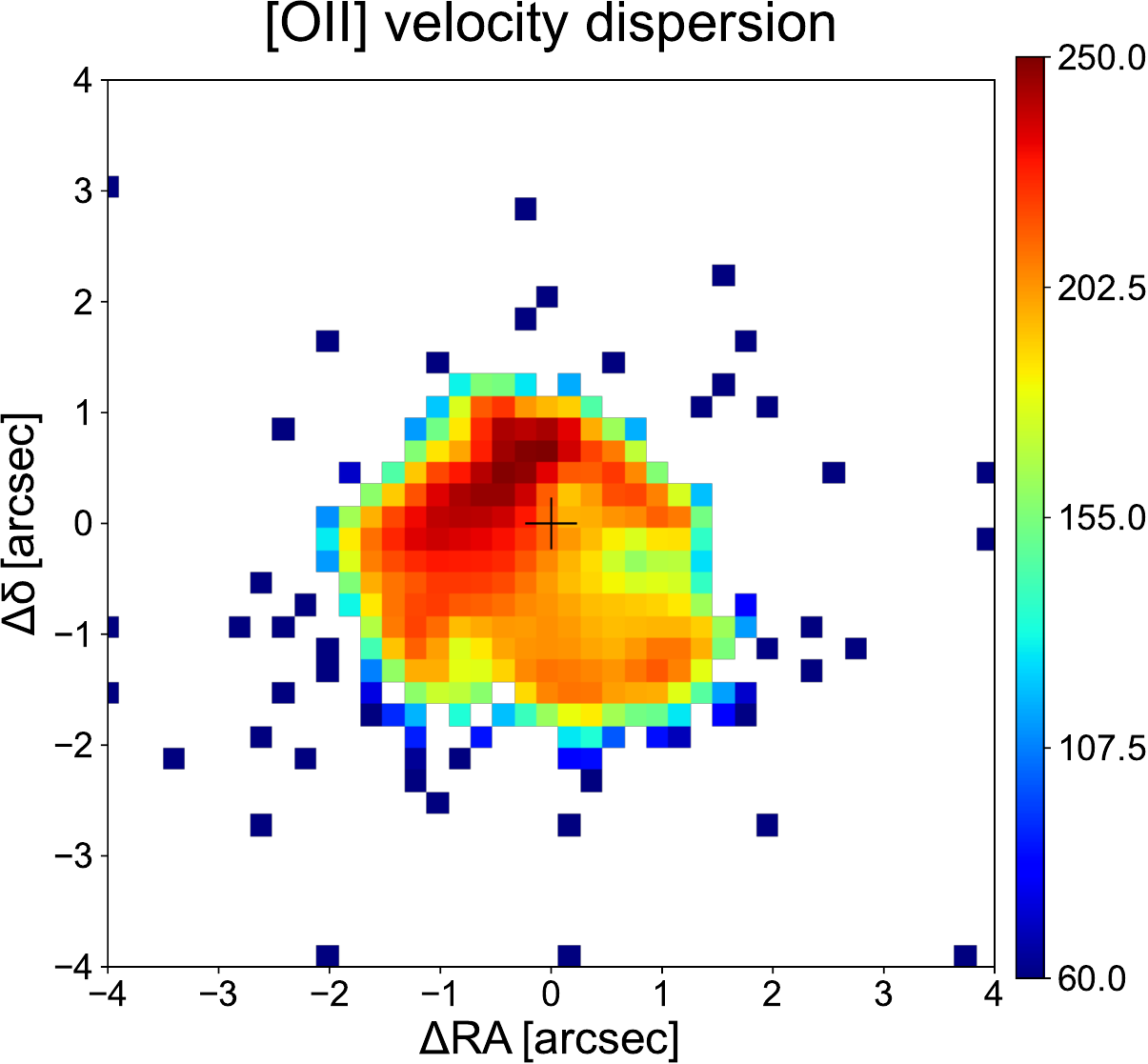}
\caption{- continued}
    \label{fig1}
\end{figure*}

\addtocounter{figure}{-1}
\begin{figure*}
\centering

\includegraphics[width=0.205\linewidth]{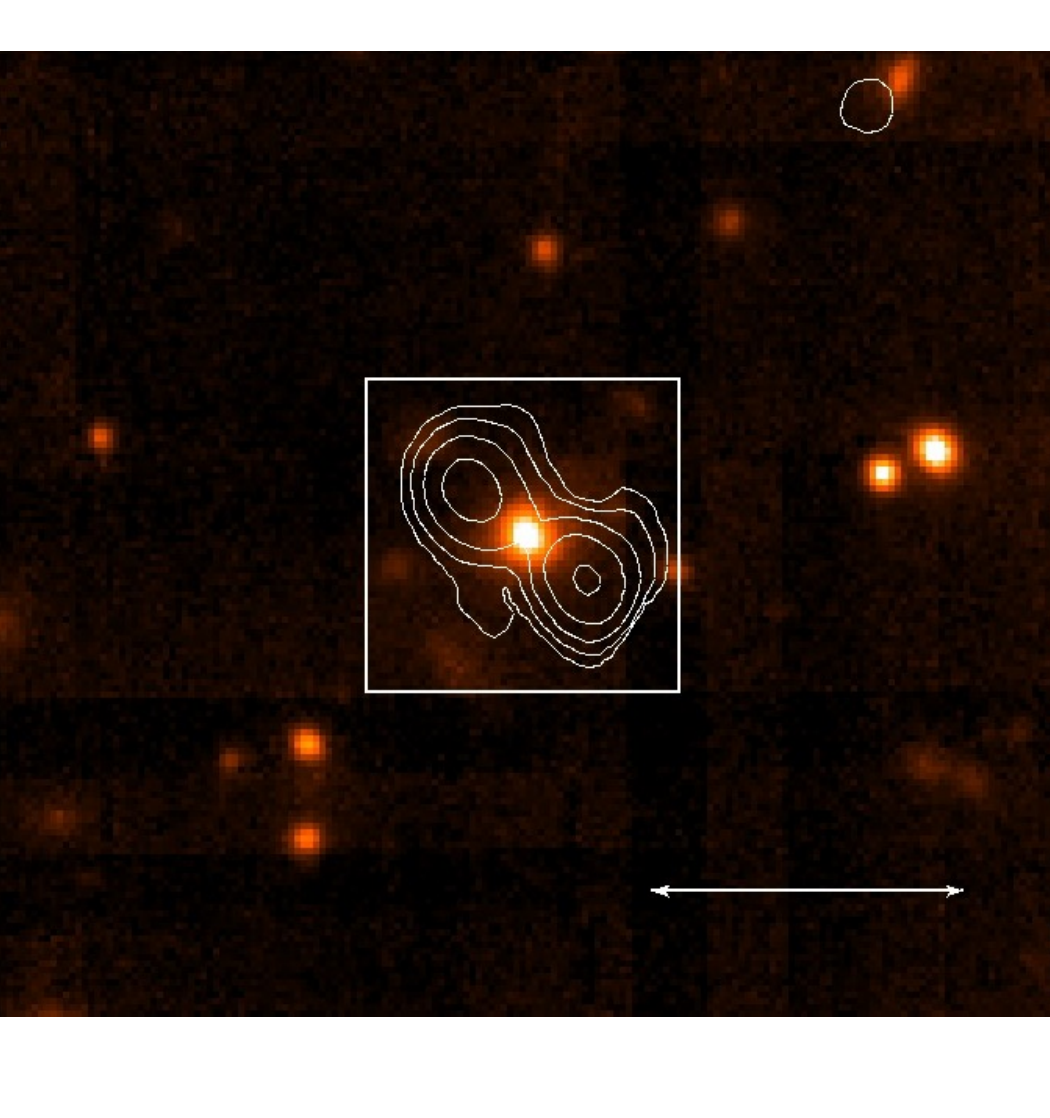}
\includegraphics[width=0.255\linewidth]{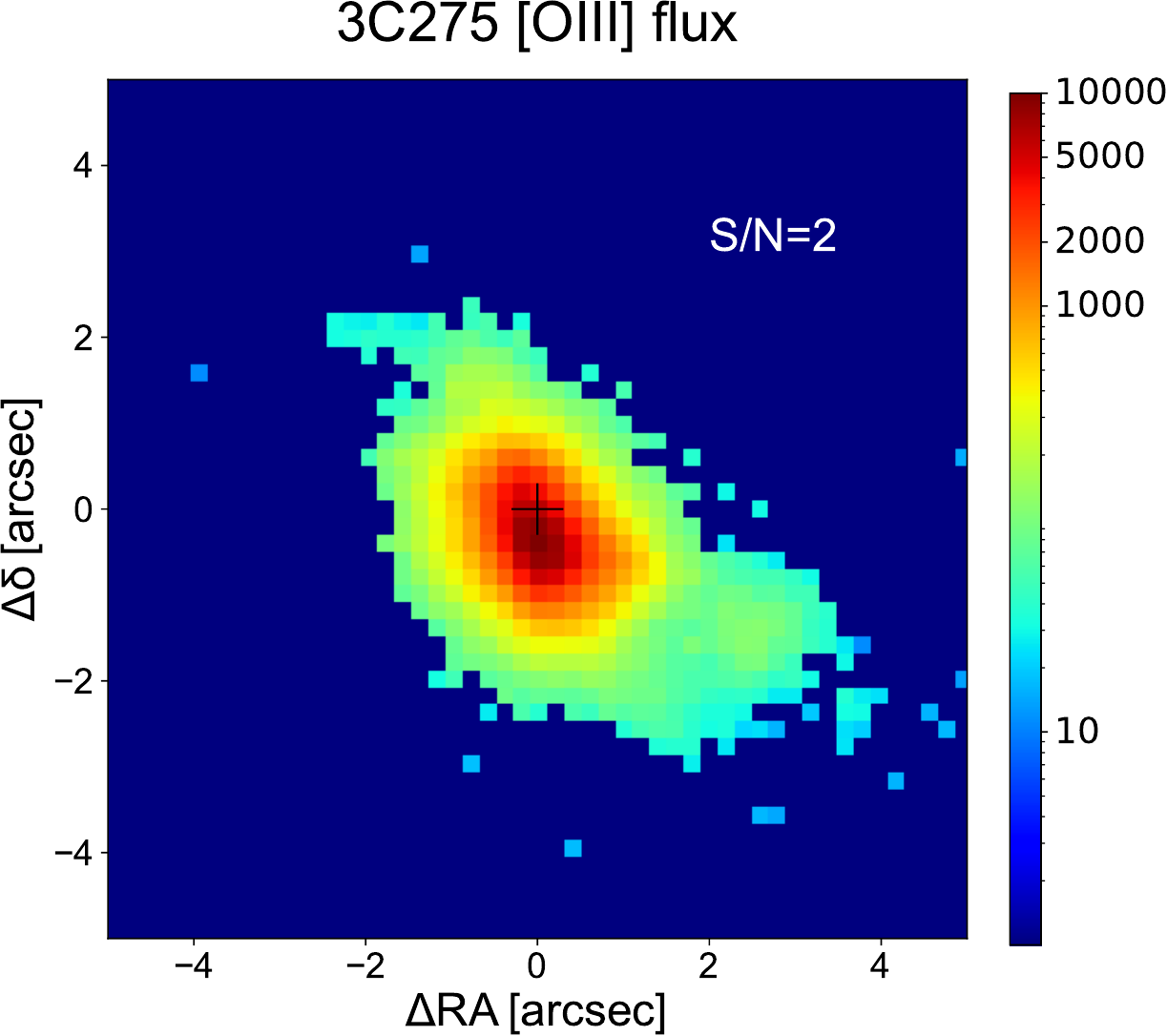}
\includegraphics[width=0.25\linewidth]{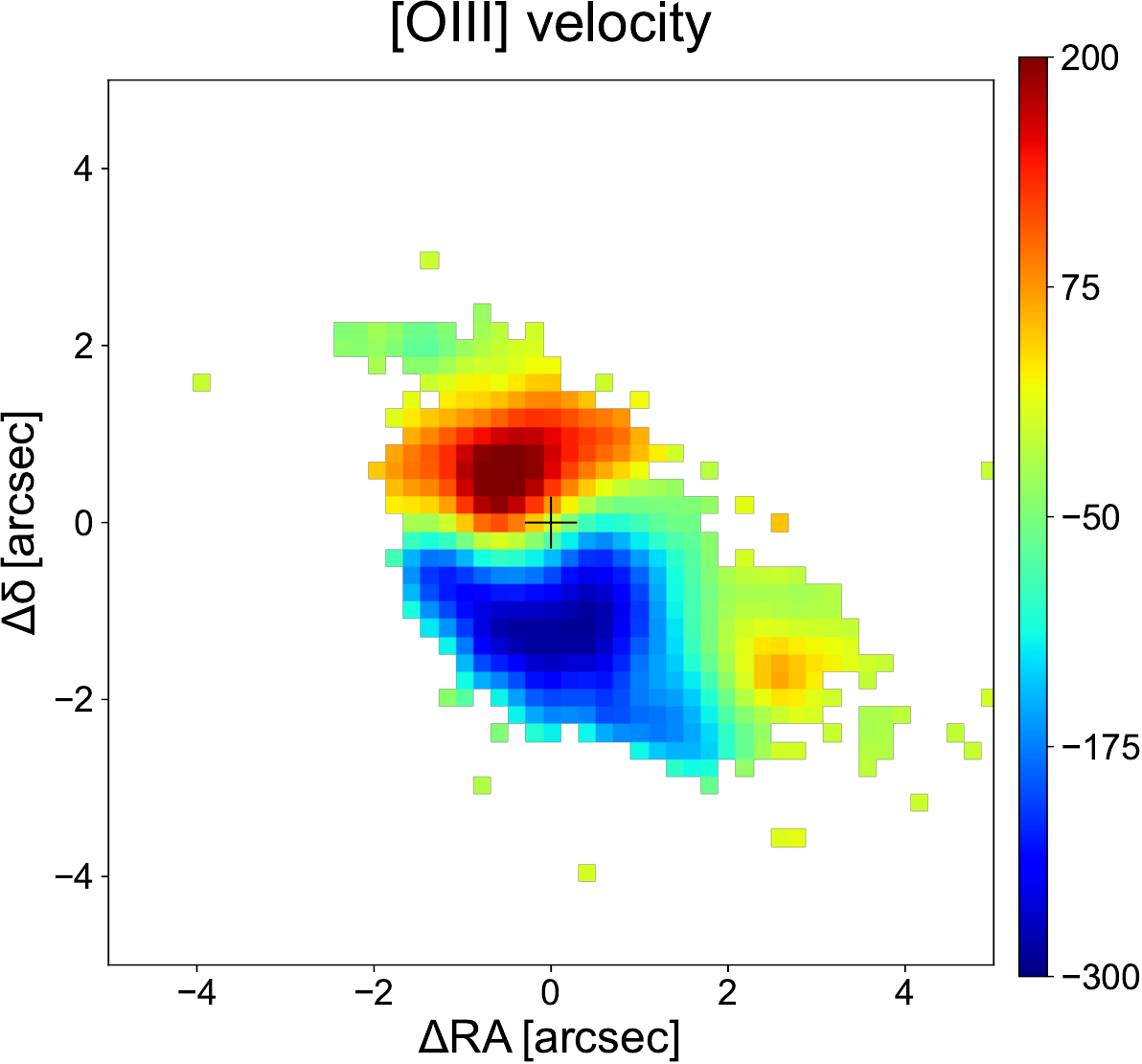}
\includegraphics[width=0.25\linewidth]{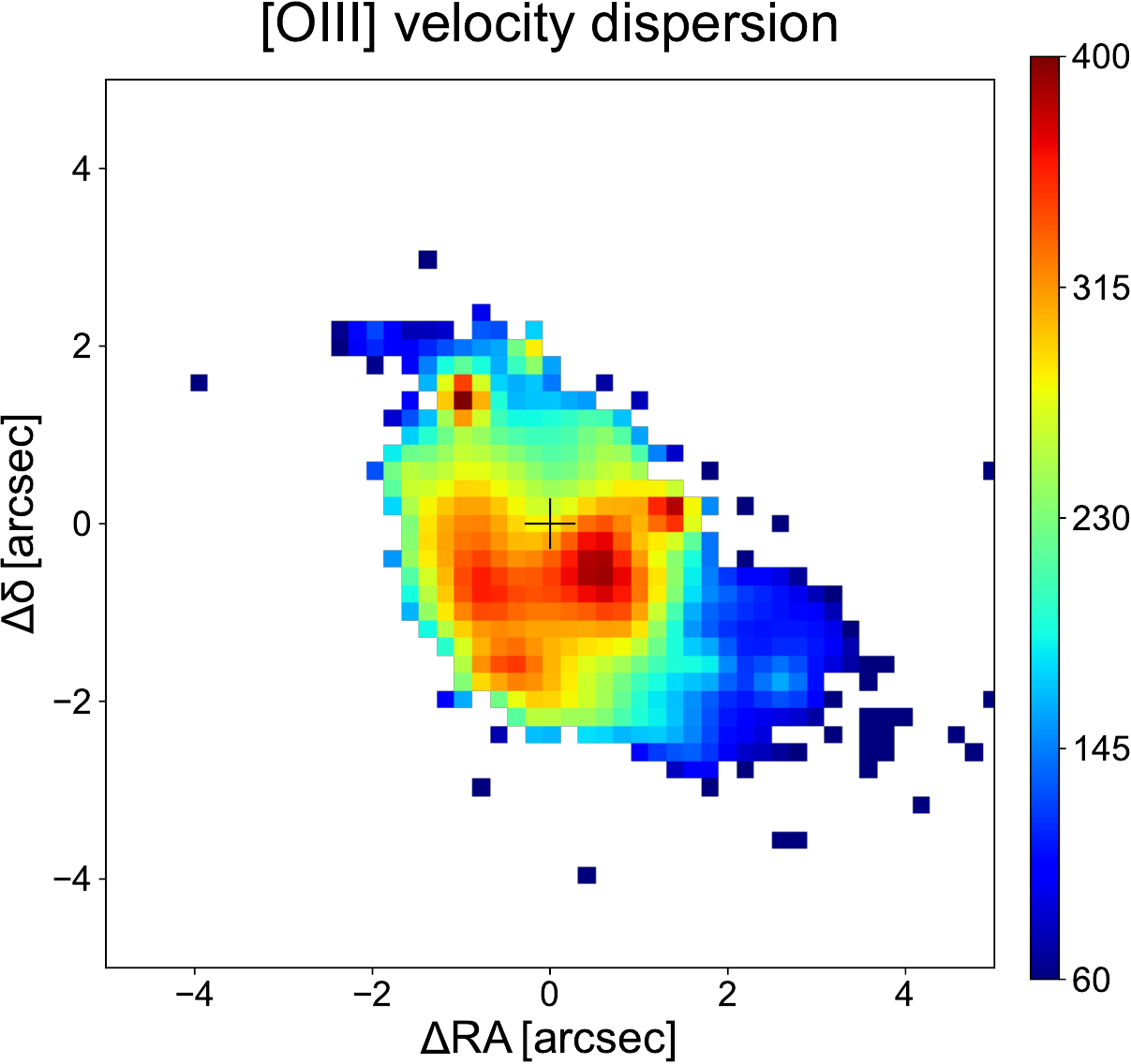}

\includegraphics[width=0.20\linewidth]{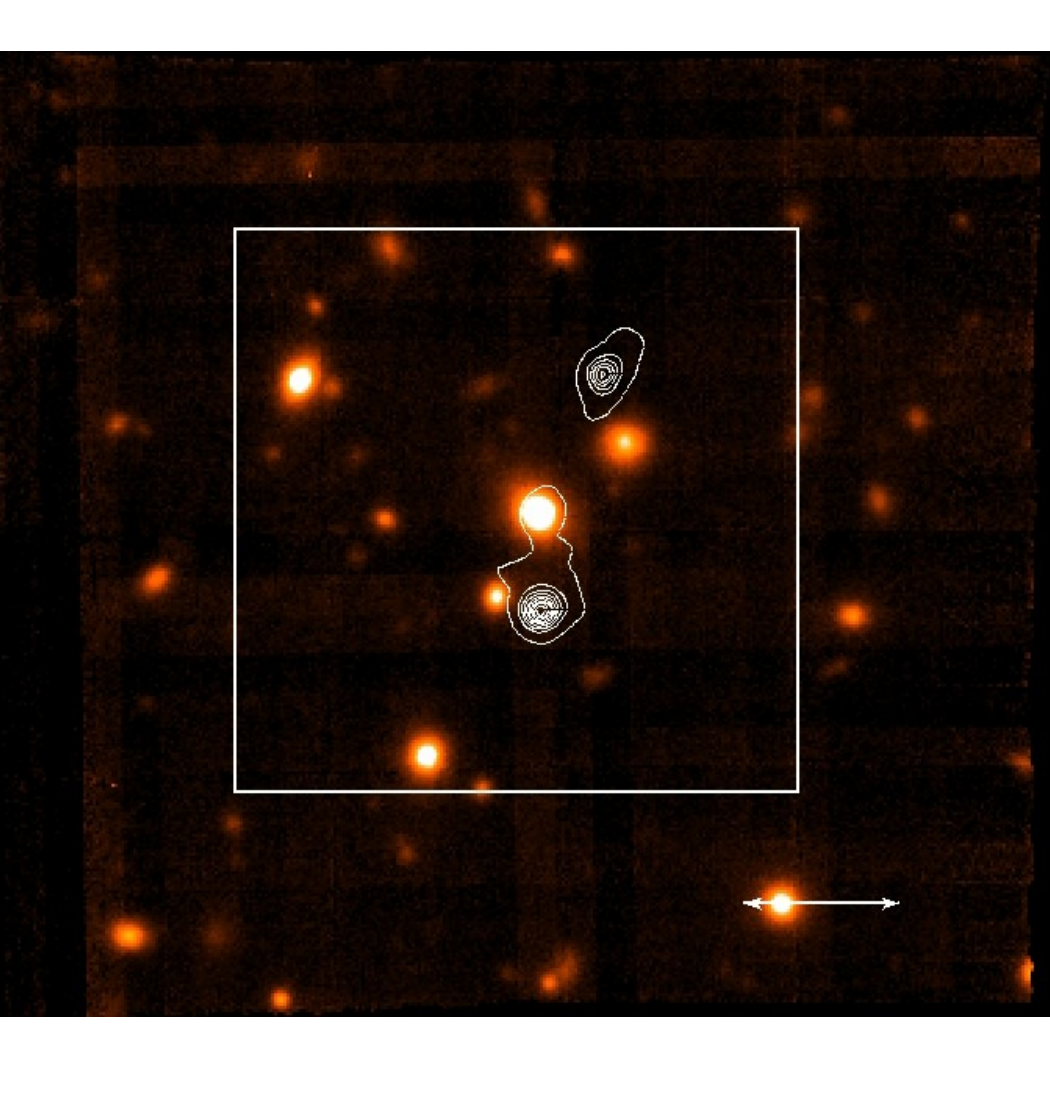}
\includegraphics[width=0.255\linewidth]{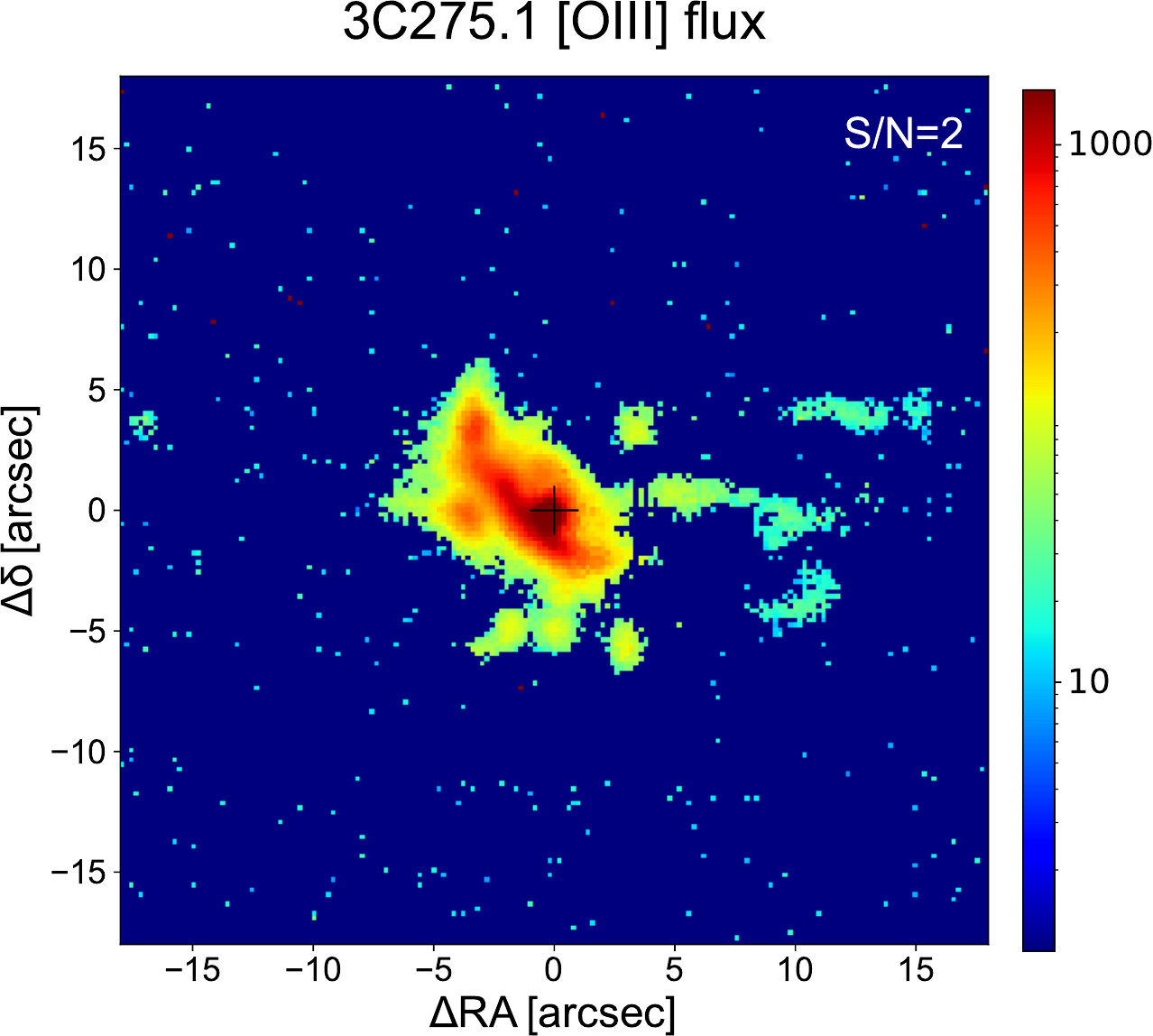}
\includegraphics[width=0.25\linewidth]{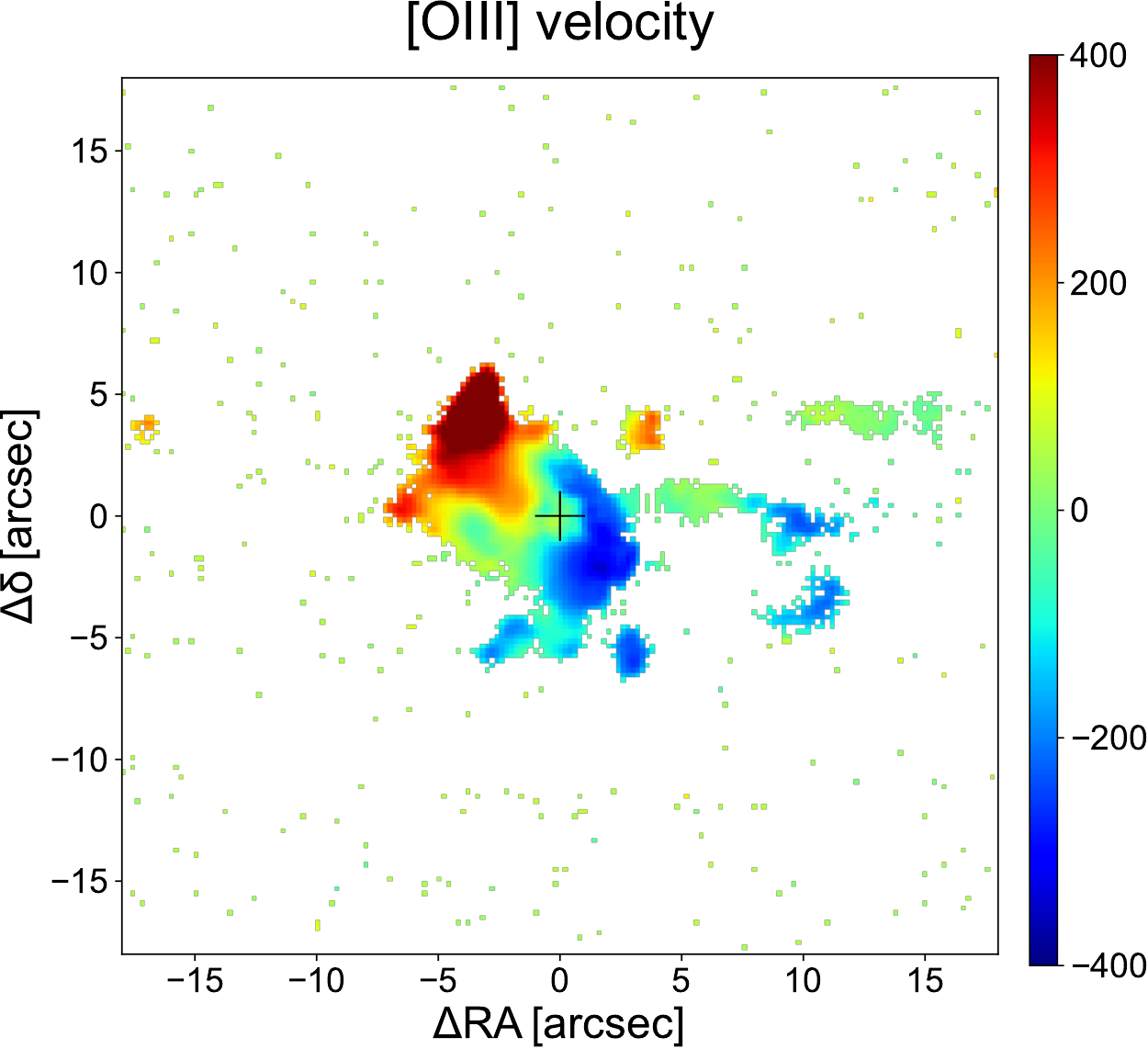}
\includegraphics[width=0.25\linewidth]{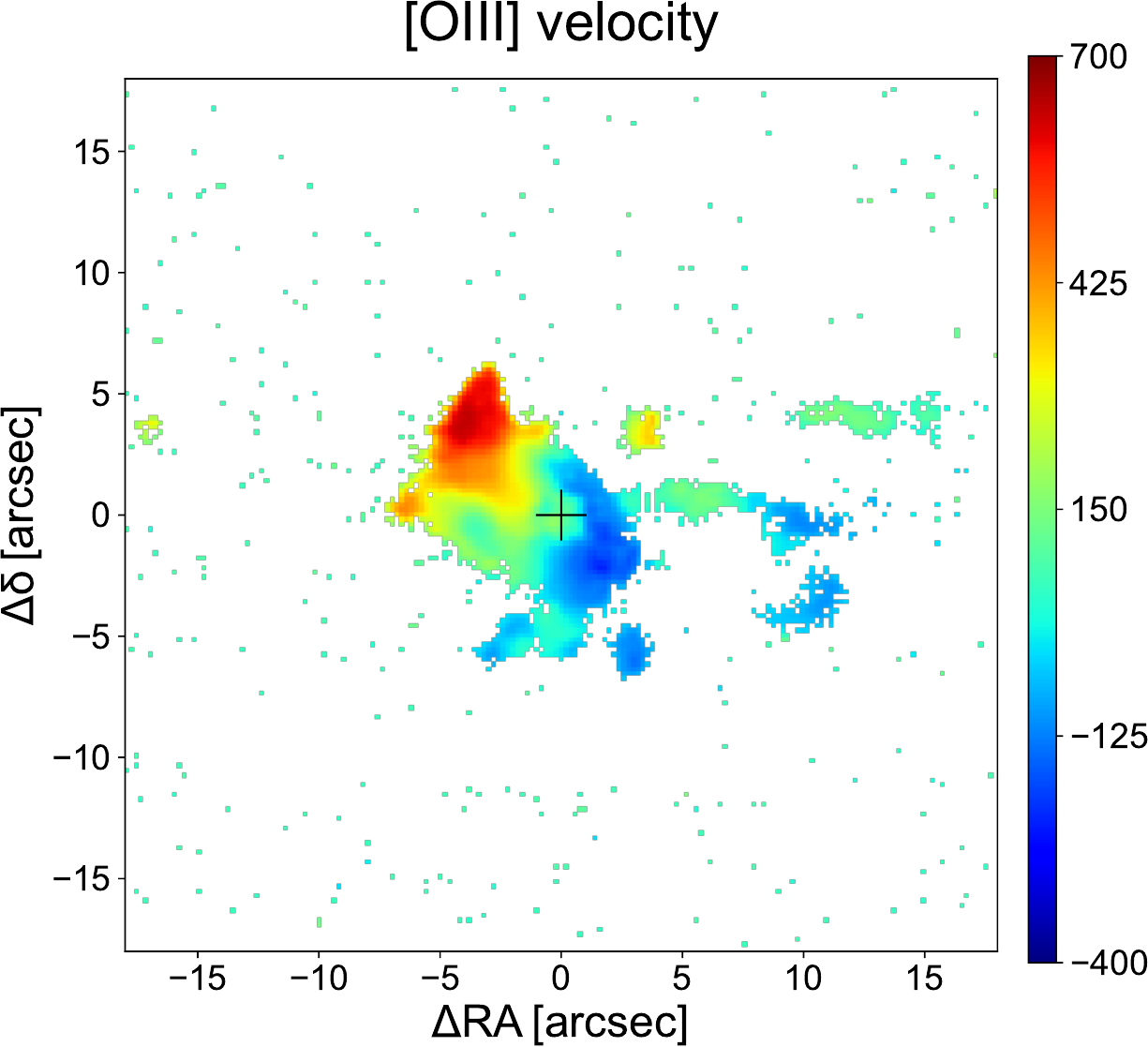}

\includegraphics[width=0.205\linewidth]{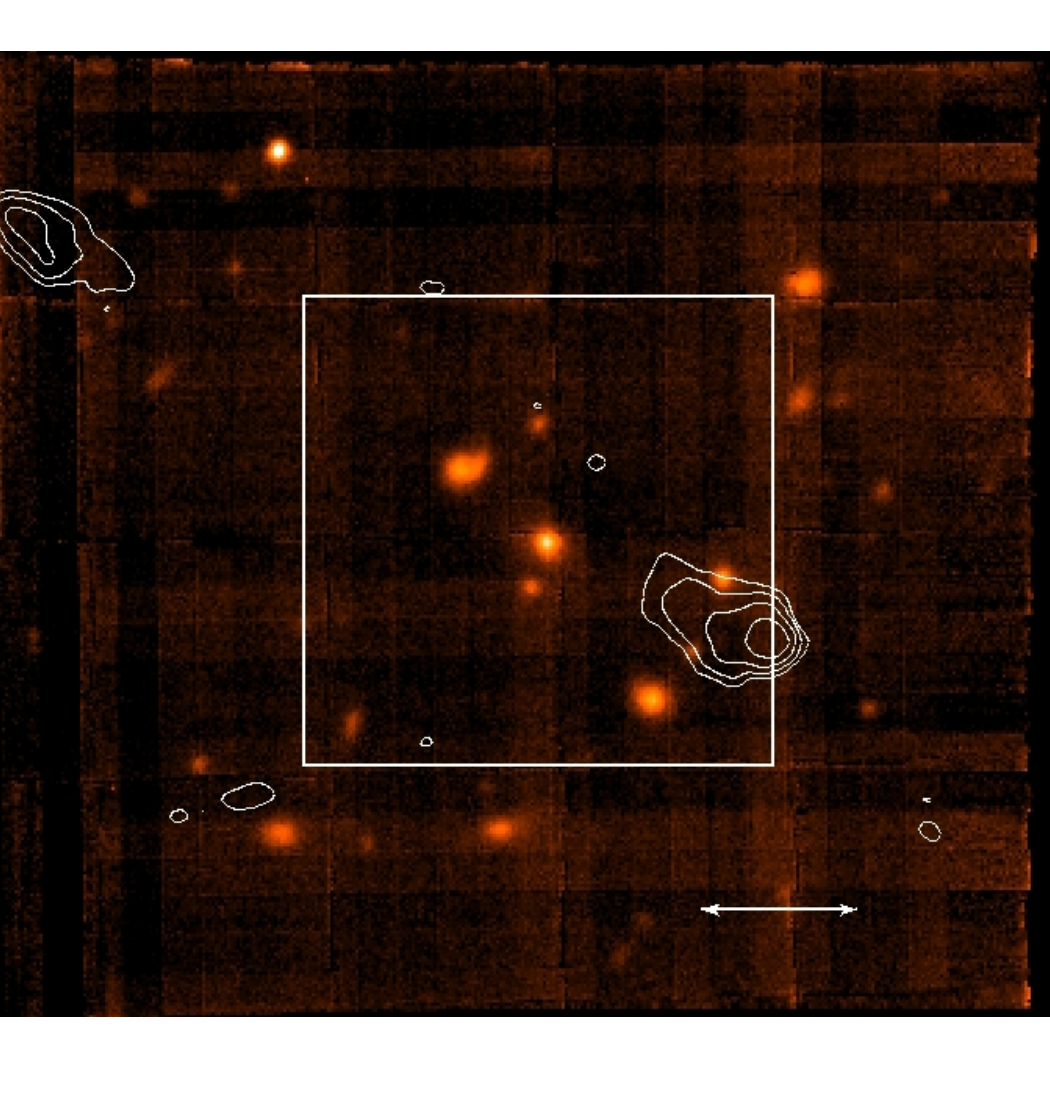}
\includegraphics[width=0.25\linewidth]{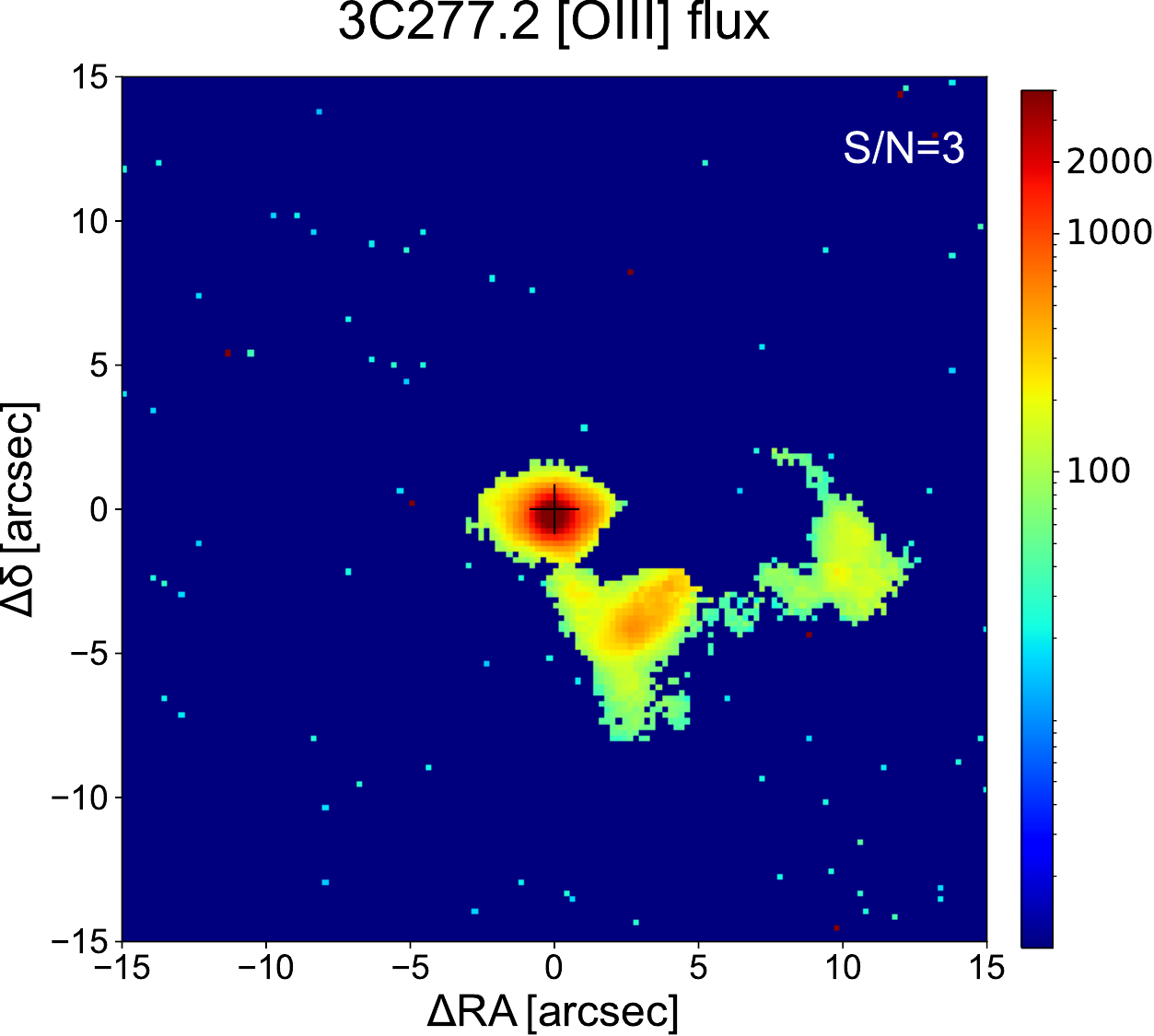}
\includegraphics[width=0.25\linewidth]{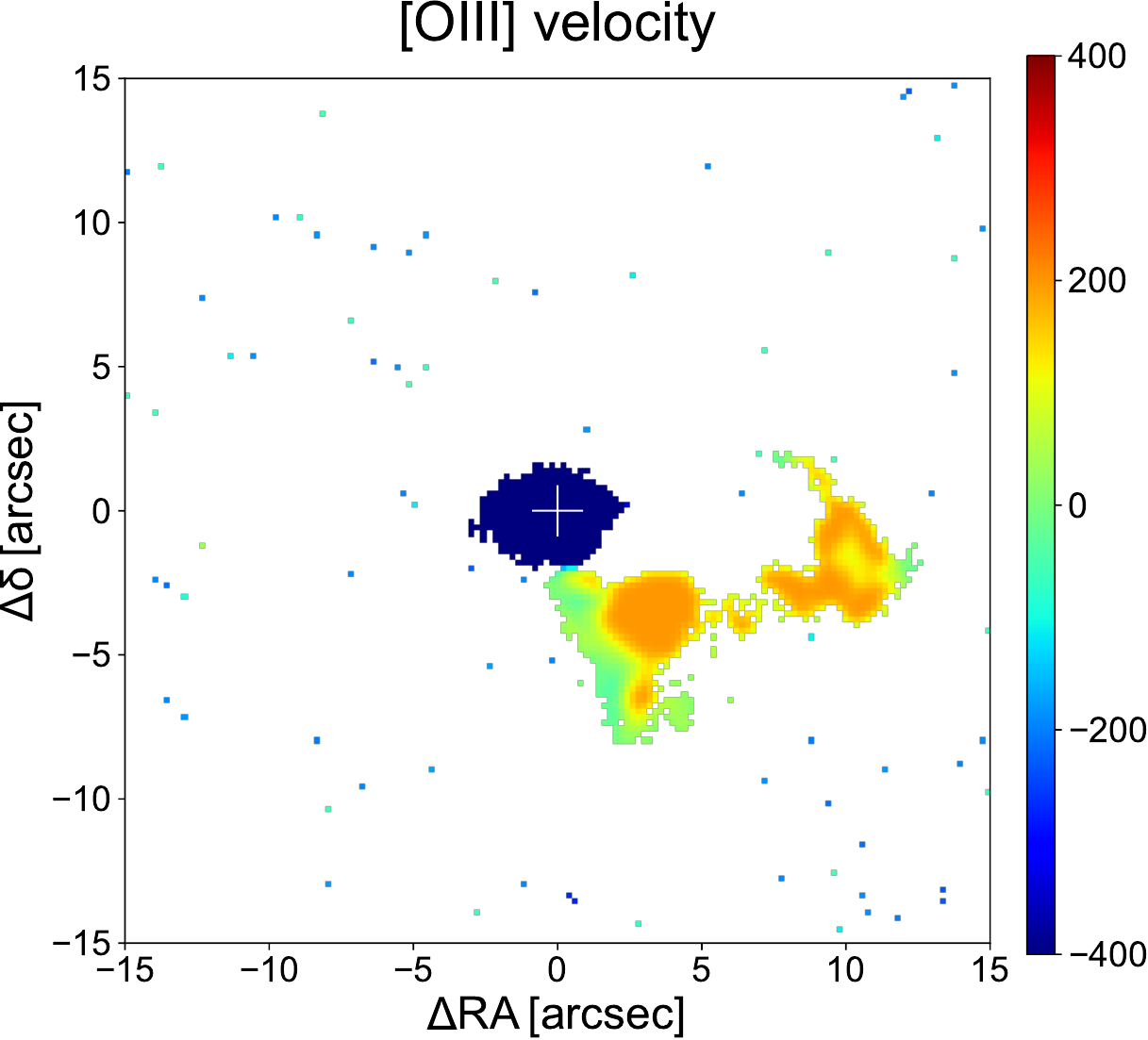}
\includegraphics[width=0.25\linewidth]{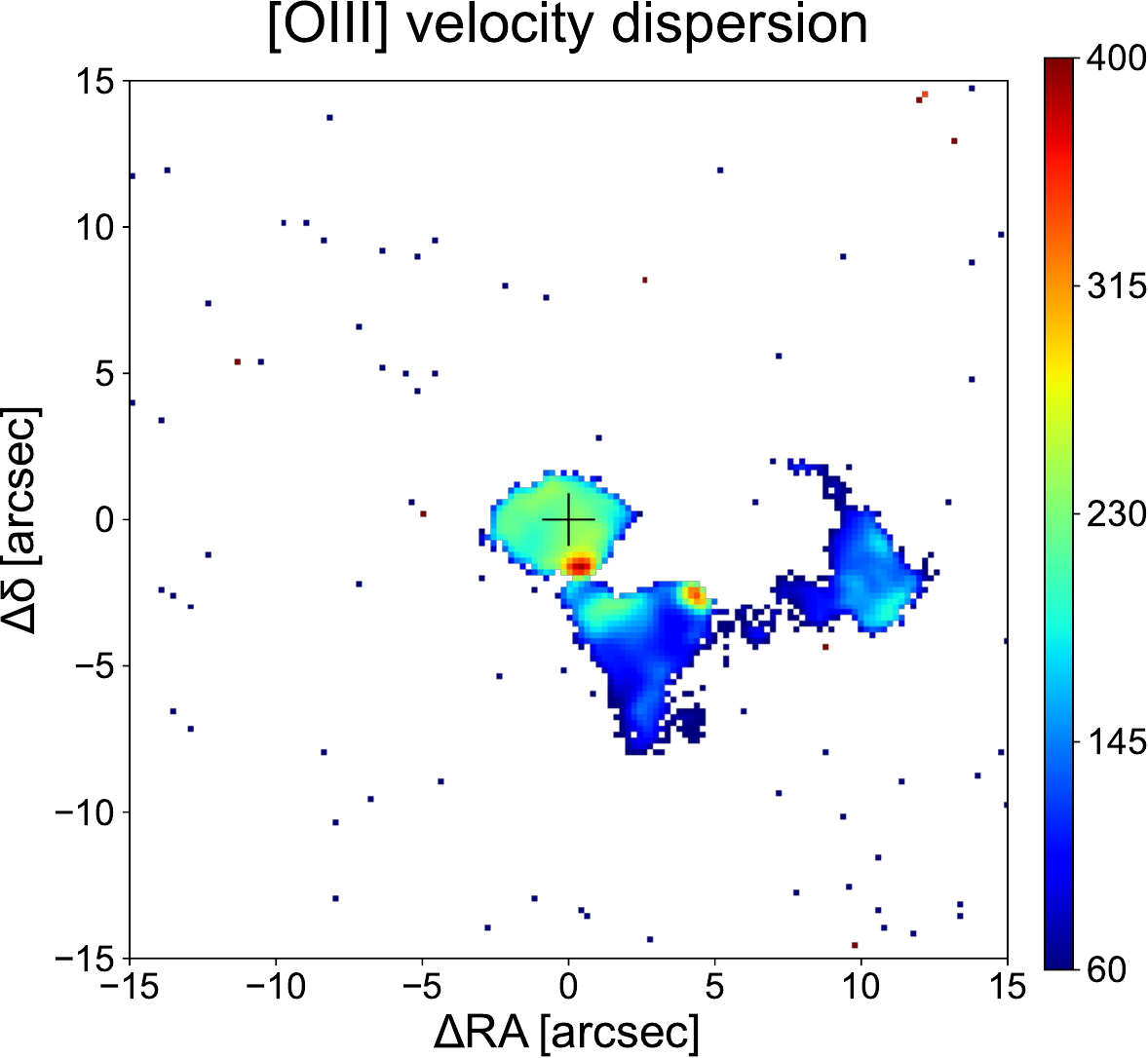}

\includegraphics[width=0.205\linewidth]{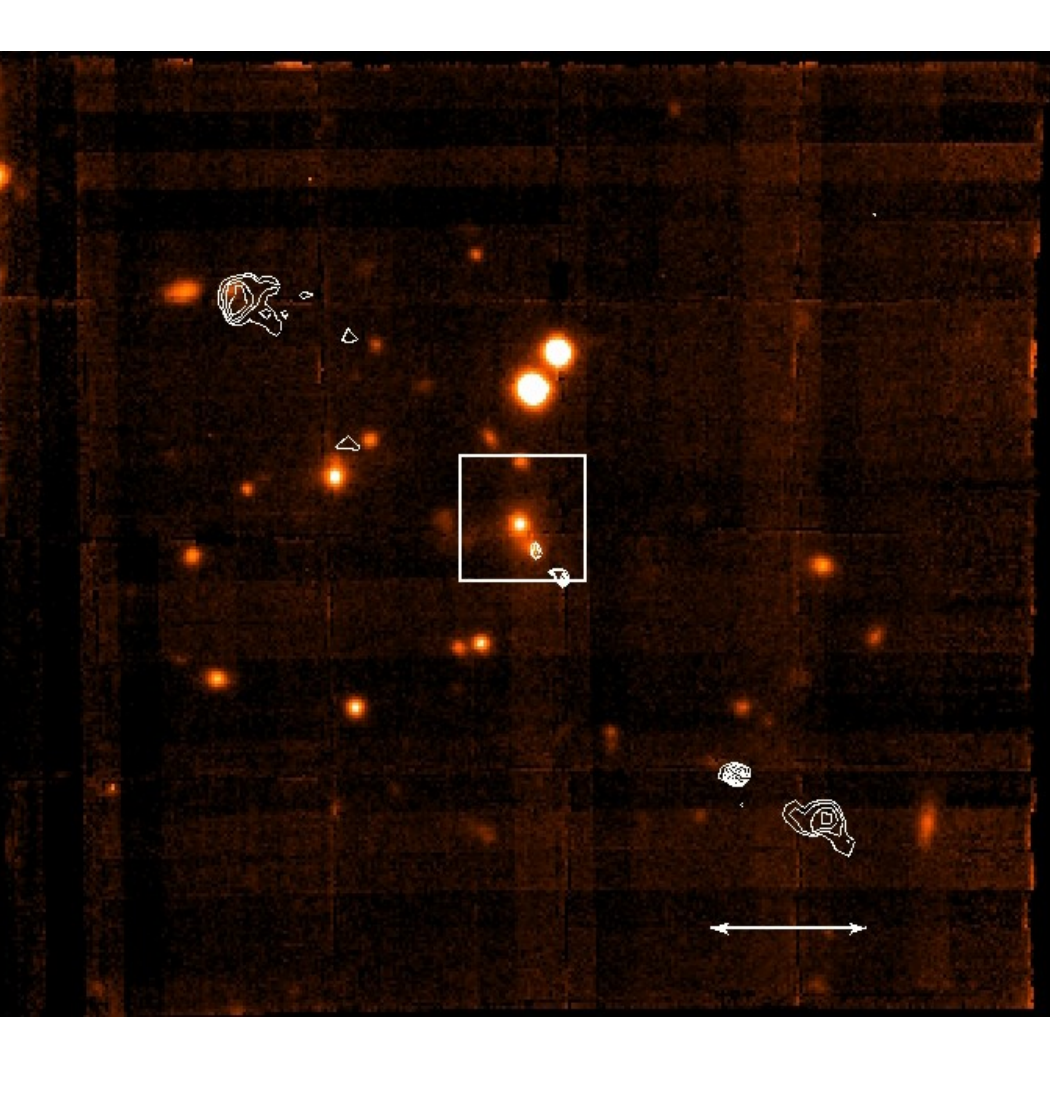}
\includegraphics[width=0.25\linewidth]{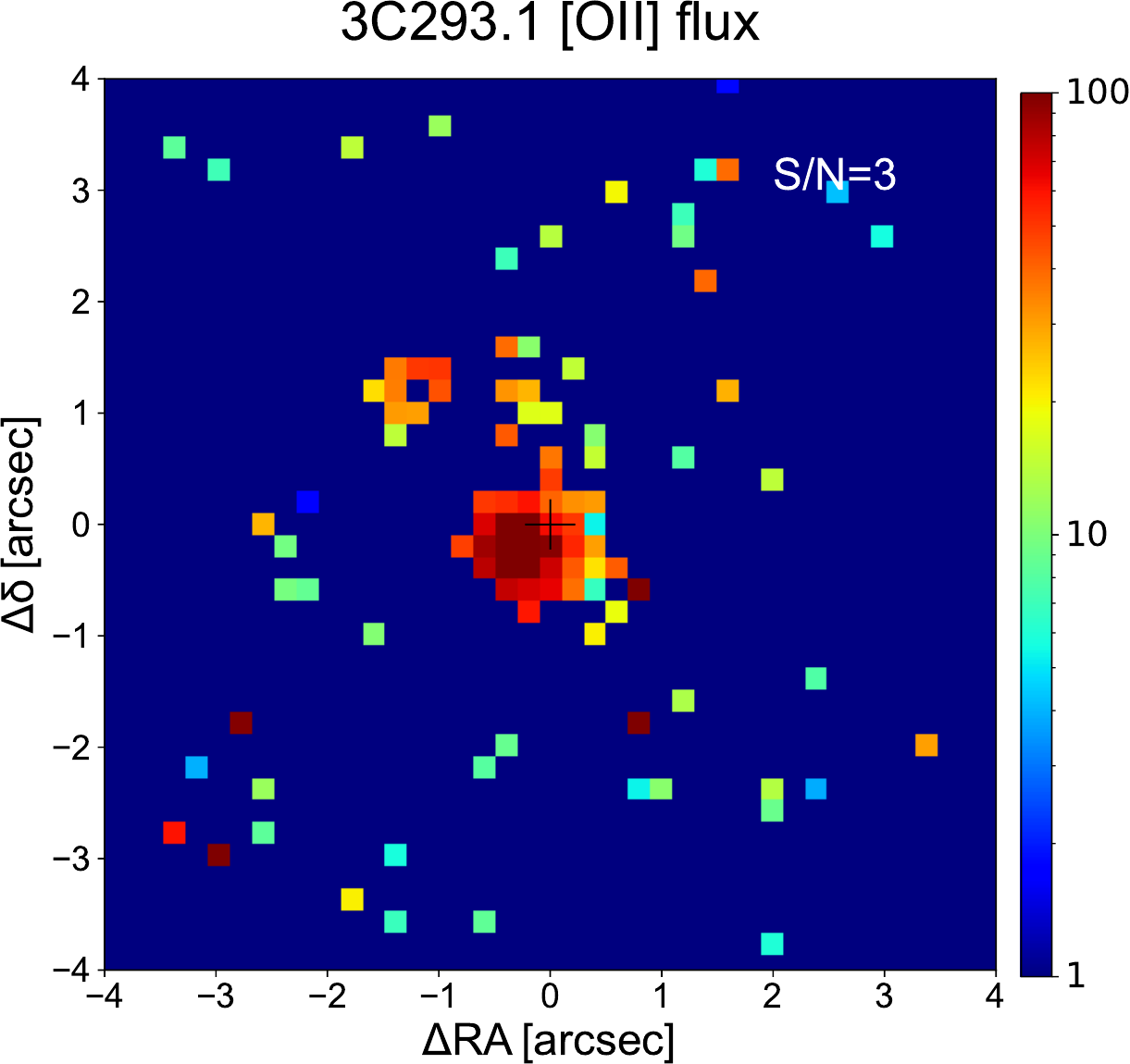}
\includegraphics[width=0.25\linewidth]{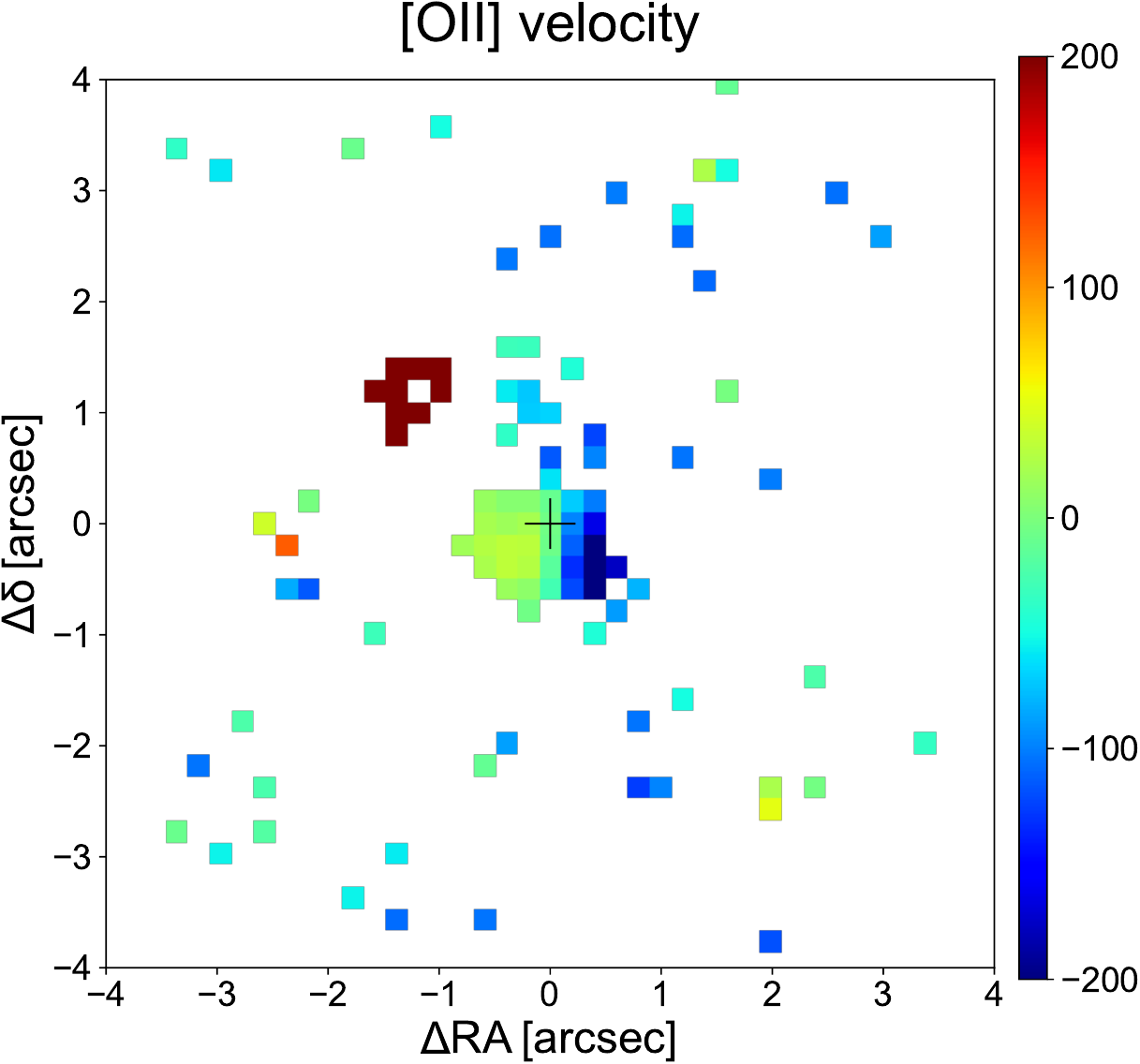}
\includegraphics[width=0.25\linewidth]{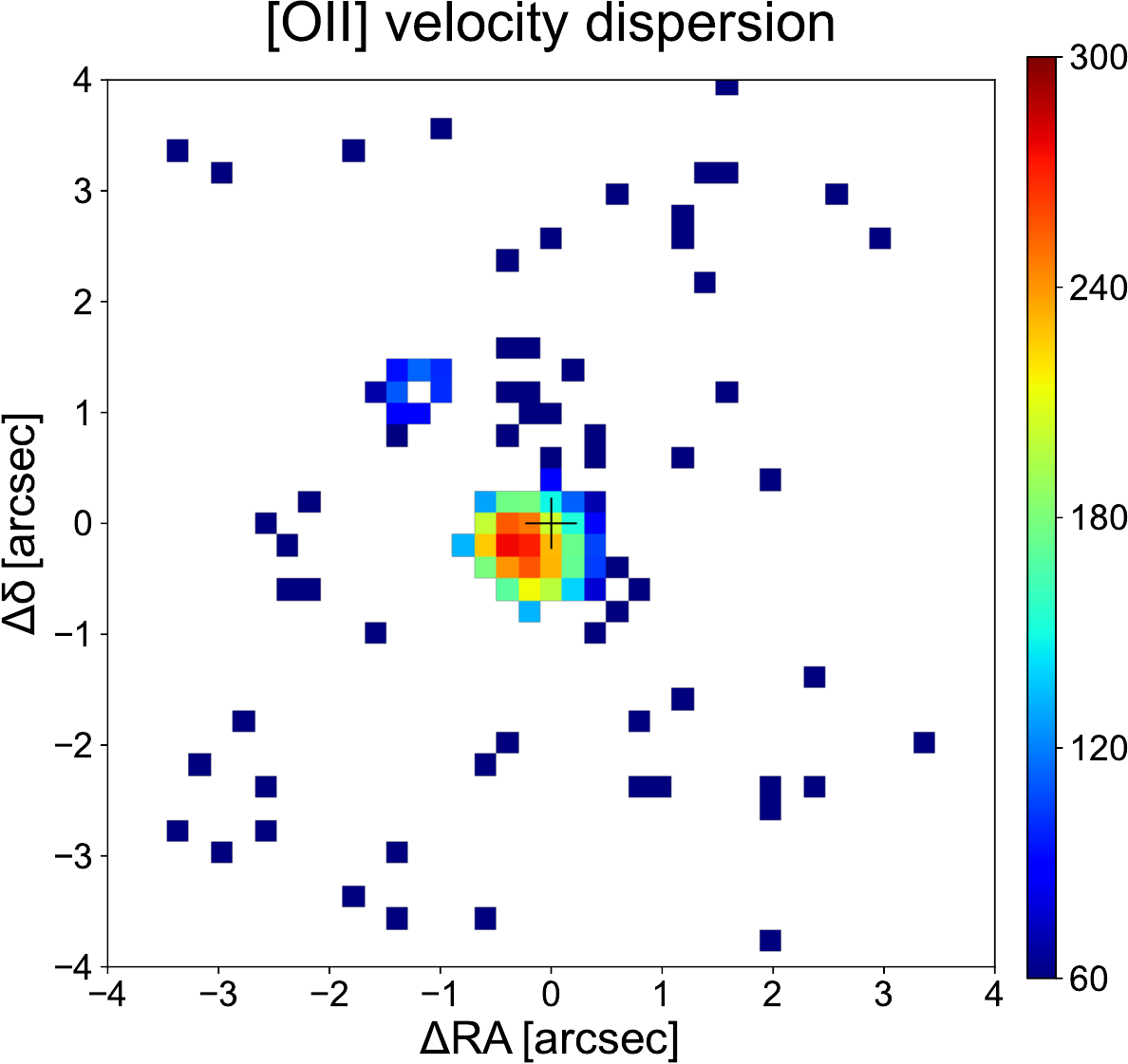}

\includegraphics[width=0.215\linewidth]{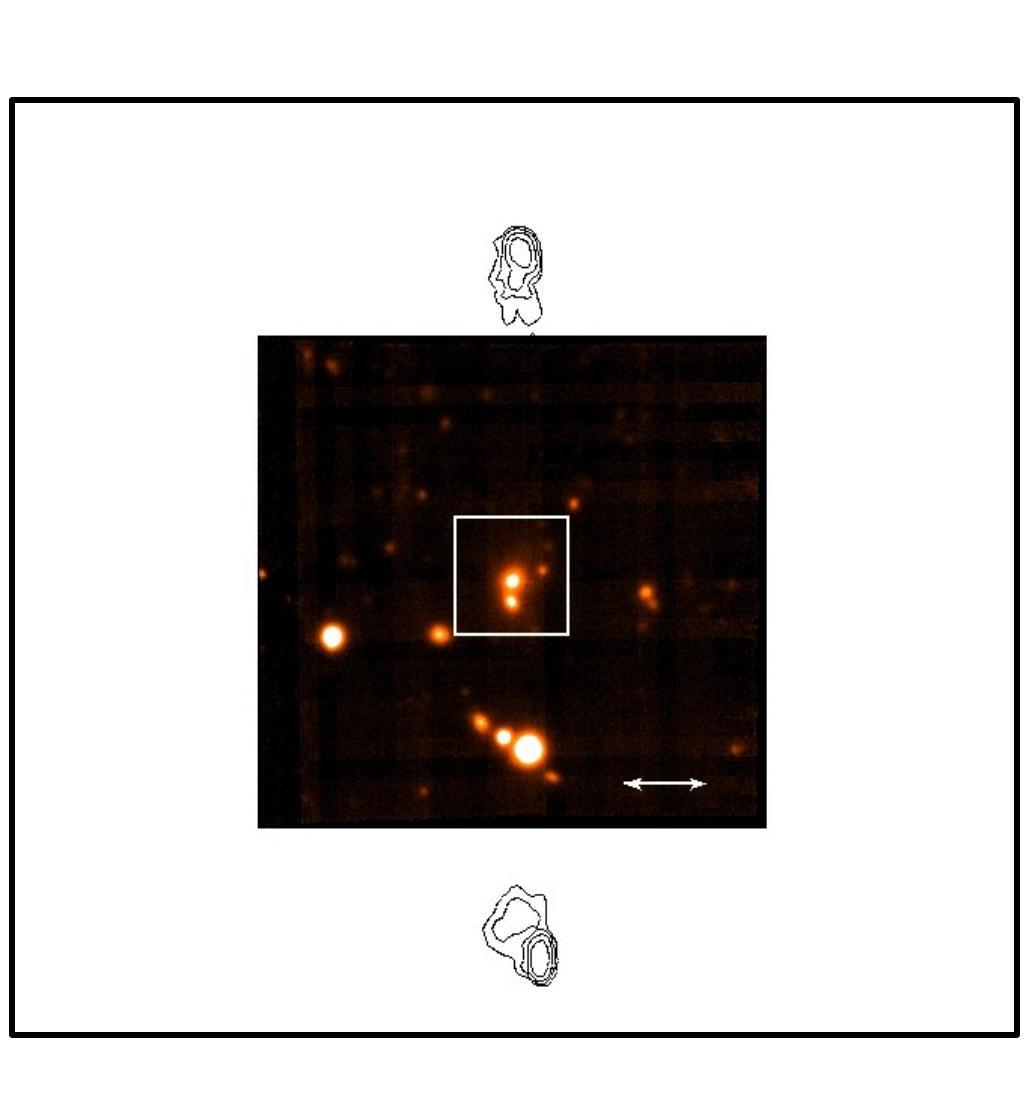}
\includegraphics[width=0.25\linewidth]{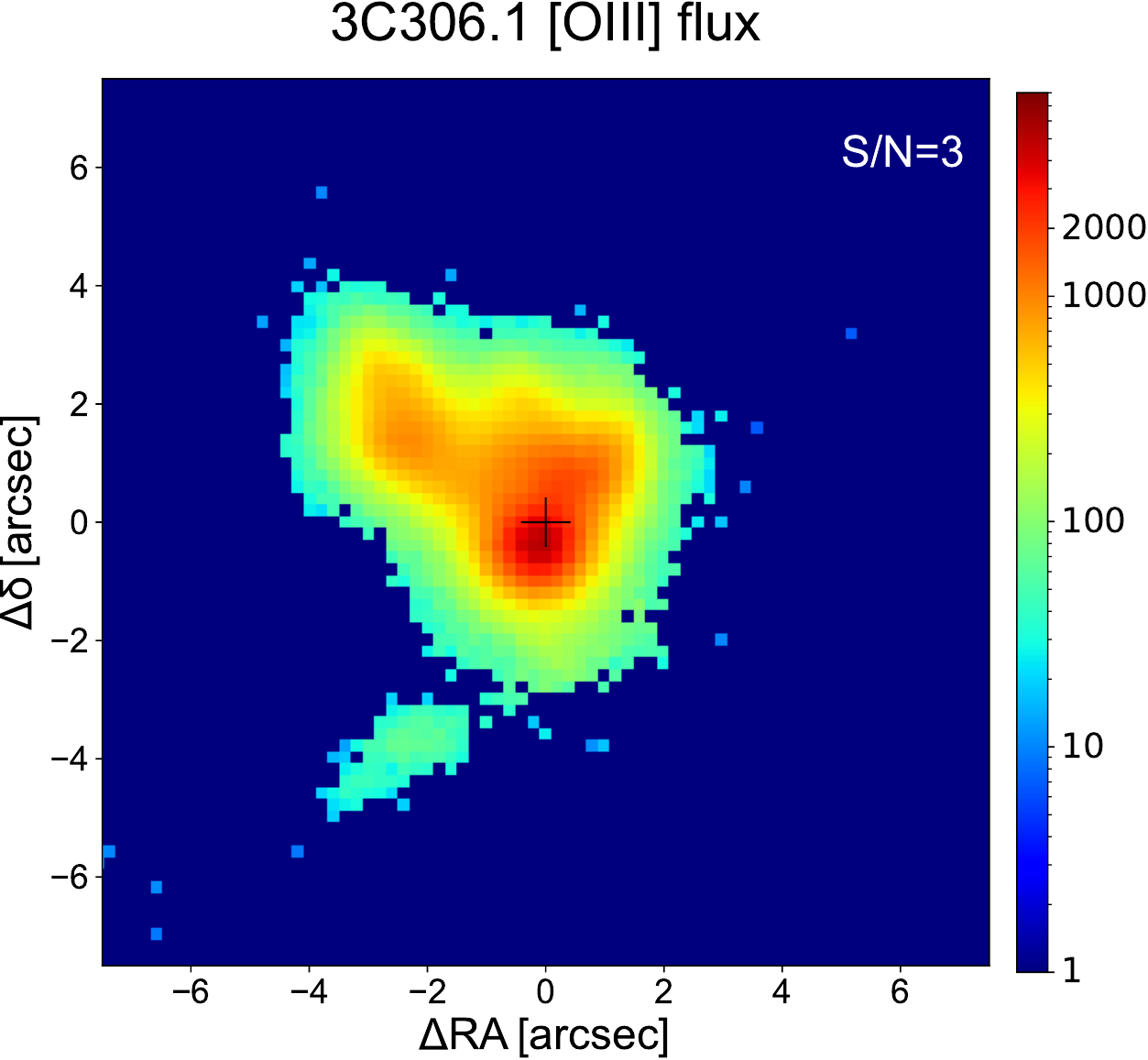}
\includegraphics[width=0.25\linewidth]{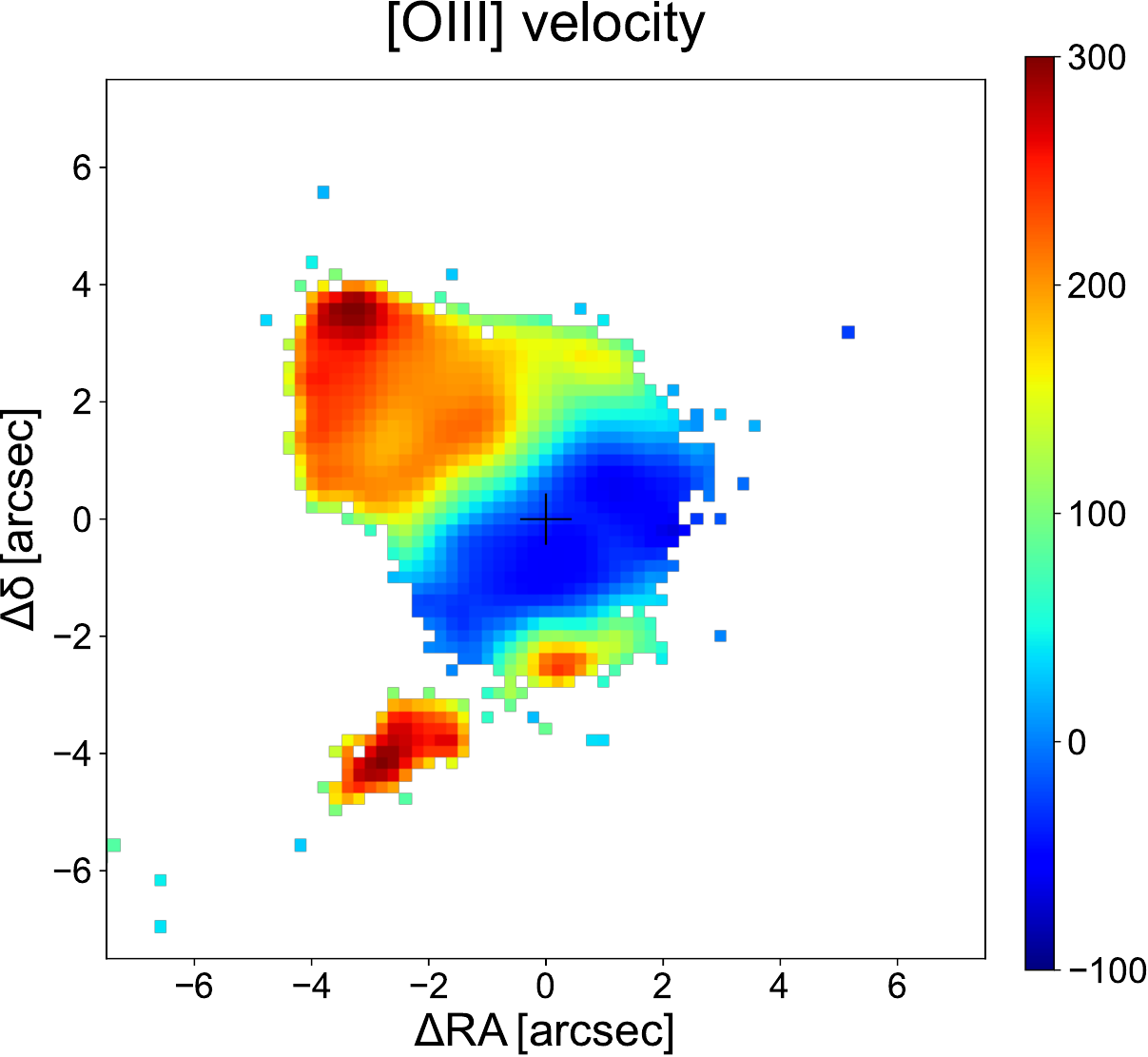}
\includegraphics[width=0.25\linewidth]{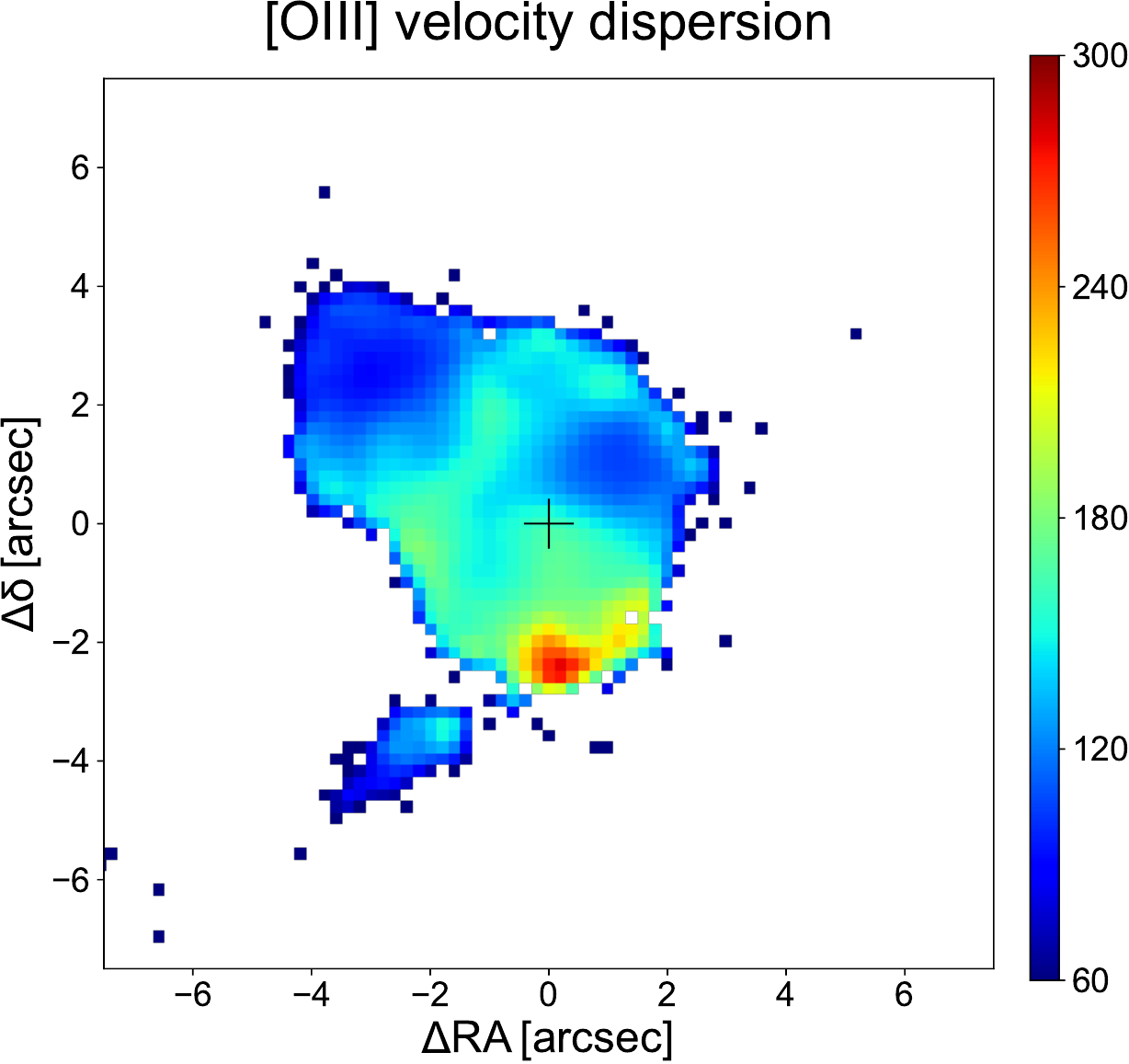}

\caption{- continued}
    \label{fig1}
\end{figure*}

\addtocounter{figure}{-1}
\begin{figure*}
\centering
\includegraphics[width=0.21\linewidth]{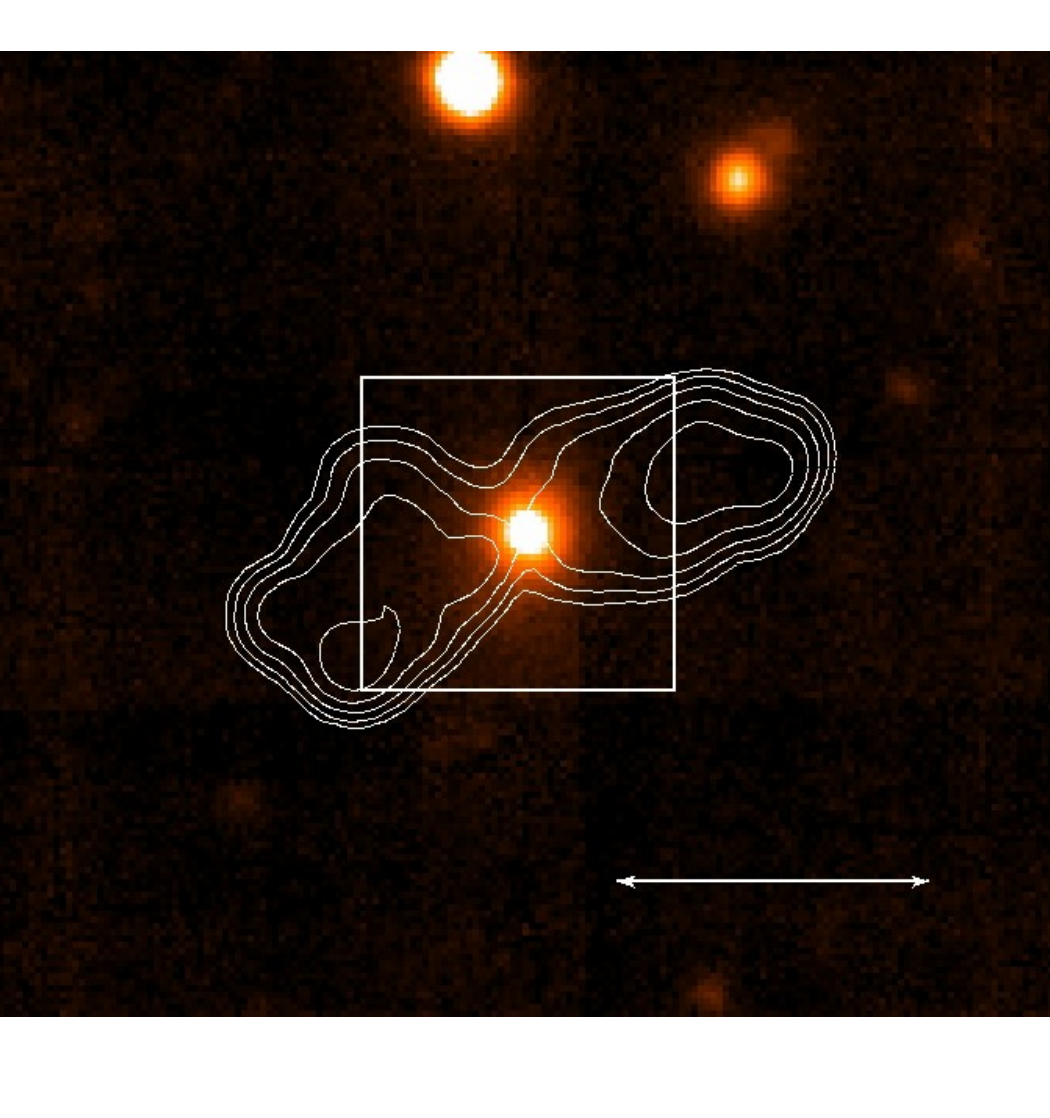}
\includegraphics[width=0.255\linewidth]{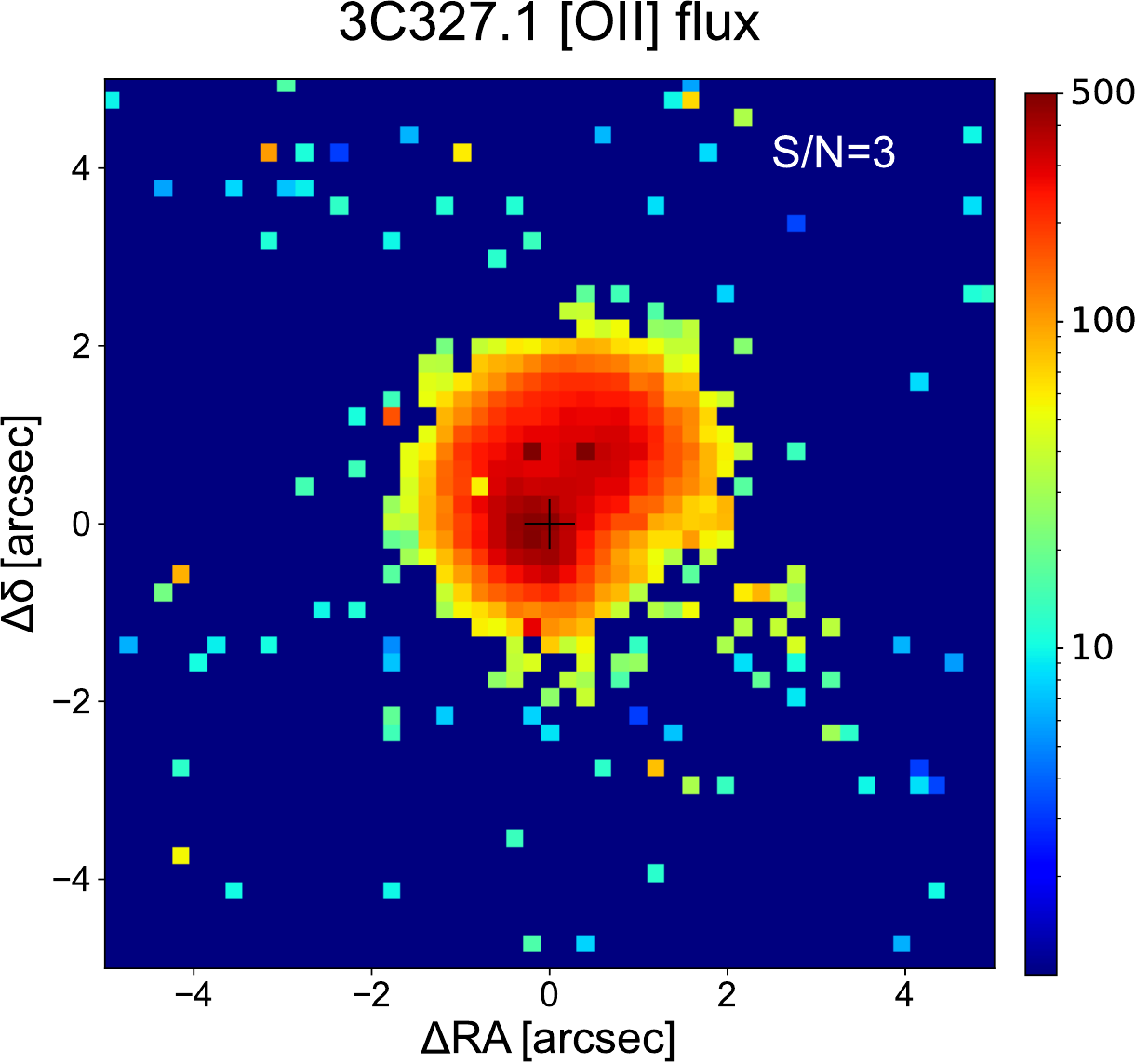}
\includegraphics[width=0.25\linewidth]{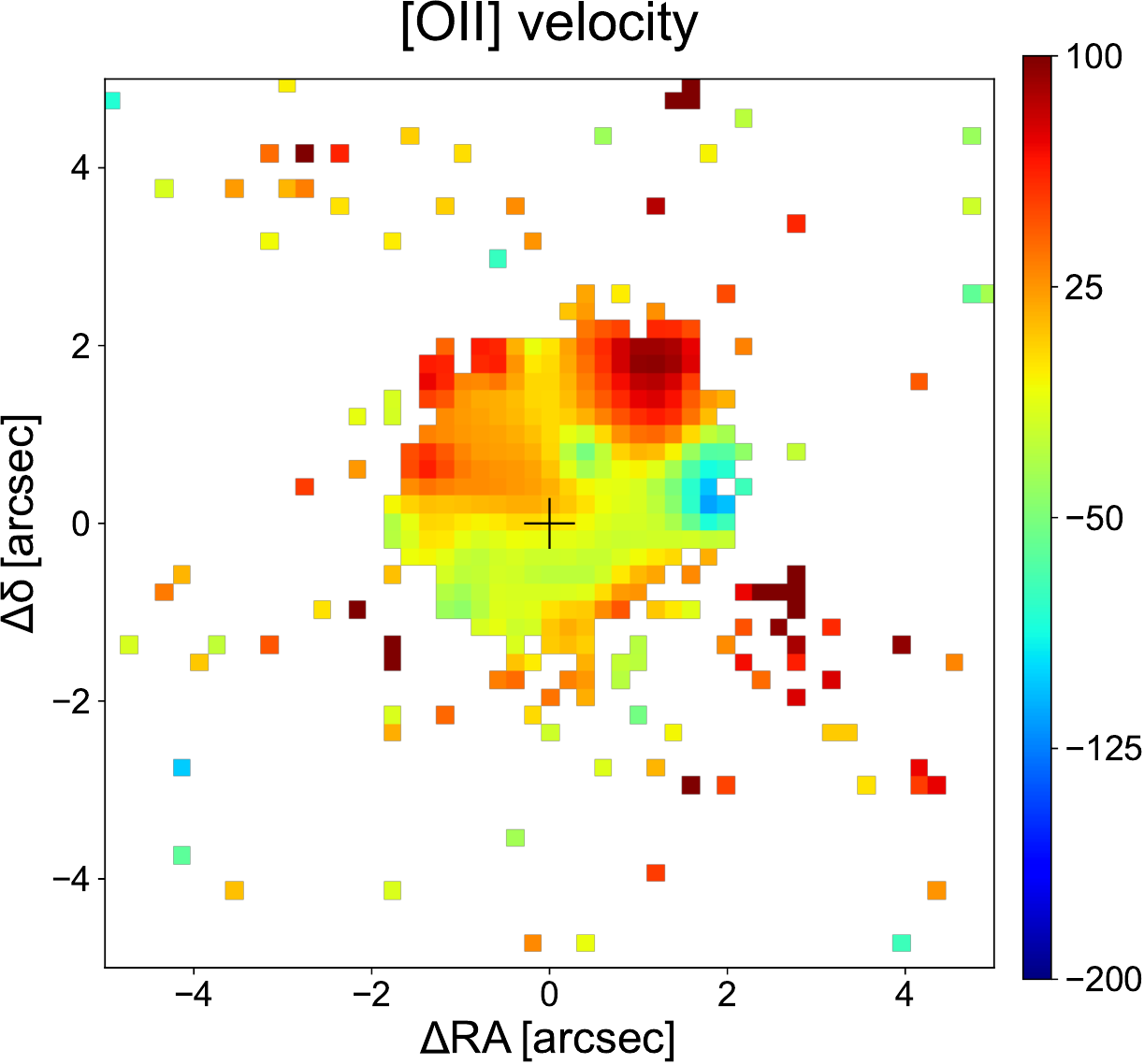}
\includegraphics[width=0.25\linewidth]{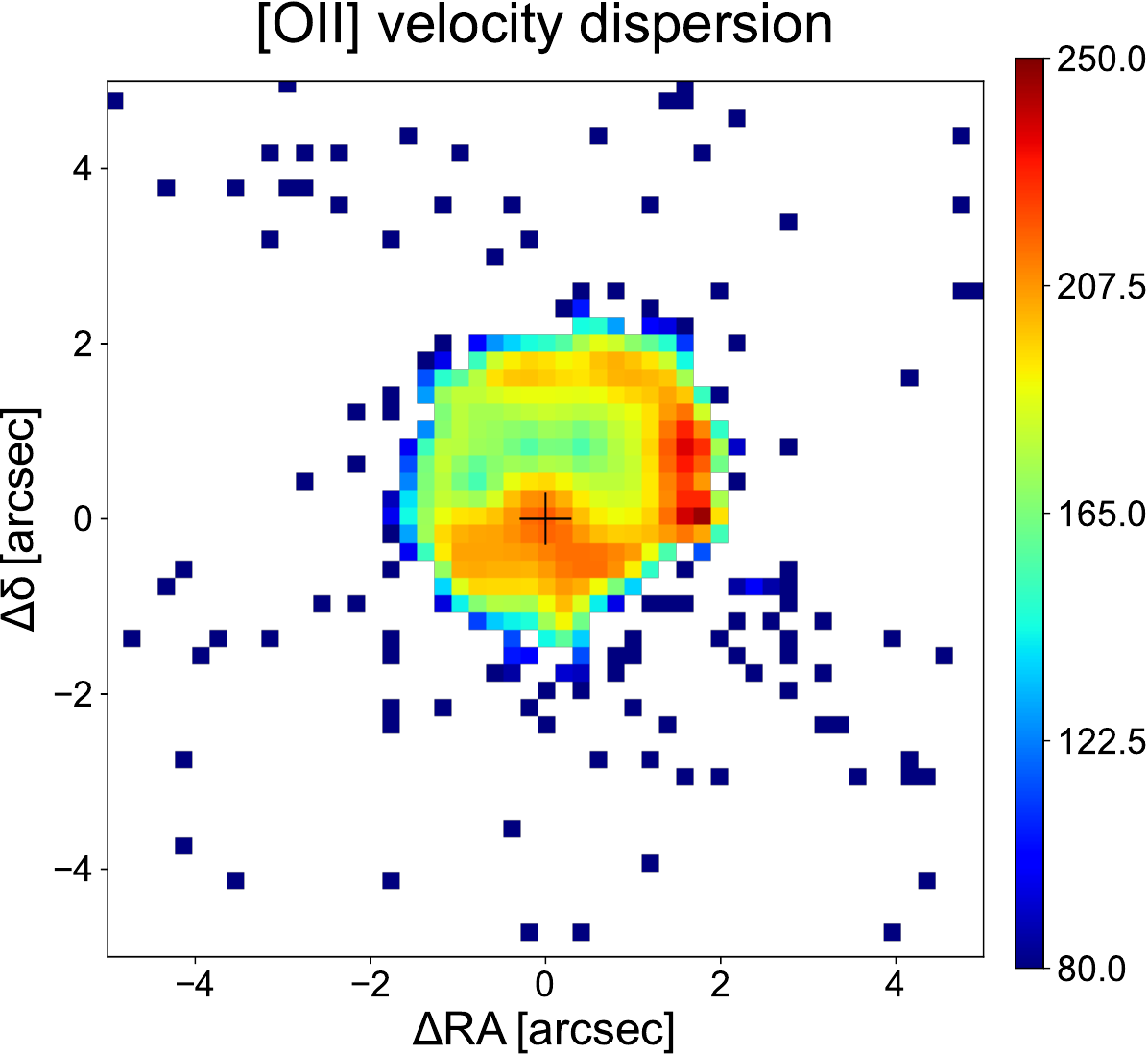}

\includegraphics[width=0.21\linewidth]{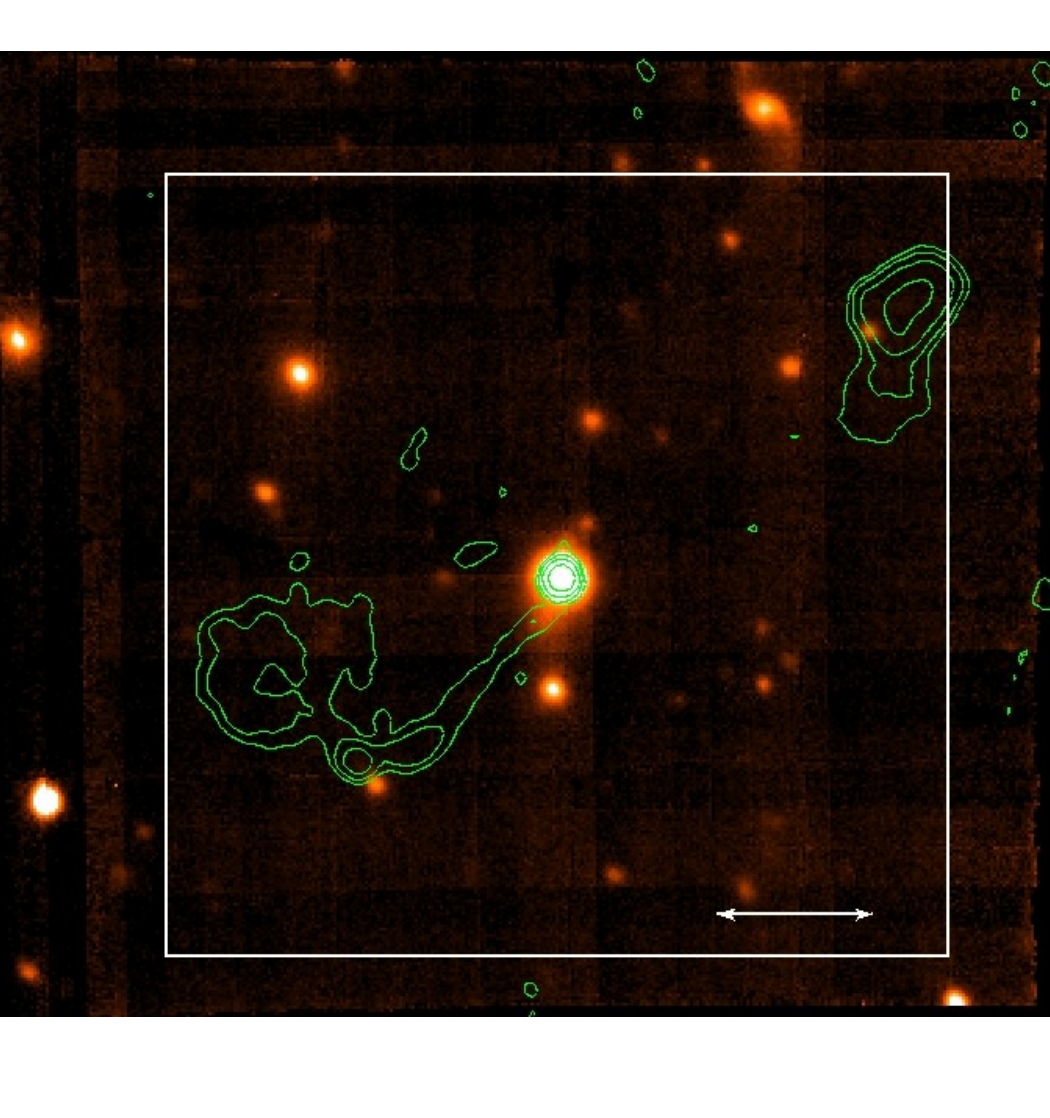}
\includegraphics[width=0.25\linewidth]{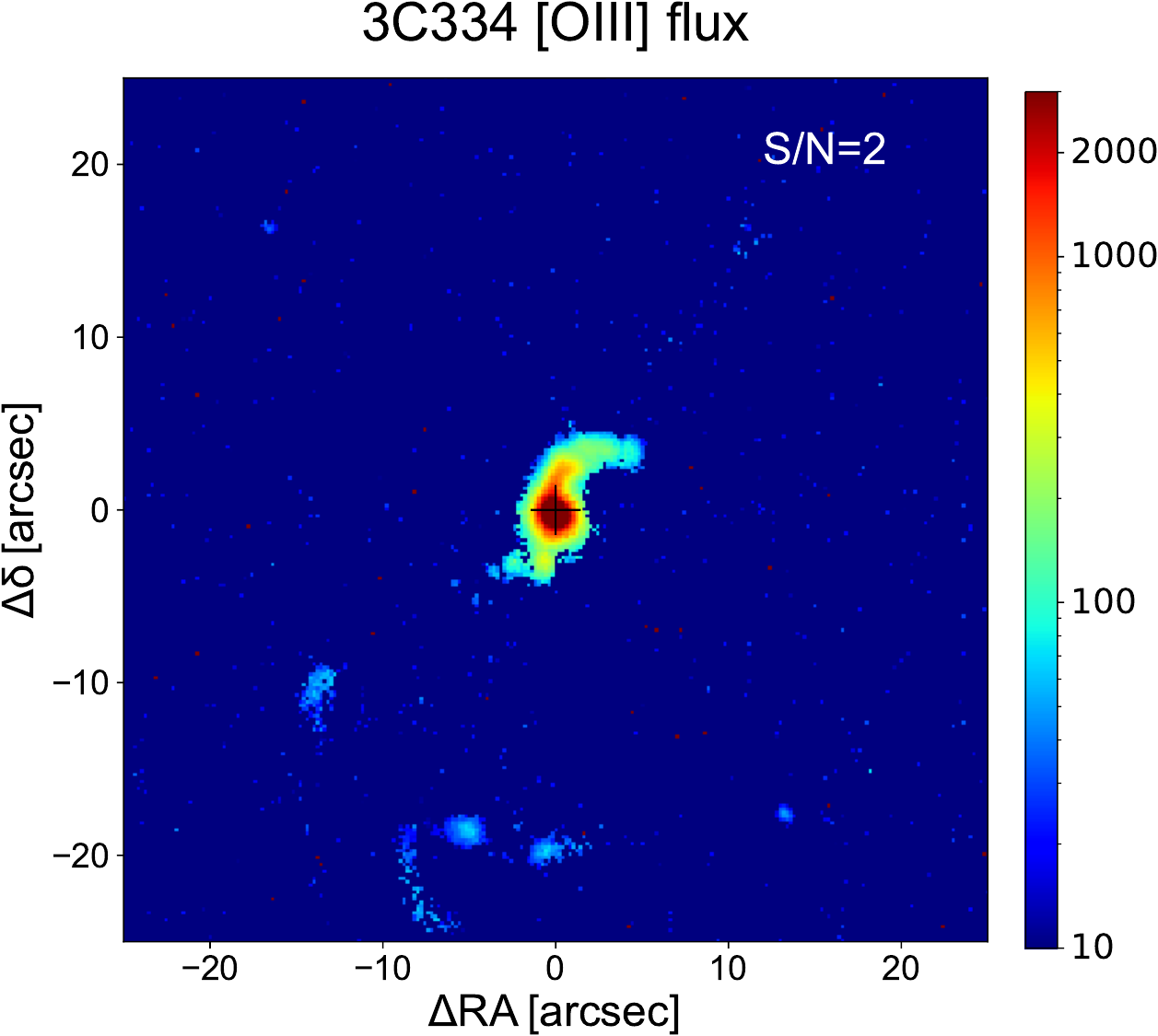}
\includegraphics[width=0.25\linewidth]{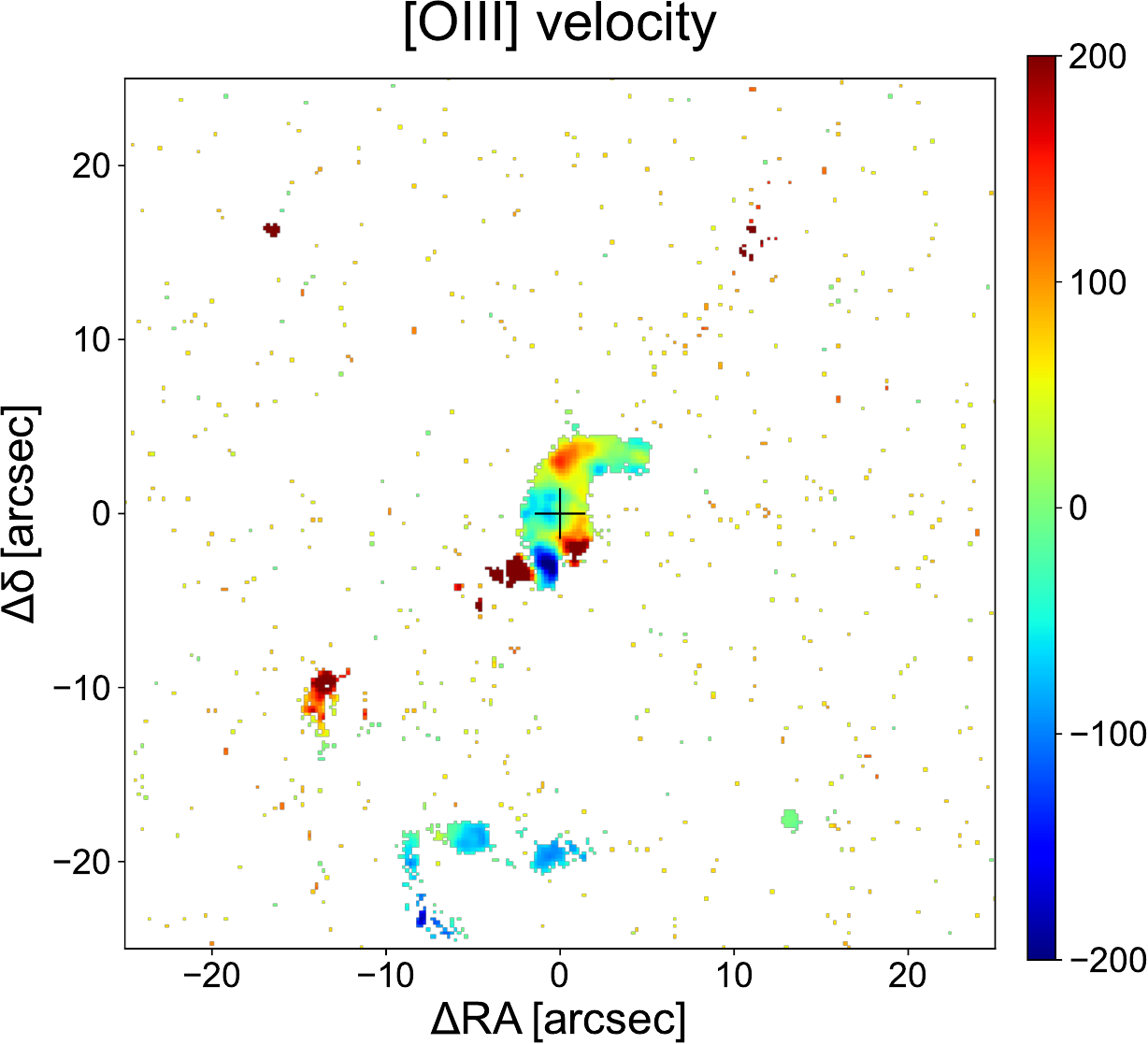}
\includegraphics[width=0.25\linewidth]{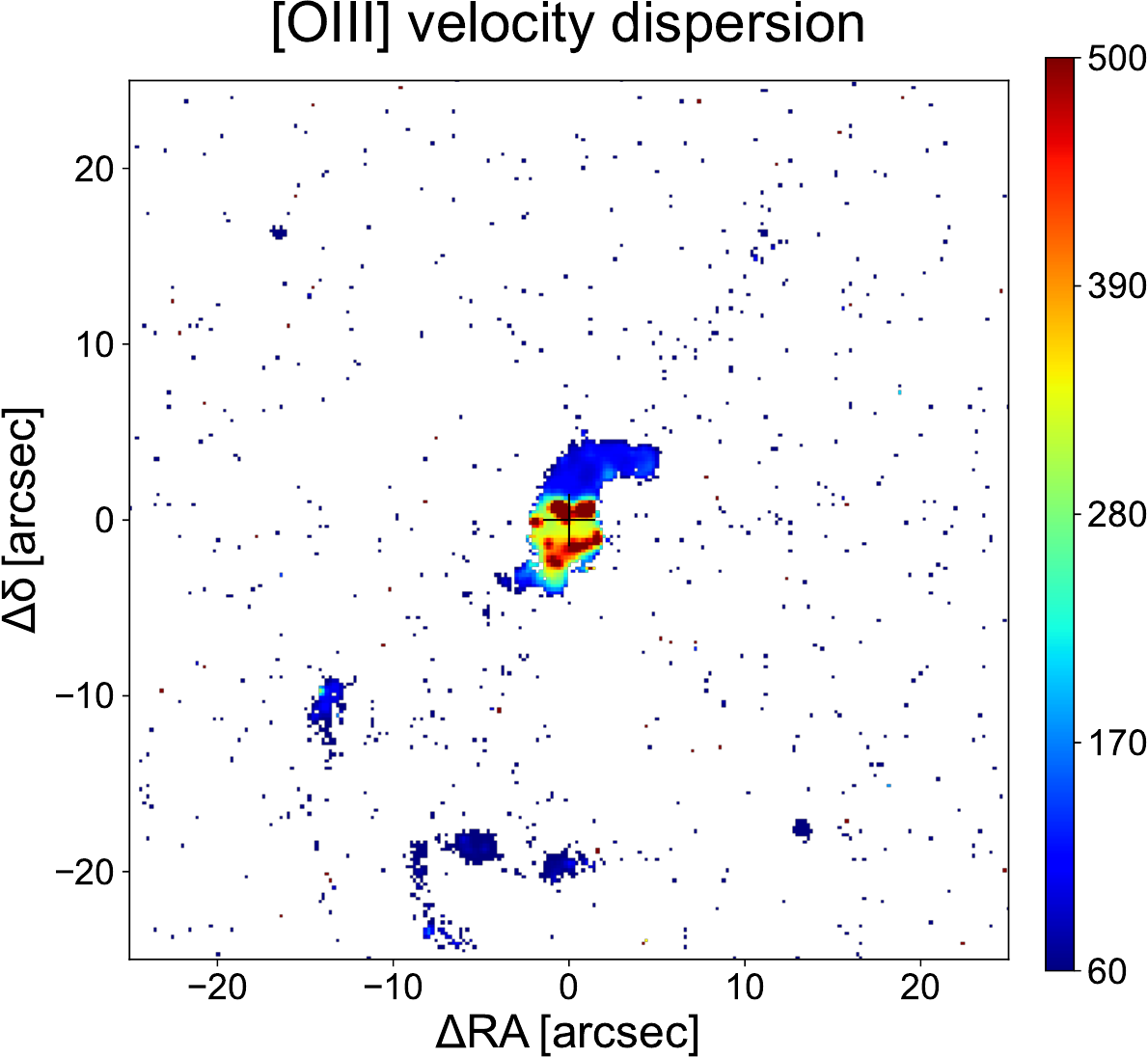}

\includegraphics[width=0.21\linewidth]{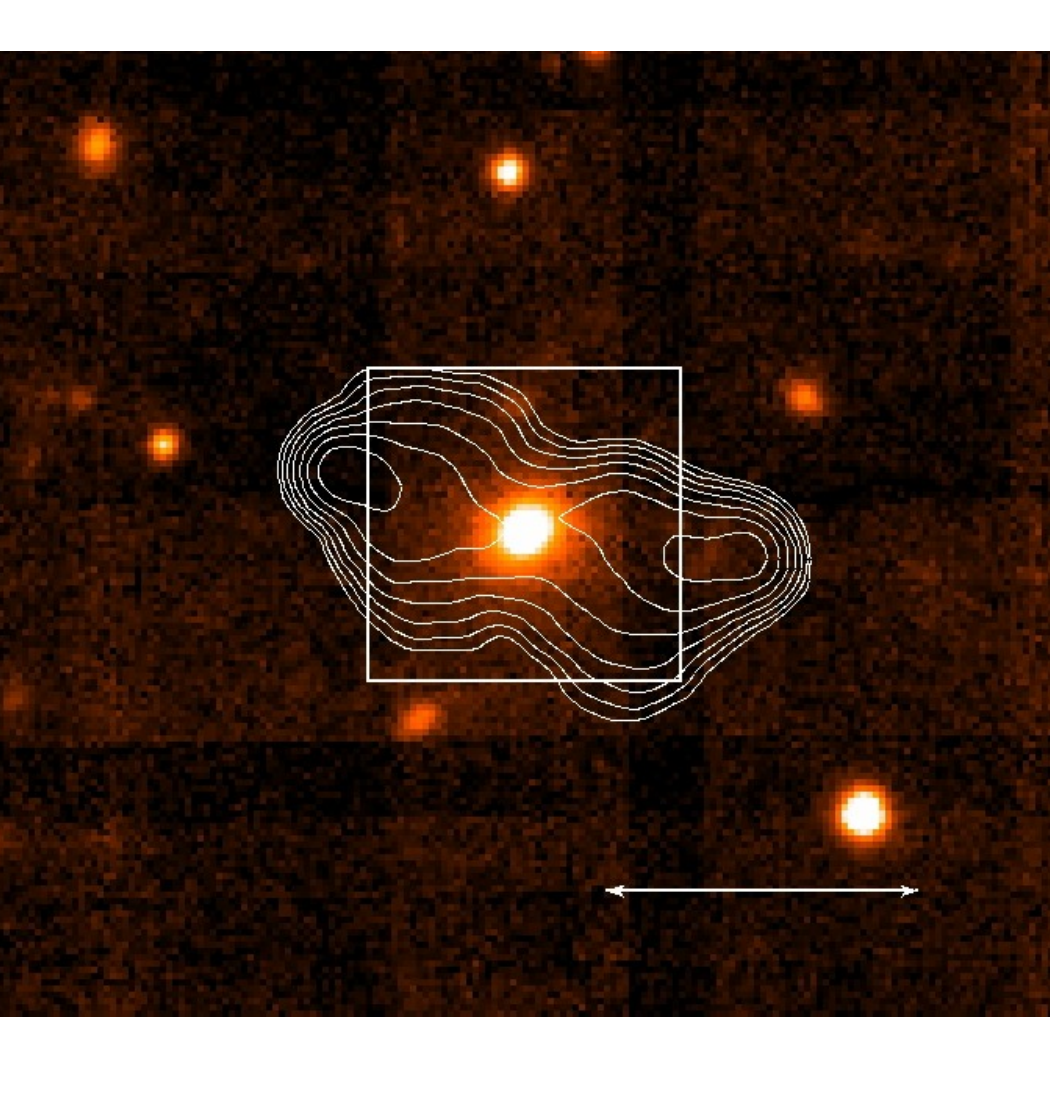}
\includegraphics[width=0.255\linewidth]{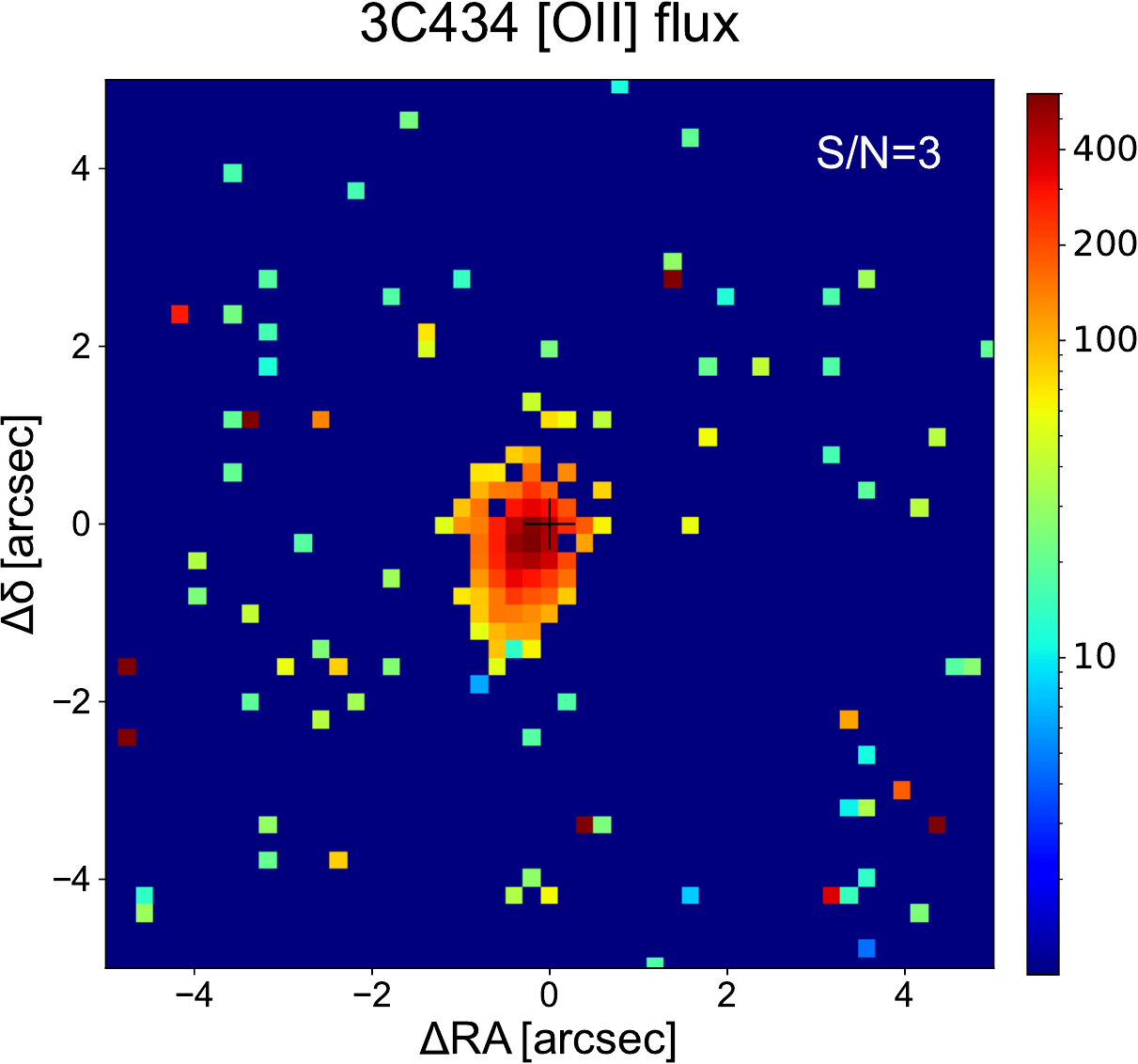}
\includegraphics[width=0.25\linewidth]{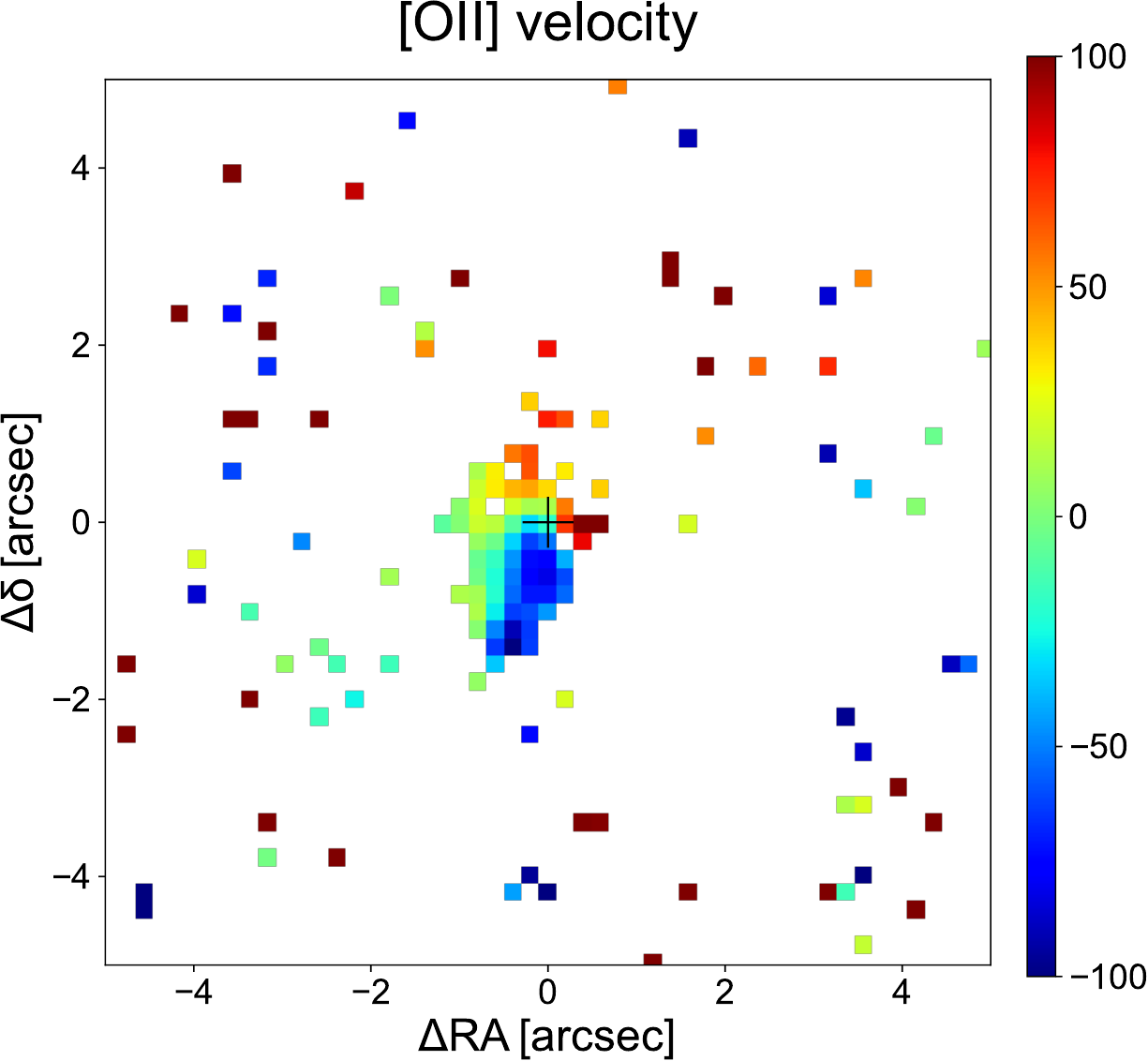}
\includegraphics[width=0.25\linewidth]{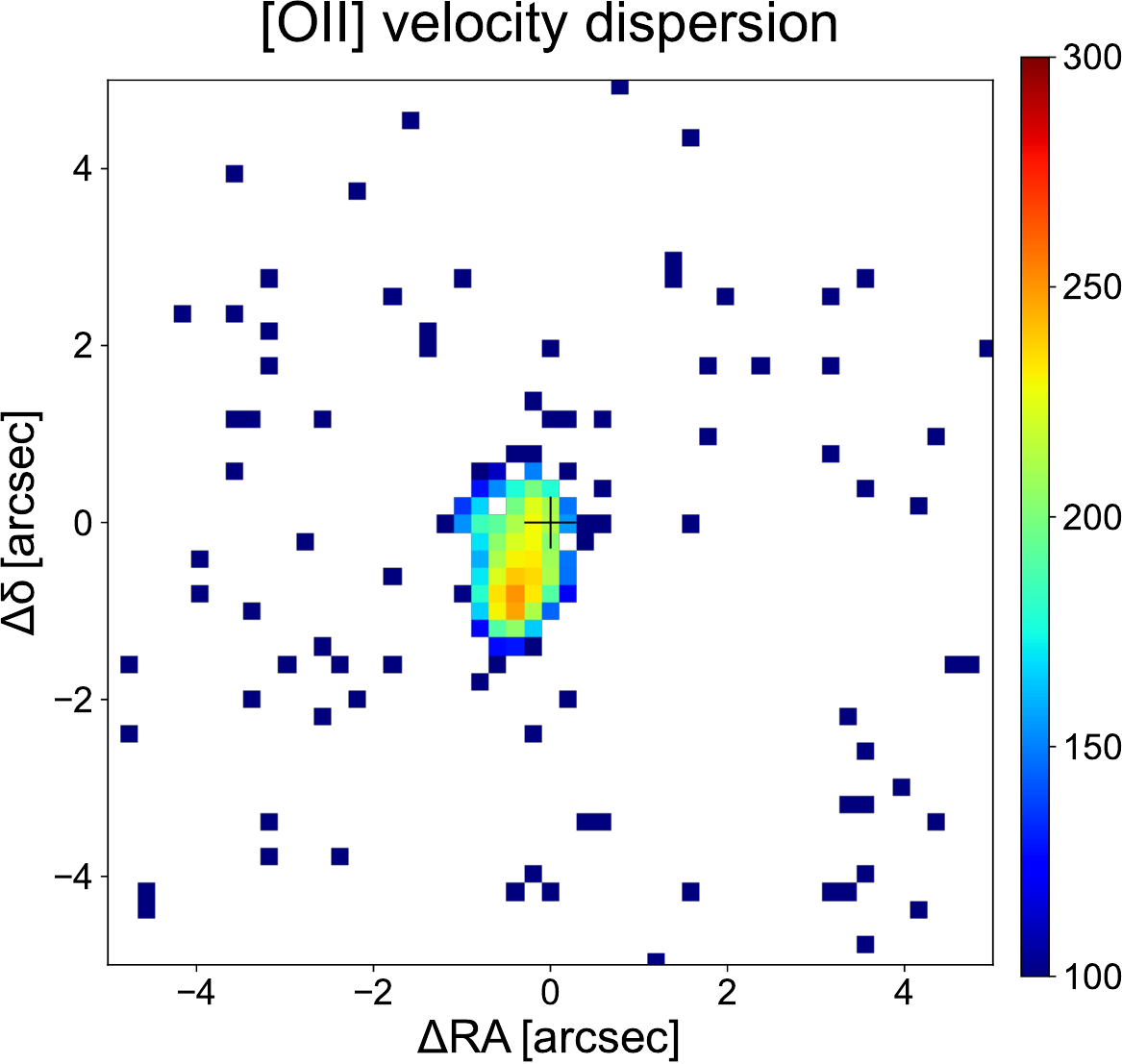}

\includegraphics[width=0.20\linewidth]{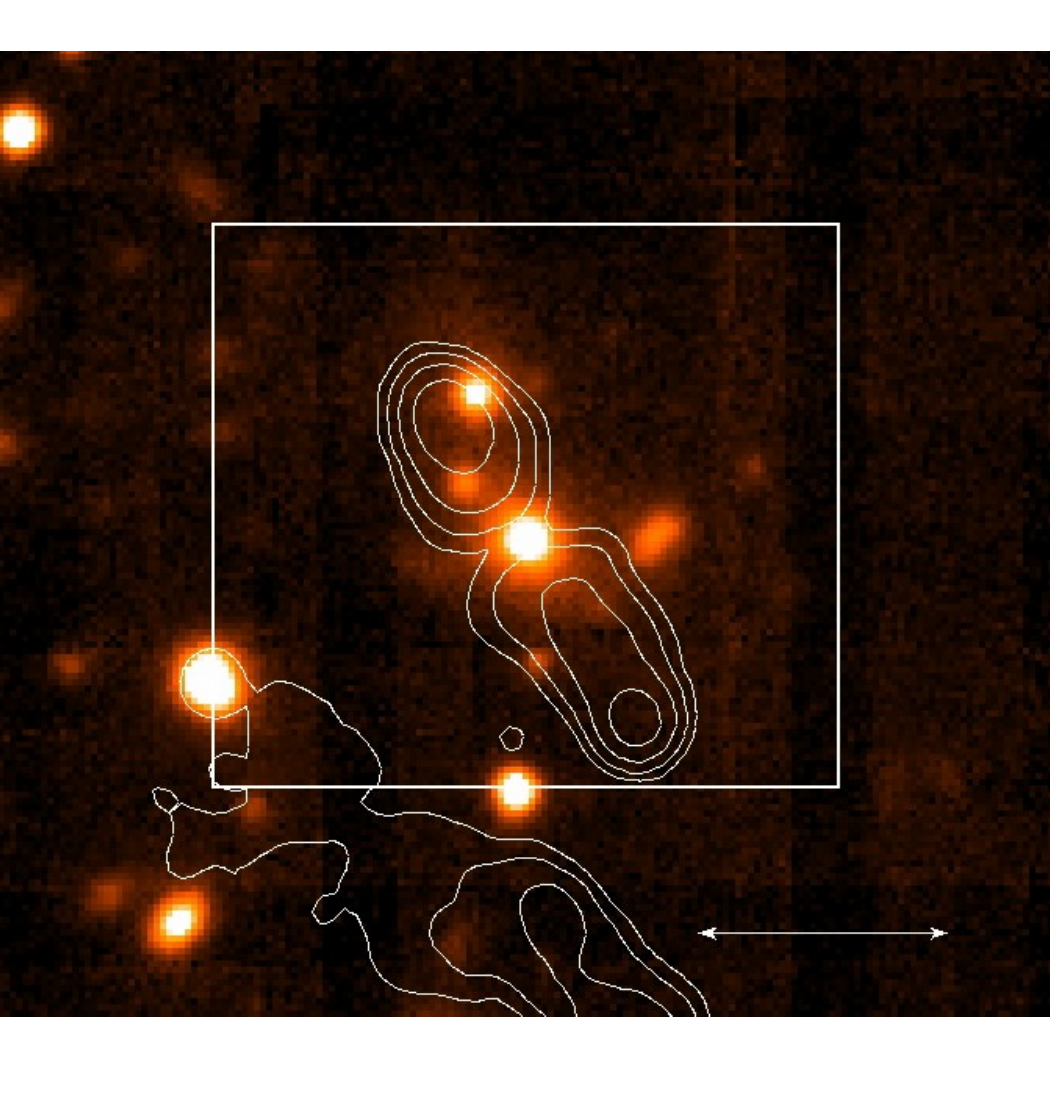}
\includegraphics[width=0.25\linewidth]{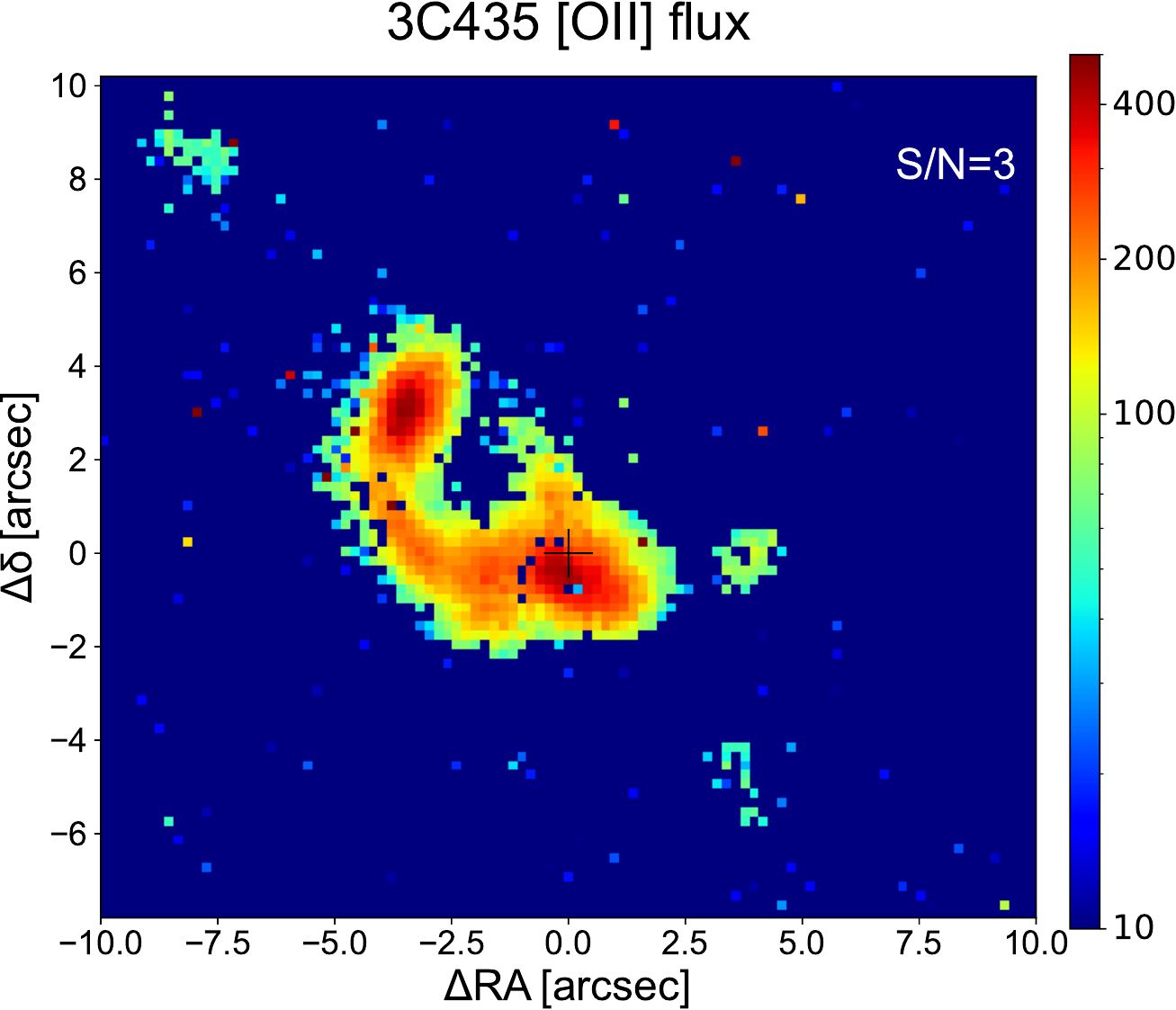}
\includegraphics[width=0.26\linewidth]{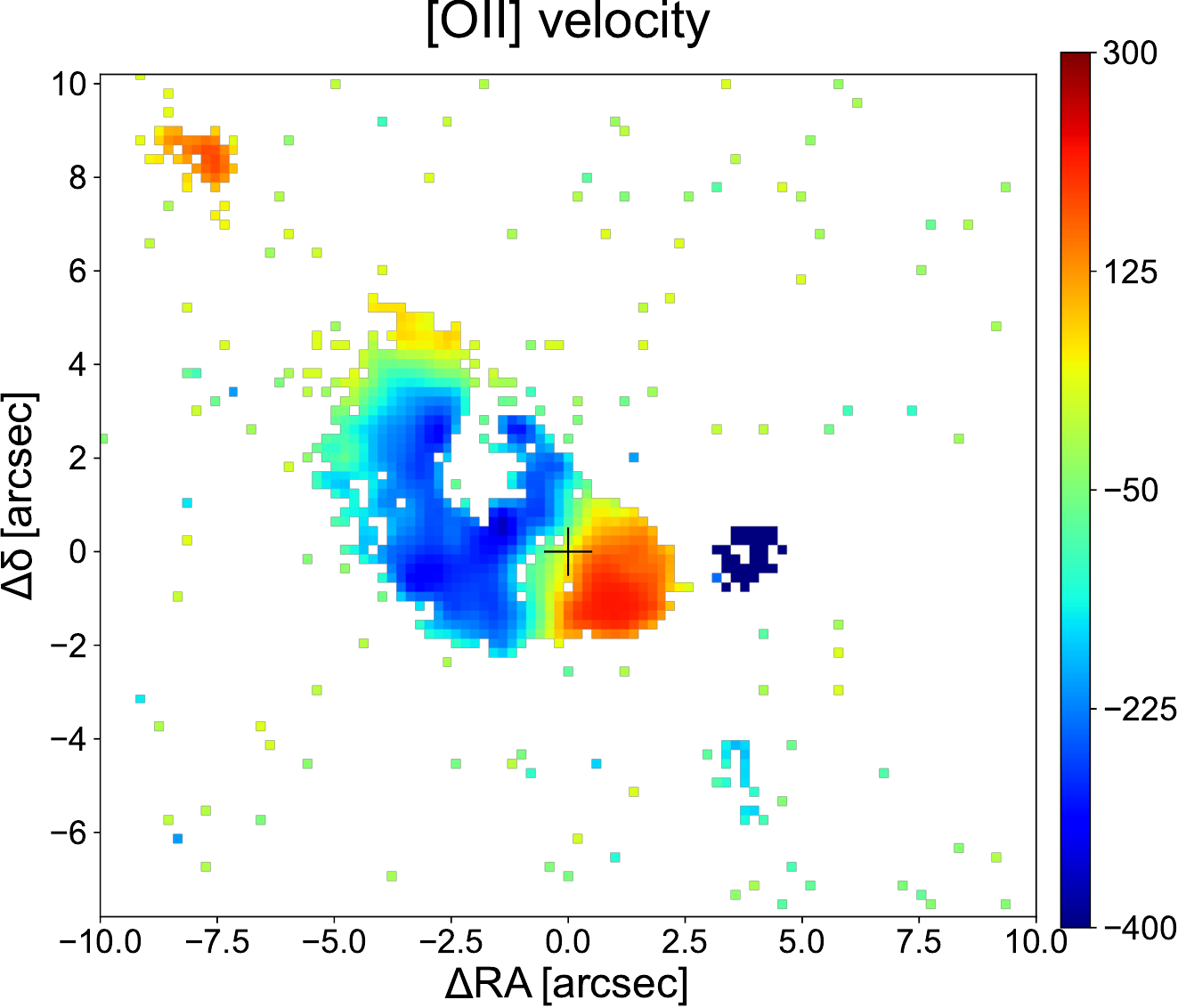}
\includegraphics[width=0.26\linewidth]{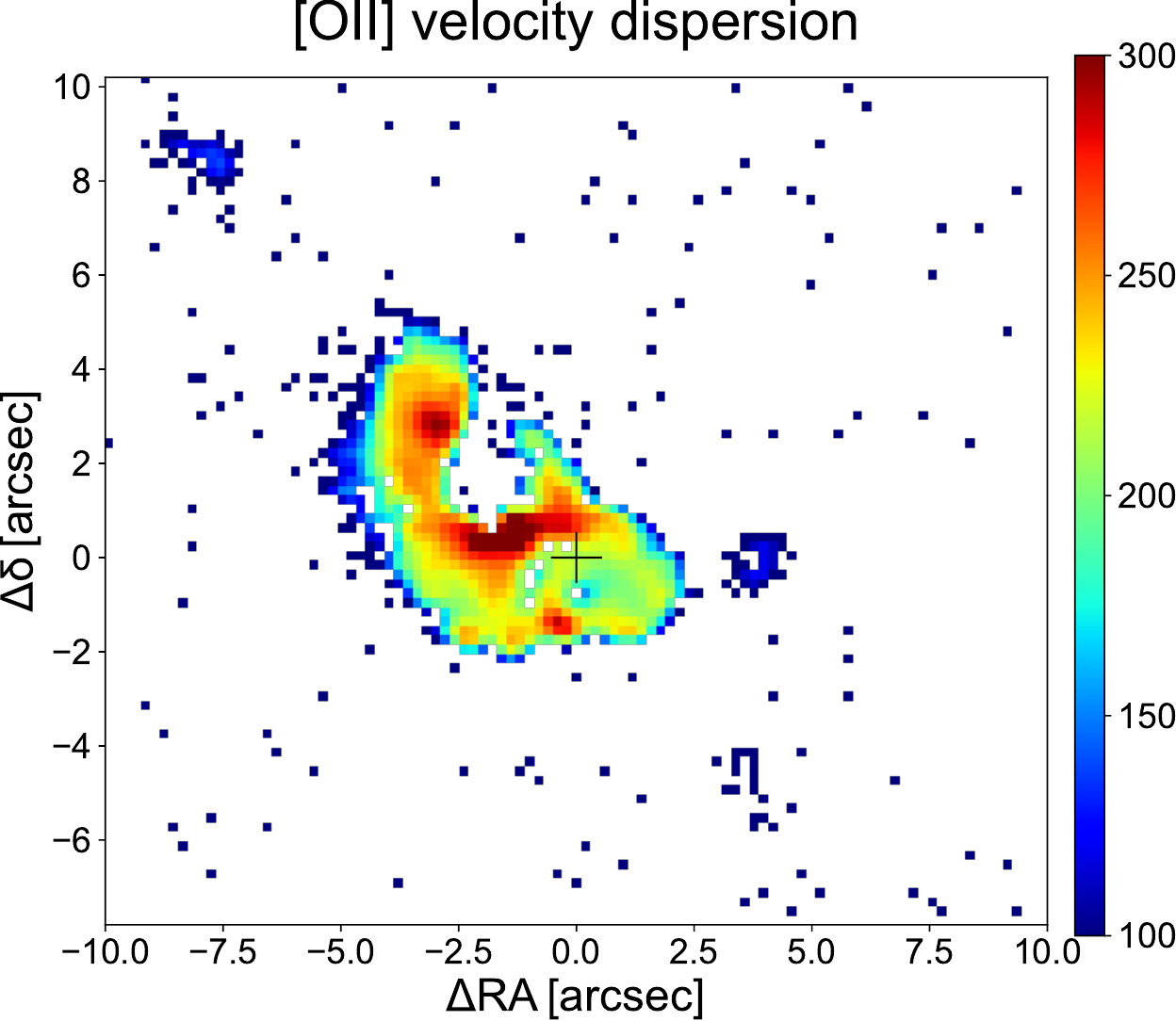}

\includegraphics[width=0.21\linewidth]{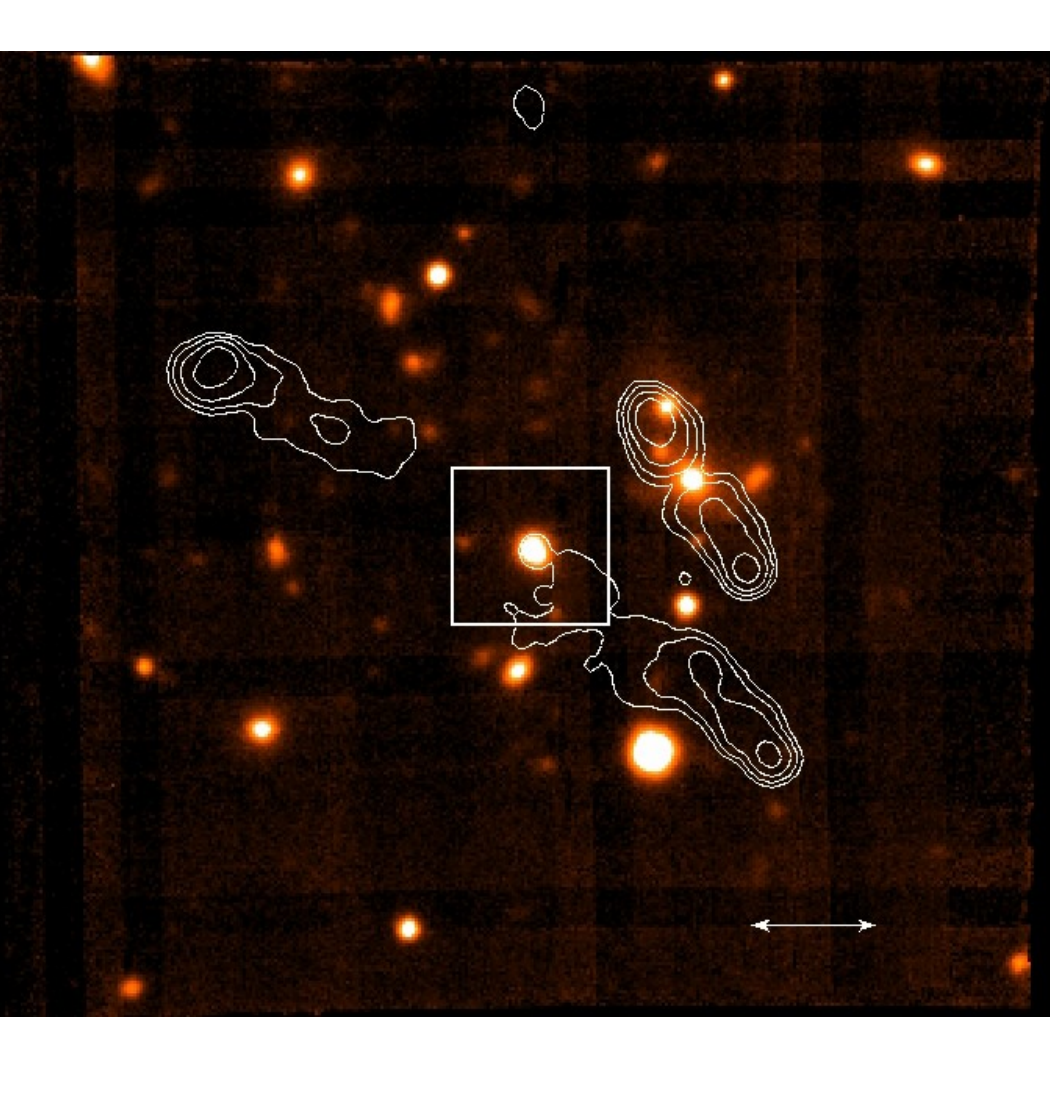}
\includegraphics[width=0.25\linewidth]{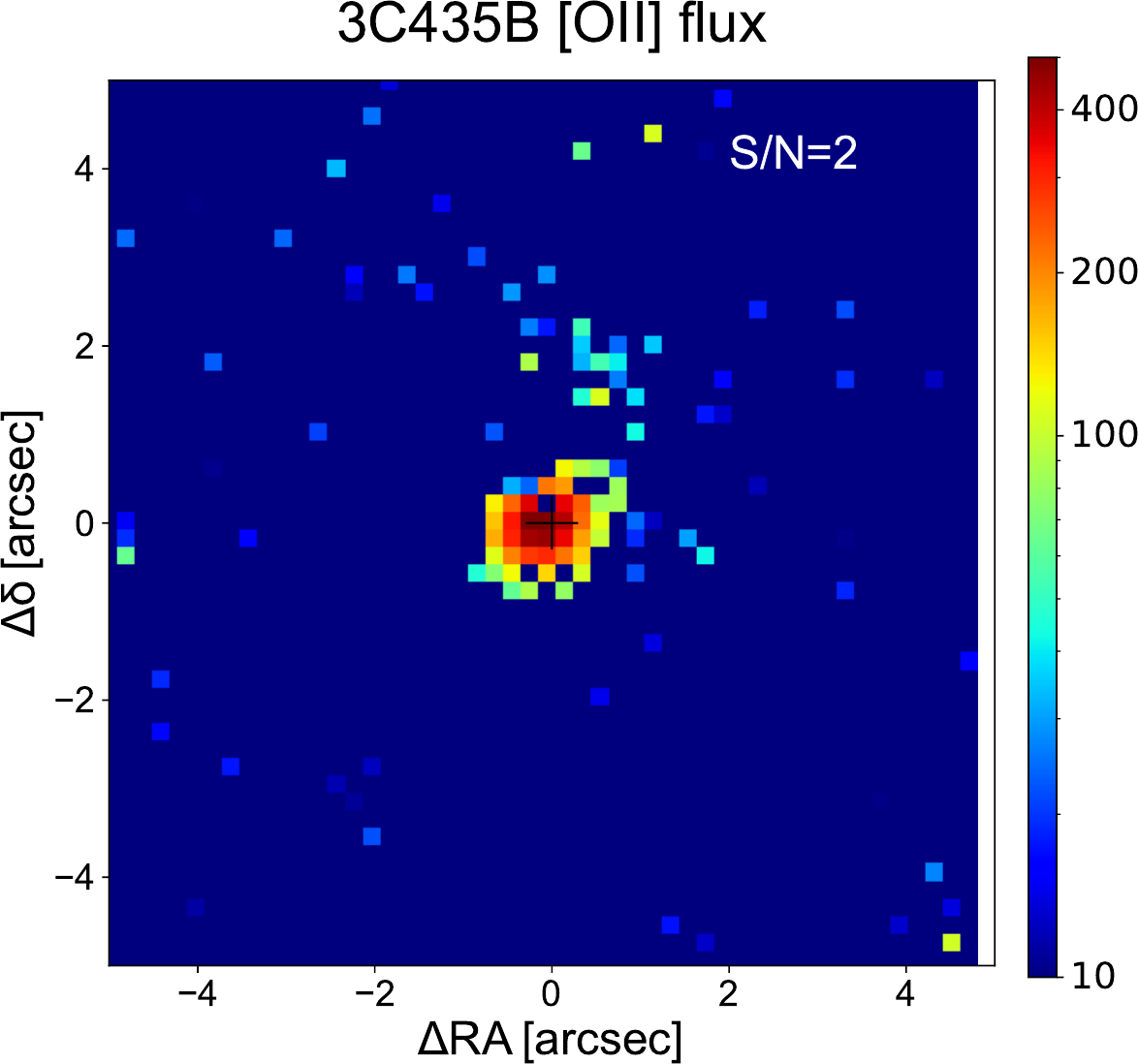}
\includegraphics[width=0.25\linewidth]{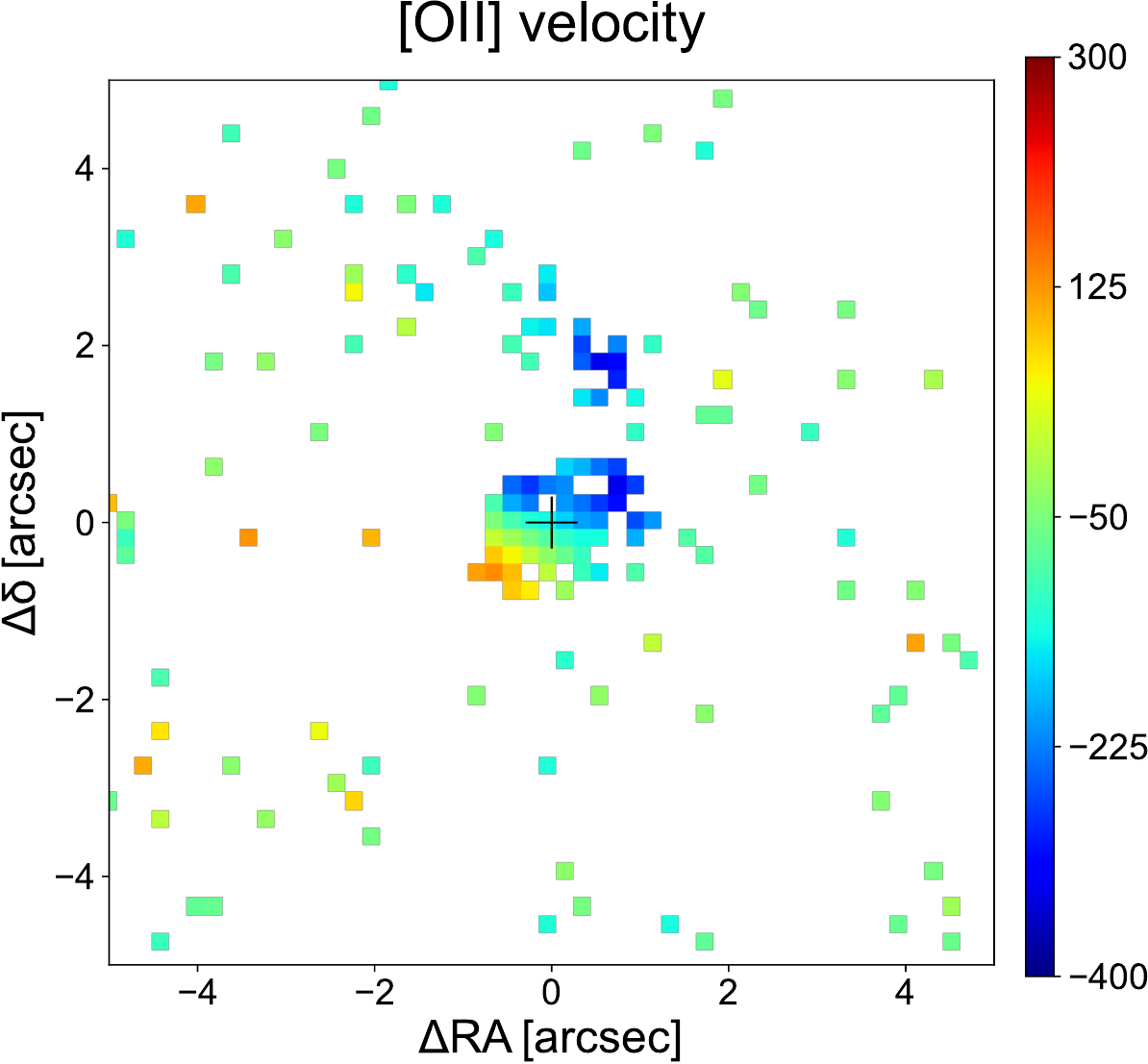}
\includegraphics[width=0.25\linewidth]{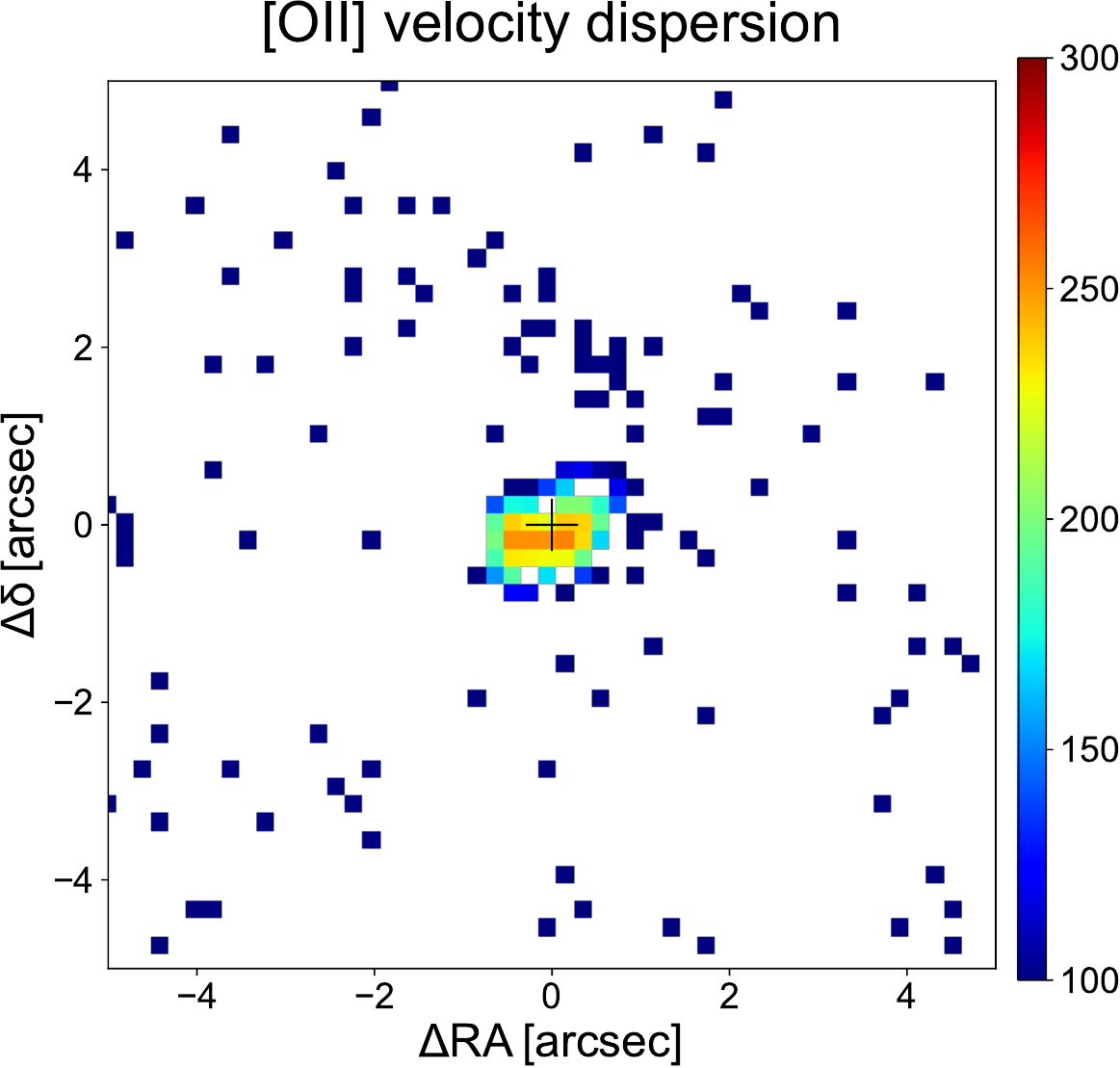}

\includegraphics[width=0.21\linewidth]{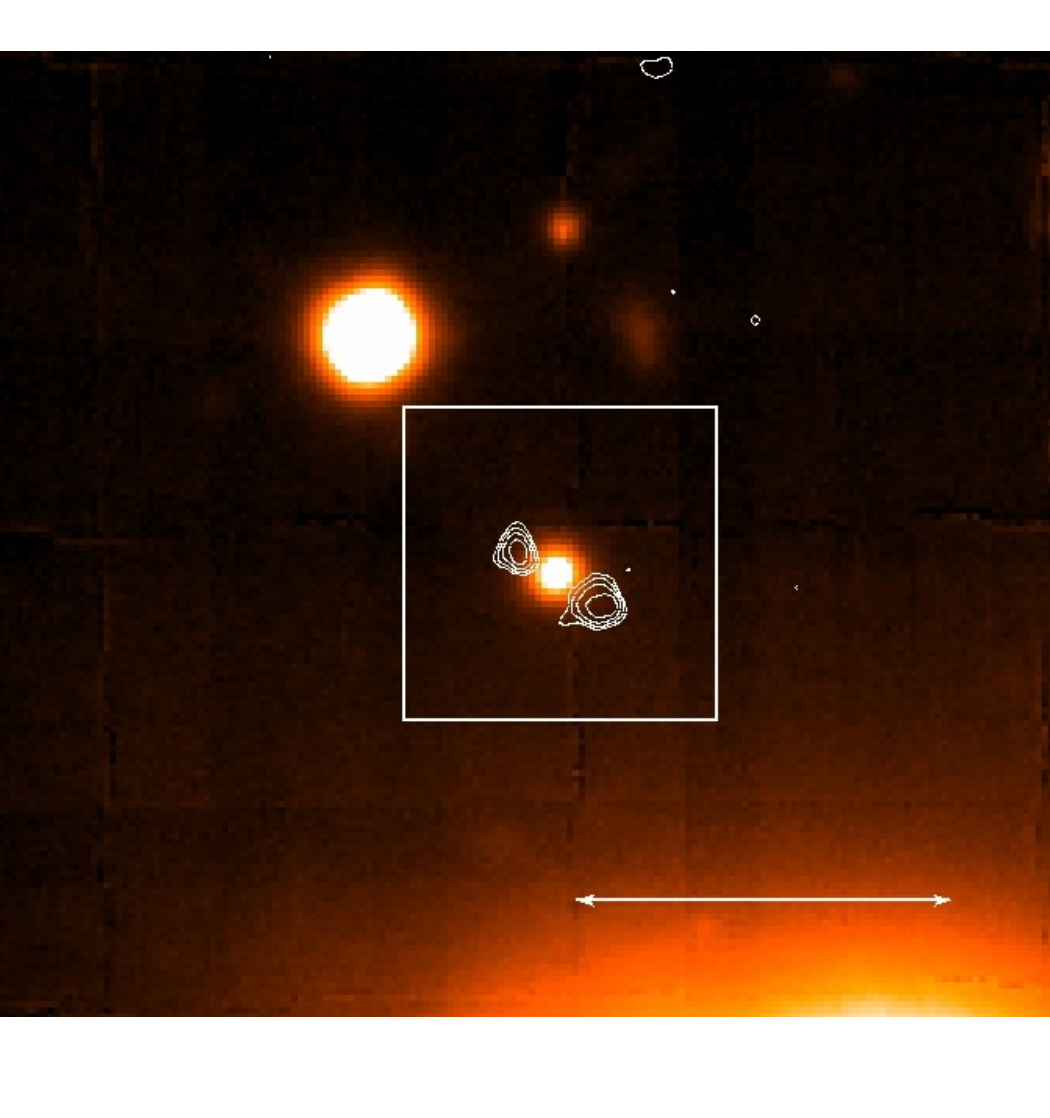}
\includegraphics[width=0.25\linewidth]{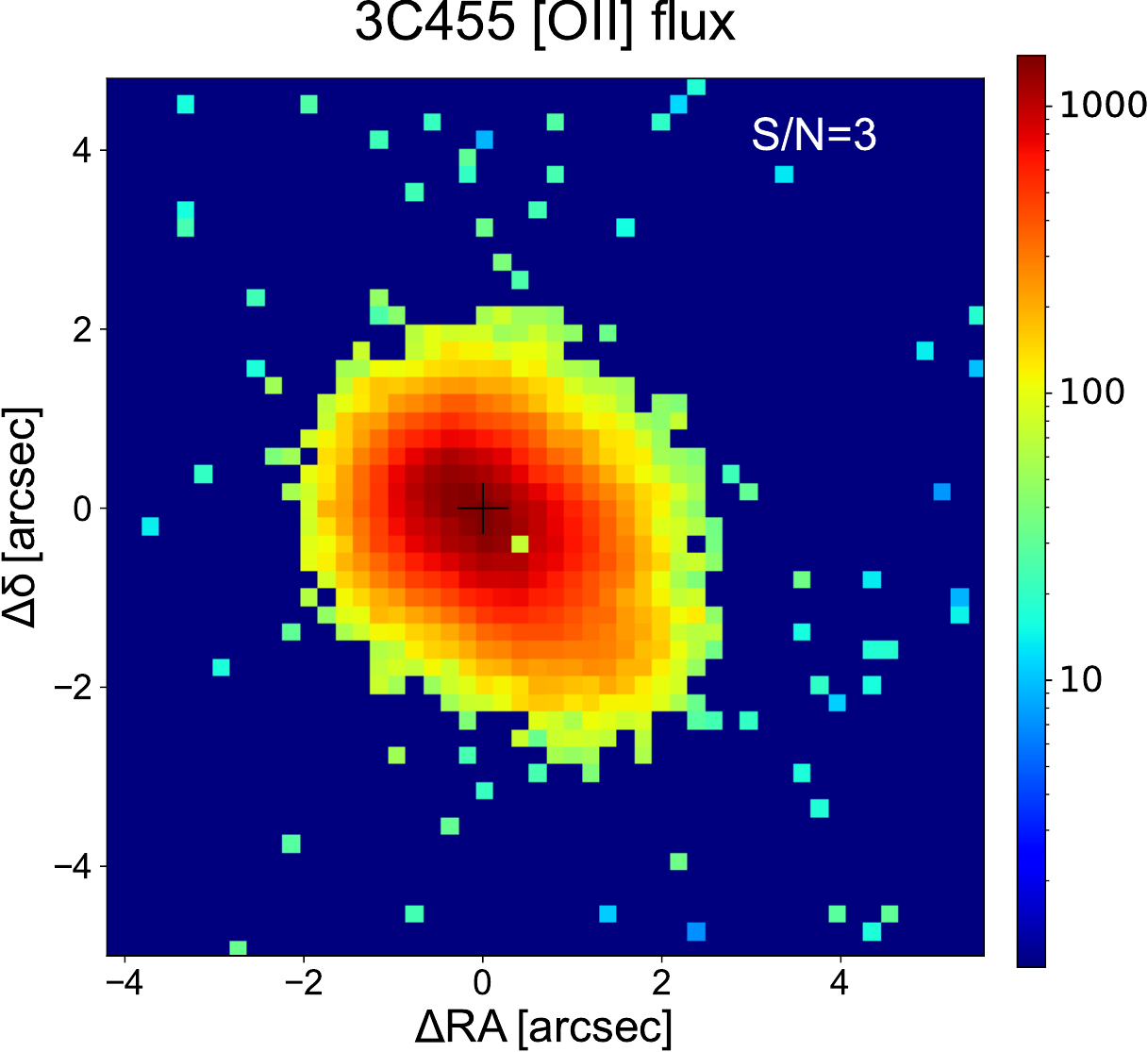}
\includegraphics[width=0.25\linewidth]{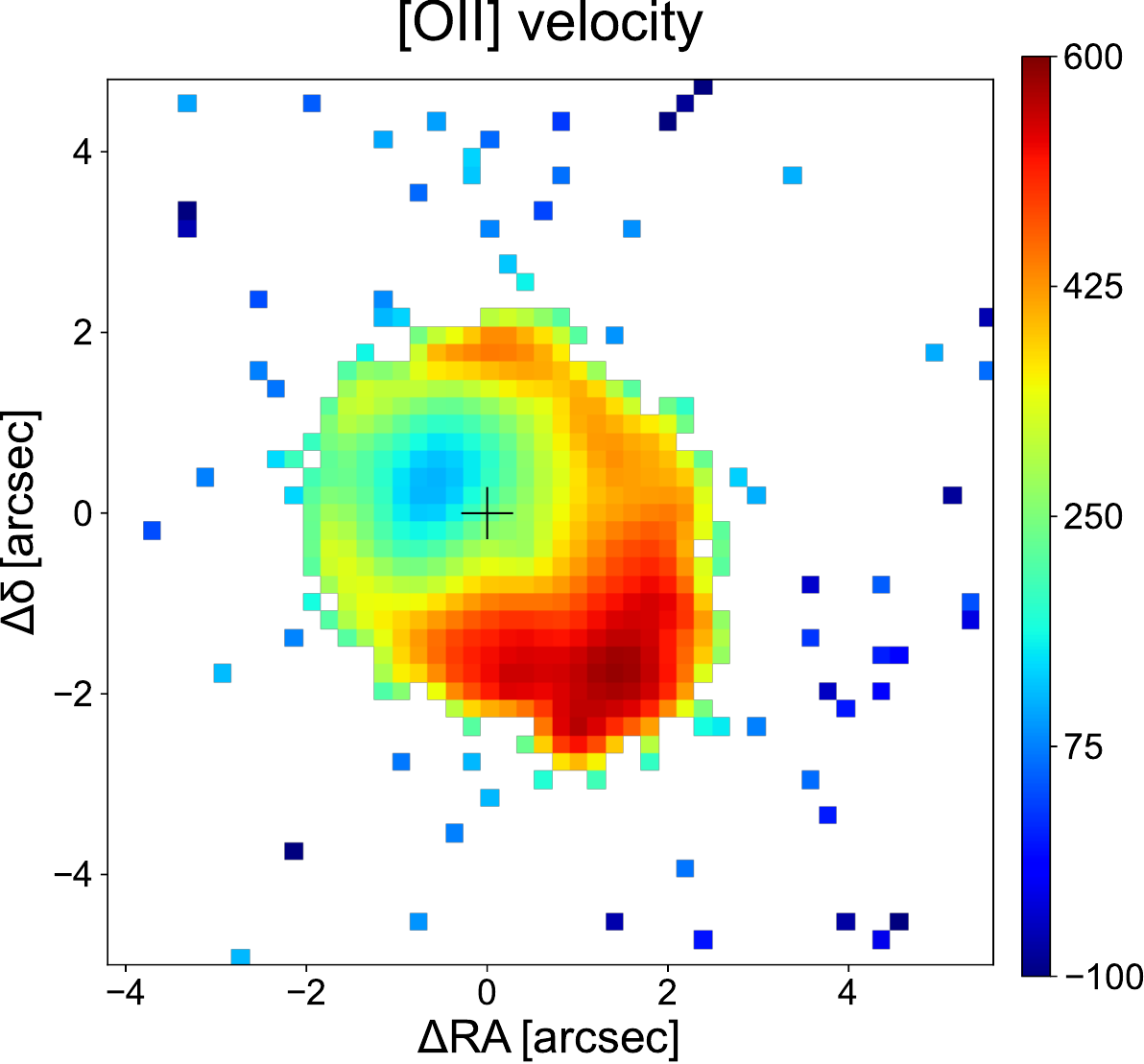}
\includegraphics[width=0.25\linewidth]{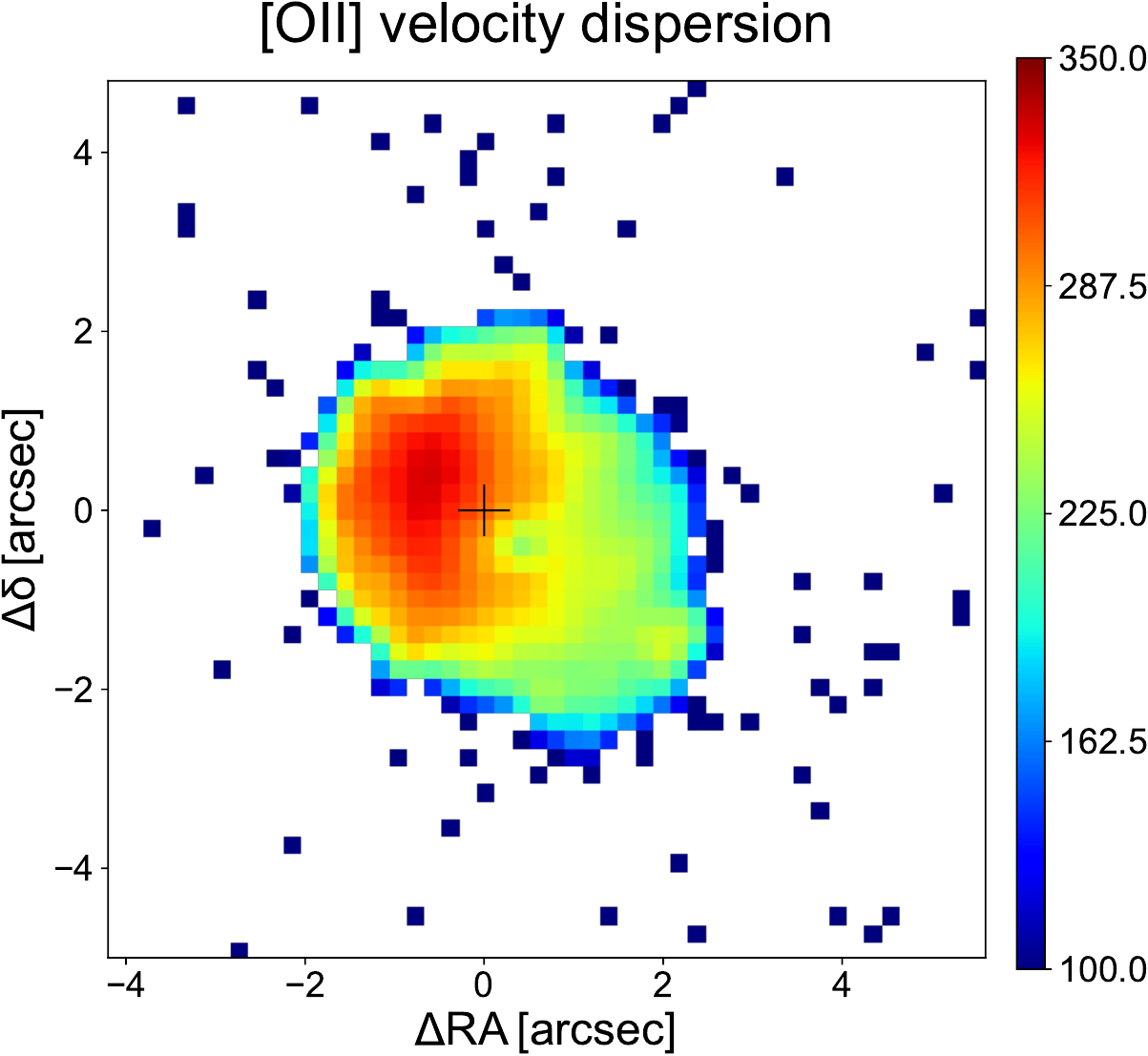}
\caption{- continued}
    \label{fig1}
\end{figure*}

\clearpage

\section{Parameters of the radio images used.}

\FloatBarrier

  \begin{table}
    \caption{Main properties of the observed galaxies.}             
    \label{table:radio}      
    \centering          
    \begin{tabular}{l | c c clc }
        \hline\hline       
Name  & Band & Config. & Beam & rms (mJy beam$^{-1}$)\\
        \hline                    
3C~016   & L  & A & 1\farcs35  & 0.144  \\  
3C~093   & L  & A & 1\farcs31  & 0.189  \\
3C~099   & C  & A & 0\farcs41  & 0.124  \\
3C~107   & C  & A & 0\farcs42  & 0.053  \\
3C~109   & L  & B &  3\farcs80 & 7.04  \\
3C~114   & L  & A & 1\farcs24 &  0.189 \\
3C~138   & L & A & 1\farcs26 & 0.122  \\
3C~142.1 &  L & A & 1\farcs40 & 0.287  \\
3C~175   &  L & A &  1\farcs37& 0.211  \\
3C~187   & L  & A &  1\farcs36 & 0.322  \\
3C~207   & L  & A & 1\farcs14 & 0.143  \\
3C~215   & L  & A & 1\farcs44 & 0.148 \\
         & C  & A & 0\farcs39 & 0.296  \\
3C~225B  & L  & A & 1\farcs20 & 1.15  \\
3C~226   & L  & A & 1\farcs29 &  6.99 \\
3C~228   & X  & B/C &  2\farcs25& 1.14  \\
3C~275   & L  & A & 1\farcs36 & 0.118  \\
3C~275.1 & L  & A & 1\farcs35 & 0.116  \\
3C~277.2 & L  & A & 1\farcs35 &  0.192 \\
3C~293.1 & C & A &  0\farcs40 &  0.038 \\
3C~306.1 & L  & A & 1\farcs50 & 0.519  \\
3C~327.1 & L  & A & 1\farcs47 & 6.40  \\
3C~334   & L  & A & 1\farcs26 &  0.245 \\
3C~434   & L  & A & 1\farcs38 & 0.106  \\
3C~435A  & L  & A & 1\farcs33& 0.266  \\
         & C  & A & 0\farcs39 & 0.187   \\
3C~455   & C  & A & 0\farcs43 & 0.118  \\
\hline                   
\end{tabular}
\end{table} 

\end{appendix}

\end{document}